\documentclass[%
 reprint,
nofootinbib,
 amsmath,amssymb,
 aps,
prb,
]{revtex4-2}

\usepackage{graphicx}
\usepackage{dcolumn}
\usepackage{bm}
\usepackage{mathrsfs}
\usepackage[T1]{fontenc}
\usepackage{scalerel}

\usepackage{siunitx}
\usepackage{physics}
\AtBeginDocument{\RenewCommandCopy\qty\SI}
\DeclareSIUnit\rydberg{\text{Ry}}
\DeclareSIUnit{\au}{a.u.}
\usepackage[normalem]{ulem} 
\usepackage[dvipsnames]{xcolor}

\newcommand*{\paral}{\stretchrel*{\parallel}{\perp}}
\DeclareMathOperator{\ee}{e}
\DeclareMathOperator{\ii}{Im}

\def\f{\phi} 

\def\d{\mathrm{d}} 
\def\br{\mathbf{r}}
\def\bq{\mathbf{q}}

\def\bk{\mathbf{k}}

\def\bG{\mathbf{G}^{\paral}}
\def\bg{\mathbf{G}}

\def\bkp{\mathbf{k}^{\paral}}

\begin{document}
\preprint{APS/123-QED}

\title{\textit{Ab initio} study of angle-resolved electron reflection spectroscopy of few-layer graphene}

\author{Ale\v{s} Pat\'{a}k}
\email{patak@isibrno.cz}
\author{Martin Zouhar}
\author{Ivo Konvalina}
\author{Eli\v{s}ka Materna Mikmekov\'{a}}
\author{Luk\'{a}\v{s} Pr\r{u}cha}
\author{Ilona M\"{u}llerov\'{a}}
\affiliation{
Institute of Scientific Instruments of the Czech Academy of Sciences, Kr\'{a}lovopolsk\'{a} 147, 612 00 Brno, Czech Republic
}%
\author{Anna Charv\'{a}tov\'{a} Campbell}
\author{Miroslav Valtr}
\affiliation{
 Czech Metrology Institute, Okru\v{z}n\'{i} 31, 638 00 Brno, Czech Republic
}%
\author{Michal Hor\'{a}k}
\author{Vlastimil K\v{r}\'{a}pek}
\affiliation{
Central European Institute of Technology, Brno University of Technology, Purky\v{n}ova 123, 612 00 Brno,
Czech Republic
}
\author{Eugene Krasovskii}
\affiliation{
Departamento de Pol\'{i}meros y Materiales Avanzados: F\'{i}sica, Qu\'{i}mica y Tecnolog\'{i}a, Universidad del Pais Vasco/Euskal Herriko Unibertsitatea, 20080 Donostia/San Sebasti\'{a}n, Basque Country, Spain}
\affiliation{
Donostia International Physics Center (DIPC), 20018 Donostia/San Sebastián, Basque Country, Spain.}
\affiliation{
IKERBASQUE, Basque Foundation for Science, 48013 Bilbao, Basque Country, Spain.
}

\date{\today}

\begin{abstract}
We present \textit{ab initio} theory for electron reflection spectroscopy of few-layer graphene for arbitrary angles of incidence.
The inelastic effects are included in a consistent way using the optical potential retrieved from \textit{ab initio} simulations of electron energy-loss spectra.
We demonstrate a significant impact of inelastic effects even for single-layer graphene.
Next, we address the ability of the electron reflection spectroscopy to determine specific parameters of graphene including not only the number of layers in the few-layer graphene but also the stacking type in the graphene multilayers, and to resolve moir\'{e} patterns in twisted graphene bilayers. 
We show that the predicted contrast, although significantly reduced by inelastic effects, is sufficient for the experimental detection of all considered parameters.
Our findings are corroborated by a fair correspondence of our theoretical predictions with experimental data, both our own and recently published by other authors.
\end{abstract}

\maketitle

\section{Introduction}\label{sec:intro}

Graphene with its remarkable properties has attracted enormous interest since its discovery~\cite{novoselov_electric_2004}.
The unique physical characteristics of graphene, such as high tensile strength, flexibility of the sheets, high conductivity and transparency to visible light, are attractive not only for basic research but for applied science and industry as well. Significant effort has been invested in opening a band gap in graphene, with the aim of developing graphene-based nanoelectronics in the future.
There are well known methods to achieve this goal. For example chemical doping~\cite{Adorno_2018} is one of the best studied methods, which quite often reduces
the superior charge carrier mobility in graphene.
Density functional theory (DFT) simulations show another possibility~\cite{guo_tuning_2008} to open the energy band gap by adding a second layer of graphene. 
Recently, stacking two sheets of graphene with the interlayer twist corresponding to the so-called magic angle of about \SI[number-unit-product={}]{1.1}{\degree} led to the discovery of superconductivity of the twisted bilayer graphene~(TBG)~\cite{cao_unconventional_2018} for a critical temperature up to \SI{1.7}{K} as predicted in Ref.~\cite{bistritzer_moire_2011} several years before the experiment.
Interesting new physics also arises when graphene stacks of three or more layers are considered.
It was reported~\cite{cao_pauli-limit_2021} that twisted trilayer graphene exhibits superconductivity for a specific magic angle as well; furthermore, it occurs way above the Pauli limit.
Twisted double bilayer graphene~(TDBG), i.e., two mutually twisted sheets of AB-stacked bilayer graphene, has been investigated~\cite{haddadi_moire_2020} and there is
evidence of the presence of spin-polarized ground state~\cite{cao_tunable_2020}.
Different correlated and topological phases can be studied and controlled in these systems.
These unique properties demonstrate that few-layer graphene (FLG) is an interesting subject of scientific investigation.
Consequently, the field of two-dimensional (2D) materials has expanded from monolayers to encompass multilayer systems and heterostructures, exhibiting electronic, mechanical, and optical properties that are significantly influenced by the number of layers, their stacking order, interlayer spacing, and twist.
Hence, high-precision characterization of these structural parameters is of utmost importance.

A plethora of characterization methods exists, each with their advantages and disadvantages.
Optical microscopy and Raman spectroscopy are widely used.
To overcome limitations in spatial resolution given by the wavelength of light, one can utilize atomic force microscopy or scanning tunneling microscopy~(STM).
These are powerful techniques, which enable studies of twisted bilayer graphene, its moir\'{e} patterns and topographic corrugations at atomic resolution~\cite{brihuega_unraveling_2012}.
However, scaling up the size of graphene for graphene-based nanoelectronics introduces further challenges for analytical techniques.
The resulting areas of interest (larger than \SI{1}{\um}) are typically too extensive to be assessed with STM~\cite{ndiaye_structure_2008}.
Therefore, one can try to exploit small de Broglie wavelength of electrons and use the versatility of electron microscopes. 

Transmission electron microscopy (TEM) and electron energy loss spectroscopy (EELS) have been used to 
differentiate the number of graphene layers~\cite{eberlein_plasmon_2008}, their stacking order~\cite{ping_layer_2012} and twist~\cite{sung_stacking_2019}.
Standard TEMs operate at the primary voltage of several hundreds of keV which can cause the knock-on radiation damage of 2D materials.
The radiation damage can be avoided by lowering the landing energy of electrons or by using the aloof spectroscopy, the latter in the case of EELS.
Nowadays, the TEMs primary voltage can be lowered to $\approx\,$\SI{60}{\kilo \volt} while keeping atomic resolution and energy resolution of EELS sufficient to detect phonons ($\approx\,$\SI{10}{\milli \eV})~\cite{krivanek_atom-by-atom_2010, krivanek_vibrational_2014}. 
The atomic resolution is typically restricted to small areas (lateral dimension of $\approx\,$\SI{20}{\nano \metre}). 

To overcome the small area limitation, one can employ methods based on electron diffraction to cover larger regions of interest. 
It is possible to measure the unoccupied electronic band structure of 2D materials with a technique based on low energy electron microscopy (LEEM) and still have a spatial resolution $\approx\,$\SI{10}{\nano \metre}~\cite{jobst_nanoscale_2015,jobst_quantifying_2016}.
We note that such measurements were done in ultra-high vacuum (UHV).
Fortunately, these measurements are feasible, at least in the case of normal incidence of electrons, even in a scanning electron microscope (SEM) after a proper sample preparation to avoid contamination by crosslinked hydrocarbons~\cite{materna_mikmekova_low-energy_2019}. 
Pixelated detectors, which are currently under development for SEM~\cite{STEM2DPAD}, are expected to enable the determination of the angle of incidence and the in-plane momentum of electrons obtained by tilting the electron beam from shifts of the central spot registered by the detector. 
Thus all labs equipped with a SEM can do this characterization with good spatial resolution (\SI{0.8}{\nano \metre} at \SI{200}{\eV}~\cite{frank_imaging_2015}) even if they do not posses a special UHV LEEM device, since the stage bias based on the cathode lens principle~\cite{mullerova_approaches_1992} has become a standard feature in SEMs.
We stress that the flexibility of electron microscopes does not end here. 
Electron vortex beams (see, e.g., Ref.~\cite{patak_visualizing_2019} for an overview in the field of electron microscopy and other research areas) which have already been implemented in SEM~\cite{Tom.UM.2021} may be employed for an analysis or nanomanipulation with precision that is inaccessible to optical beams.

To efficiently employ these techniques, it is important to understand the relation between the experimental response and the parameters of FLG.
Several groups have already measured and analyzed very low-energy ($\lessapprox\,$\SI{30}{\eV}) electron specular reflectivities of FLG, which show interesting quantized energy and thickness-dependent oscillations~\cite{hibino_microscopic_2008,sutter_epitaxial_2008,riedl_quasi-free-standing_2009,locatelli_corrugation_2010,feenstra_low-energy_2013,geelen_ev-tem_nodate}.
These oscillations were theoretically explained at different levels of sophistication--by Fabry-P\'{e}rot type interference, using tight-binding calculations or DFT simulations. 
The \textit{ab initio} momentum-resolved low-energy electron spectrum of graphene was obtained with a full-potential linear augmented plane waves method~\cite{nazarov_scattering_2013}.
Low-energy reflectivities of normally-incident electrons from graphene and FLG were calculated 
with the projector-augmented wave
method and the generalized-gradient approximation for the
density functional employing VASP in Ref.~\cite{feenstra_low-energy_2013}.
These spectra were simulated over the energy range of 
0 to \SI{8}{\eV} 
where the layer-dependent oscillations
are most prominent. The energy range was enlarged in Ref.~\cite{gao_inelastic_2015}, which demonstrates how the inclusion of inelastic effects suppresses a second set of  layer-dependent oscillations in the \SIrange{14}{21}{\eV} range.
A computationally efficient method based on DFT~\cite{mcclain} was applied to calculate the reflectivity spectra of FLG recorded by LEEM for off-normal angles of electron incidence within the local density approximation (LDA) using an ultrasoft pseudopotential in \textsc{Quantum ESPRESSO}.

An understanding of the interaction of low-energy electrons with FLG may also be relevant in other fields beyond the field of electron microscopy.
The spectral and angular distributions, especially those of
secondary and elastically reflected electrons, are of practical interest
in quantitative electron-cloud simulations, since these low energy electron clouds may cause instabilities in accelerators.
Different strategies to suppress secondary electron yield (SEY) are investigated, e.g., in the Large Hadron Collider~\cite{Cimino,Schulte}.
This may be achieved by suitable conditioning, scrubbing or coating of relevant vacuum components. Carbon-based coatings produced by scrubbing reduce SEY almost to values measured for Highly Ordered Pyrolytic Graphite~\cite{Cimino}.
Similarly, simulating the consequences of electron impacts on accelerator grids in fusion reactors, such as the International Thermonuclear Experimental Reactor, requires the knowledge of the energy and spatial distribution of reflected and secondary electrons~\cite{Fubiani}.
Therefore, angle-resolved low-energy electron simulations can improve 
assumptions on the spectral and angular distributions of the emitted electrons in the aforementioned facilities.

However, we are not aware of any DFT study of FLG for general incidence reflectivity spectra, where the inelastic effects are included.
Therefore the main purpose of this paper is to simulate these spectra and to include the inelastic effects. It should be noted that 
the electron-phonon coupling is not considered in the present paper, as it is considerably less important than the
electron-electron scattering~\cite{Echenique_CP251_2000}. 
The optical potential included in the presented method
phenomenologically 
treats the electron-electron scattering,
i.e., the finite lifetime of the quasi-particles.

The structure of this paper is the following.
In Sec.~\ref{sec:normal}, the theoretical normal incidence reflectivities of FLG are presented.
The reflectivities were also simulated using a rigorous embedding method, 
which for brevity we will refer to as the APW method for its central ingredient--
augmented plane wave (APW) representation of the wave functions~\cite{Eugene_nanomaterials}.
These simulations serve as a benchmark for the simple supercell Bloch wave approach as described in Appendix~\ref{sec:matching}.
In Sec.~\ref{sec:twist}, these methods are then extended to
stackings different from the AB-stacking for bilayer and four-layer systems composed of graphene layers and a proposition is made that moir\'{e} superlattices in TBG and TDBG should be measurable in electron microscopes as well.
Subsequently, the general incidence reflectivities of FLG can be found in Sec.~\ref{sec:ares}.
Momentum-resolved EELS, obtained by the many-body perturbation theory (MBPT) and required to include the inelastic effects via the optical potential into reflectivities, is discussed in Sec.~\ref{sec:momentum.resolved.EELS}.
Our EELS measurements at the vicinity of the $\Gamma$ point are compared with the simulations in Sec.~\ref{sec:EELS.at.Gamma}.
A good agreement between calculated and measured EELS verifies the plausibility of the optical potential constructed from simulation as detailed in Sec.~\ref{sec:v.optical.and.IMFP}. In addition, this section presents a by-product of the optical potential computation, namely the \textit{ab initio} inelastic mean free path (IMFP) of graphene.
The main text inevitably ends with the summary and conclusion in Sec.~\ref{sec:conclusion}.
That section is followed by several appendixes discussing or presenting details of the involved methods and corresponding settings. 

\section{Normal incidence electron reflectivity of FLG}\label{sec:normal}

Our simulations are based on a supercell Bloch wave approach to calculate low-energy electron reflectivity~\cite{mcclain}.
The supercell of thickness $L$ consists of a finite slab surrounded by a vacuum region.
The reflectivities are given by matching plane waves corresponding to a sum of incoming and reflected ($\psi_L$) wave and transmitted ($\psi_R$) electron 
wave to the Kohn-Sham wave functions calculated for the crystal ($\psi_S$) in the supercell, see Fig.~\ref{scheme} and details in Appendix~\ref{sec:matching}.
\begin{figure}[h]\begin{center}
\includegraphics[width=0.95\linewidth]{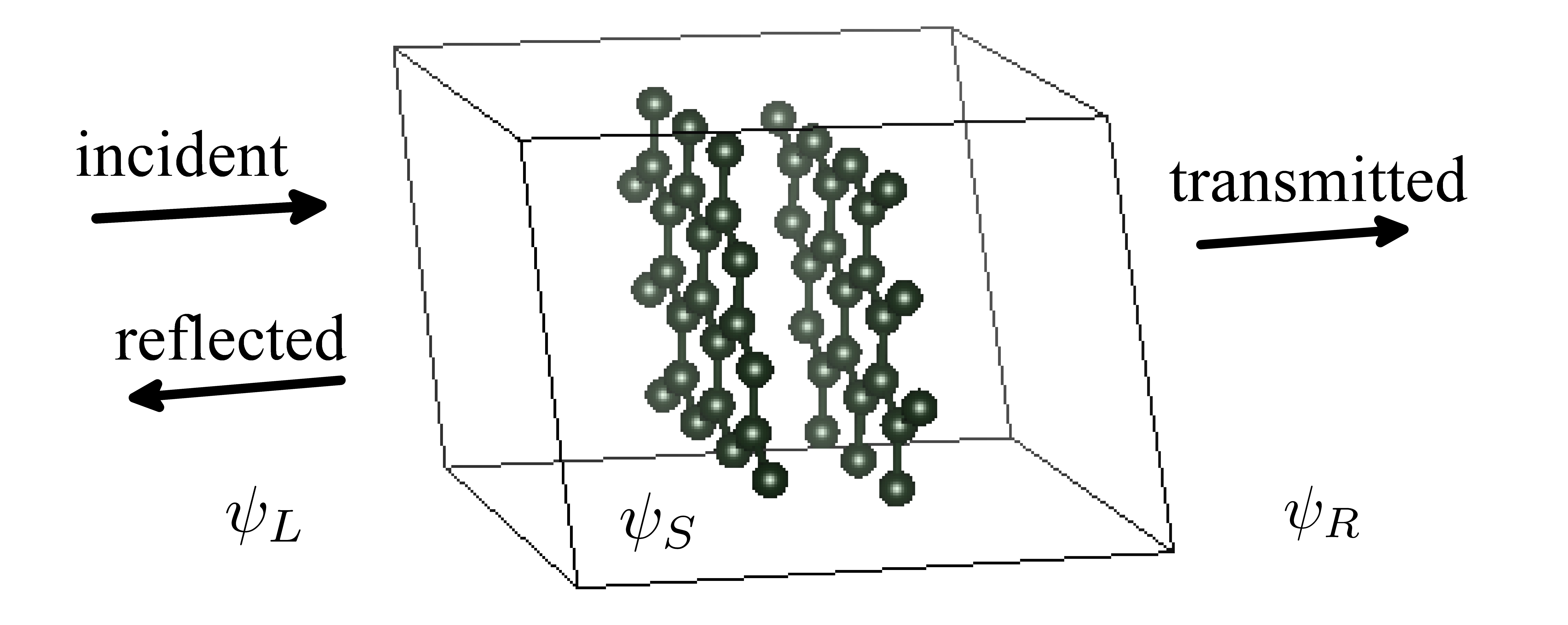}
\caption{Schematic displaying the supercell for the wave matching process. For further details regarding the waves $\psi_L$ (incoming and reflected), $\psi_S$ and $\psi_R$ (transmitted), please refer to Appendix~\ref{sec:matching}. The bilayer graphene structure is produced by \textsc{VESTA}~\cite{vesta}.}
\label{scheme}
\end{center}\end{figure}

Relaxed values of structural parameters are used in our DFT simulations instead of the experimental ones.
It provides us with a consistent way to perform DFT calculations even without having a direct access to measured structural constants, as is the case e.g. in~Sec.~\ref{sec:twist}. For further details and converged parameters see Appendixes~\ref{sec:struc_opt} and \ref{sec:compdeta_refl}.

\begin{figure}[h]\begin{center}
\includegraphics[width=0.92\linewidth]{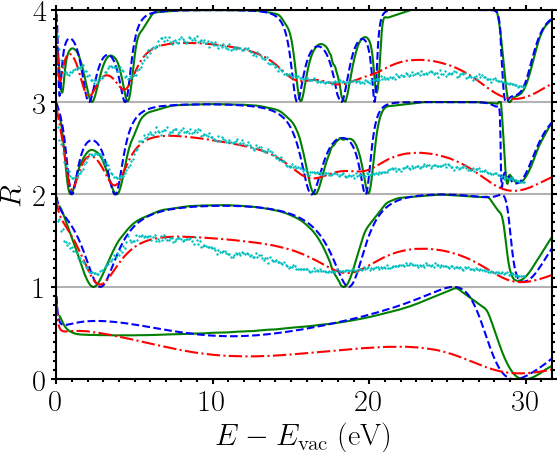}
\caption{Normal incidence FLG electron reflectivity $R$ spectra for $n = $ 1 to 4 layers, bottom to top, as a function of landing energy ($E_{\mathrm{vac}}$ being energy in vacuum), without inelastic effects using \textsc{Quantum ESPRESSO} as described in Appendix~\ref{sec:matching} (green solid lines) and the APW method (blue dashed lines), with included inelastic effects using the APW method (red dotted-dashed lines) and cyan symbols are experimental reflectivities from Ref.~\cite{jobst_nanoscale_2015} which were normalized solely by the mirror landing energy value (actually one energy step, \SI{0.1}{\eV}, below zero since the simulated reflectivities are not precisely equal to 1 at \SI{0}{\eV}). The calculated reflectivity spectra are shifted vertically by $n - 1$ for the sake of clarity and are separated by horizontal gray lines.
}\label{niflg}
\end{center}\end{figure}

Figure~\ref{niflg} displays our simulated specular (bright field) reflectivity versus energy of an electron beam normally-incident upon one to four layers of FLG. 
Pronounced local minima can be clearly seen in the reflectivity spectra. Since they correspond to transmission maxima, they are also called transmission resonances.
The number of layers, $n$, of a freestanding FLG corresponds to the count of the reflectivity minima in the low-energy range of about 
0 to \SI{6}{\eV}; 
the latter is equal to $n-1$.
This observation is of practical importance because it enables us to directly measure the number of layers $n$ in the freestanding multilayer graphene by LEEM.
The reason for this correspondence is that the wavefunctions responsible for the transmission resonances are localized in between the layers~\citep{feenstra_low-energy_2013}. 
Let us note that such a simple relationship can be disturbed by the presence of a substrate.
Such a disruption depends on the strength of the interaction of the substrate with the buffer layer (monolayer closest to the substrate having different properties). 
The agreement of simulations with the experimental data in Fig.~\ref{niflg} demonstrates that the positions of the characteristic local minima can be simulated well even with the simple supercell Bloch wave approach.
The simulated reflectivities without inelastic effects obtained using the simple approach as described in Appendix~\ref{sec:matching} and using the APW method are quite comparable.
This is no longer true when the inelastic effects are included.
The reflectivities obtained using the simple supercell Bloch wave approach appear to be overdamped while the APW method produces good agreement with the experiment reported in Ref.~\cite{jobst_nanoscale_2015}.
We refer the reader to Appendix~\ref{sec:inelastic} for details.
The FLG studied in Ref.~\cite{jobst_nanoscale_2015} was deposited on a SiC substrate.
Including the effects of substrates in \textit{ab initio} calculations is a formidable task beyond the scope of this paper.
Despite the numerous approximations employed in the theoretical calculations, it can be reasonably concluded that the agreement between the simulations and experiment is excellent.

Let us conclude this section by noting that similar characteristic oscillations in low energy electron reflectivities also appear in other 
materials.
Transition metal dichalcogenides (TMDs) are a typical example~\cite{Sergio}.
Spin-orbit interaction is strong for TMDs~\cite{mikmekova_low} and the use of fully relativistic pseudopotentials is needed; the supercell Bloch wave approach can be generalized accordingly.
Metal films are another example. The spin-dependent reflectivity measurement of low energy electrons for Fe films on W(110) has been reported in Refs.~\cite{Zdyb_2013,BAUER2020} and the normal incidence reflectivity and related density of states of free standing Fe-BCC(001) slab has been analyzed using DFT in Ref.~\cite{Fe-clanek}.

\section{Normal incidence electron reflectivity of TBG and TDBG}\label{sec:twist}

We will now address the possibility of observing and identifying TBGs and TDBGs in SEM by discussing differences in the corresponding reflectivity spectra.
TBGs exhibit a characteristic spatial distribution of their properties, such as the local interplane spacing~\cite{uchida_atomic_2014}, forming the so-called moir\'{e} lattice.
Typical lattice parameters read units of nanometers.
For instance, STM measurements give a moir\'{e} lattice distance around~\SI{14}{\nm} for a twist angle \SI[number-unit-product={}]{0.99}{\degree}~\cite{choi_electronic_2019}.
Such features shall be resolvable in SEM even at very low landing energies.
For example, the spatial resolution \SI{4.5}{\nm} is normally achieved with gold on carbon test specimens in SEMs like Magellan~400 for landing energy \SI{20}{\eV} nowadays.

TBGs are usually studied by STM which provides valuable information about electronic band structure, correlation effects, and topography of a sample~\cite{choi_electronic_2019}.
The stacking of graphene layers in TBGs continuously varies in the plane between areas with AA-like and AB-like stacking types.
The contrast observed in STM images stems from the different response of the regions with AB-stacking, where the interlayer distance is the shortest and the regions with AA stacking, where the interlayer distance is the longest.
There are some studies where transmission electron microscopes are used to study moir\'{e} structures in TBG like Refs.~\cite{lu_twisting_2013,latychevskaia_moire_2019}, where low-energy electrons enable observation of moir\'{e} diffraction peaks.

\begin{figure}[h]\begin{center}
\includegraphics[width=0.92\linewidth]{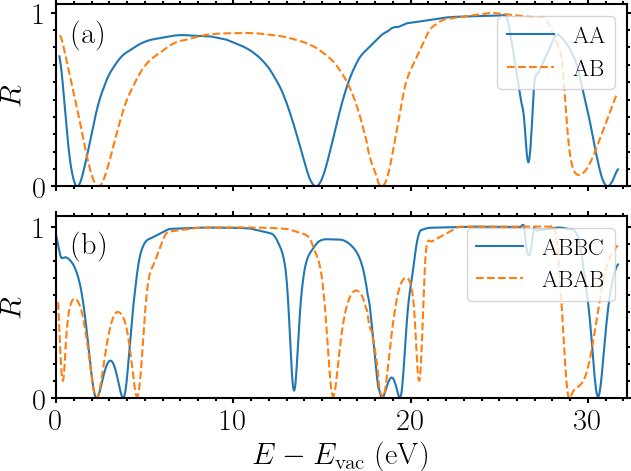}
  \caption{Comparison of the normal incidence FLG electron reflectivity $R$ spectra without inelastic effects as a function of landing energy ($E_{\mathrm{vac}}$ being energy in vacuum), simulated using the simple supercell Bloch wave approach as described in Appendix~\ref{sec:matching}, for different stacking: (a) two layers with AA (solid blue line) and AB (dashed orange line) stacking; (b) four layers with ABBC (solid blue line) and ABAB (dashed orange line) stacking.
}\label{twisted}
\end{center}\end{figure}

Based on our DFT relaxed lattice parameters, see Appendix~\ref{sec:struc_opt}, simulations of normal incidence electron reflectivity spectra of AA- and AB-stacked bilayer graphene are presented in Fig.~\ref{twisted}(a).

In this analysis, inelastic effects are not included. It is anticipated that both spectra would exhibit similar damping characteristics to those observed in the AB bilayer graphene in Fig.~\ref{niflg}.
The reflectivity spectra of AA-stacked and AB-stacked bilayer graphene exhibit
pronounced contrast for the landing energies in the ranges from \SIrange{1}{3}{\eV} and from \SIrange{13}{19}{\eV}.
For example, a pronounced contrast between the AA and AB regions is predicted for a landing
energy of approximately \SI{2.5}{\eV}, where the AA regions would be bright and the AB regions
would be dark (alike in usual STM image) when imaged using reflected electrons.

The question of whether the contrast between AA- and AB-stacked
graphene bilayers is preserved in TBG will be addressed now.
The reflectivities and the transmissivities are found to depend strongly on the interlayer distance. A straightforward rationale can be derived from the perspective of a Fabry-P\'{e}rot interferometer.
In the context of a Fabry-P\'{e}rot interferometer, maximal transmissivity is achieved when the optical path length difference between each transmitted beam is an integer multiple of the wavelength.
A quantum mechanical reasoning for a square well based on the Schr\"{o}dinger equation~\cite{merzbacher} yields the same result.
The maximum transmissivity is observed when twice the well width $a$ that a particle traverses on the way through the well and back is an integral number multiple of de Broglie wavelengths.
This results in the incident and reflected waves being in phase, thereby reinforcing each other. Consequently, the magnitude of the corresponding wave vector $k$ is proportional to~$a^{-1}$. Therefore, the corresponding kinetic energy 
 is proportional to $k^{2}\propto a^{-2}$. 
 Identifying $a$ with the interlayer distance implies that one should expect that the energy of transmission resonances will be lower for larger
interlayer distances. This straightforward line of reasoning is in agreement with the \textit{ab initio} simulations depicted in Fig.~\ref{twisted}(a).
When considering the twist angle within the following ranges 
0 to \SI{5}{\degree} and from \SIrange{55}{60}{\degree}, 
areas with interlayer distance close to homogeneous graphene bilayers with AA or AB-stacking appear~\cite{uchida_atomic_2014}.
Consequently, the contrast
in the reflectivity predicted for the AA- and AB-stacked graphene bilayers shall be
observable also in TBG, where it shall allow visualization of the moir\'{e} patterns.

Similarly, in the case of TDBG there appear two stacking regions, ABBC and ABAB (or ABCA) regions,
with the former exhibiting maximal and the later minimal interplane distances, see Ref.~\cite{haddadi_moire_2020} and its Supporting Information. 
The normal incidence electron reflectivity spectra for the two stacked domains are presented in Fig.~\ref{twisted}(b).
Again, the spectra are quite different for the considered stackings, e.g., the difference in reflectivities is 0.5 at landing energy around \SI{3.5}{\eV}.

The normal incidence reflectivities of TBG have recently been examined~\cite{deJong_NatComm_2022}.
Their theoretical simulations, see their Fig.~1(d), indicate that there is an optimal incident energy \SI{37}{\eV} at which the different stackings become easily distinguishable due to high differences in the calculated reflectivity profiles.
They predict that the reflectivities of different stacking variants are practically identical below \SI{25}{\eV}, in contrast to our results in Fig.~\ref{twisted}(a).
This discrepancy can be attributed to the assumption that the interlayer distance of the AA- and AB-stacking regions was identical in Ref.~\cite{deJong_NatComm_2022}.
Since the corrugation is dependent on the twist angle, the identical interlayer distance would correspond to a twist angle close to \SI{30}{\degree}, where the corrugation vanishes, see Fig. 1 or Fig.~9 in Ref.~\cite{uchida_atomic_2014}.
However, for smaller twist angles the corrugation is significant, e.g. the twist angle close to \SI{3.89}{\degree} leads to the corrugation \SI{0.12}{\angstrom}, the double of which gives
the difference in the interlayer distances of the AA- and AB-stacking regions.
This difference \SI{0.24}{\angstrom} is quite close to our relaxed value of \SI{0.3}{\angstrom}.
The simulations presented in Fig.~\ref{twisted}, indicate that the contrast pertains for lower energies.
The implications of our simulations are supported by the Supplemental Material of Ref.~\cite{deJong_NatComm_2022}; Figs.~3(a) and 3(b), with experimental data, which show a reasonable contrast at \SI{17}{\eV}.

\section{Angle-resolved electron spectroscopy}\label{sec:ares}

In this section, the above specular reflection regime is generalized to incident electron beams with in-plane wave vectors along the path $M\Gamma K$ in the first Brillouin zone.
It is already well established that oscillations in normal incidence reflectivities or transmissivities can be used to determine the layer count of ultrathin crystalline materials.
Moreover, 
angle-resolved reflected-electron spectroscopy~(ARRES) 
can help us study unoccupied electron band structure above the vacuum level.
This has already been experimentally examined~\cite{jobst_nanoscale_2015} but the theoretical description was rather brief relying on a tight-binding reasoning.
This provides motivation for a more thorough \textit{ab initio} study of the off-normal incidence reflectivity which can serve as a starting point for a more in-depth explanation.

Once again, as in Sec.~\ref{sec:normal}, we compare the supercell Bloch-wave matching method and the APW method.
\begin{figure*}
\centering
\includegraphics[clip,trim=0 0 {4.0cm} 0,height=0.154\textwidth]{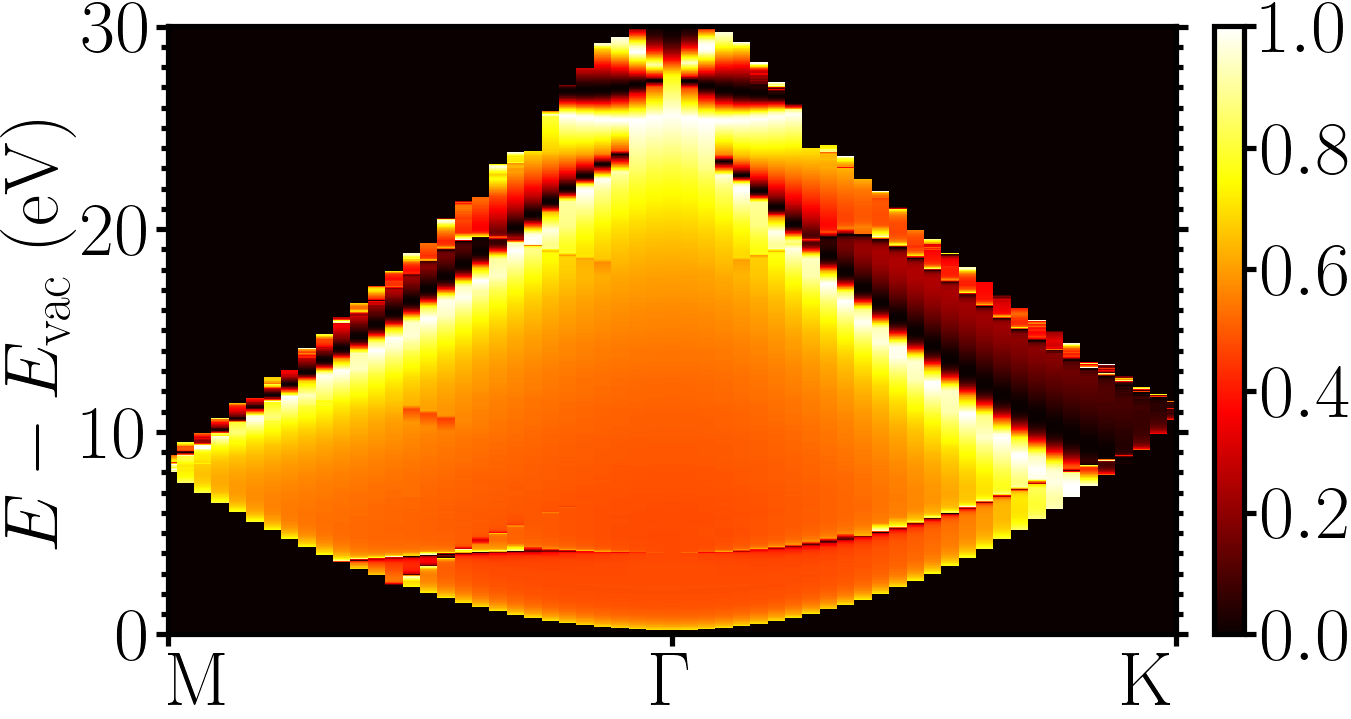}
\includegraphics[clip,trim={4.0cm} 0 {4.0cm} 0,height=0.154\textwidth]{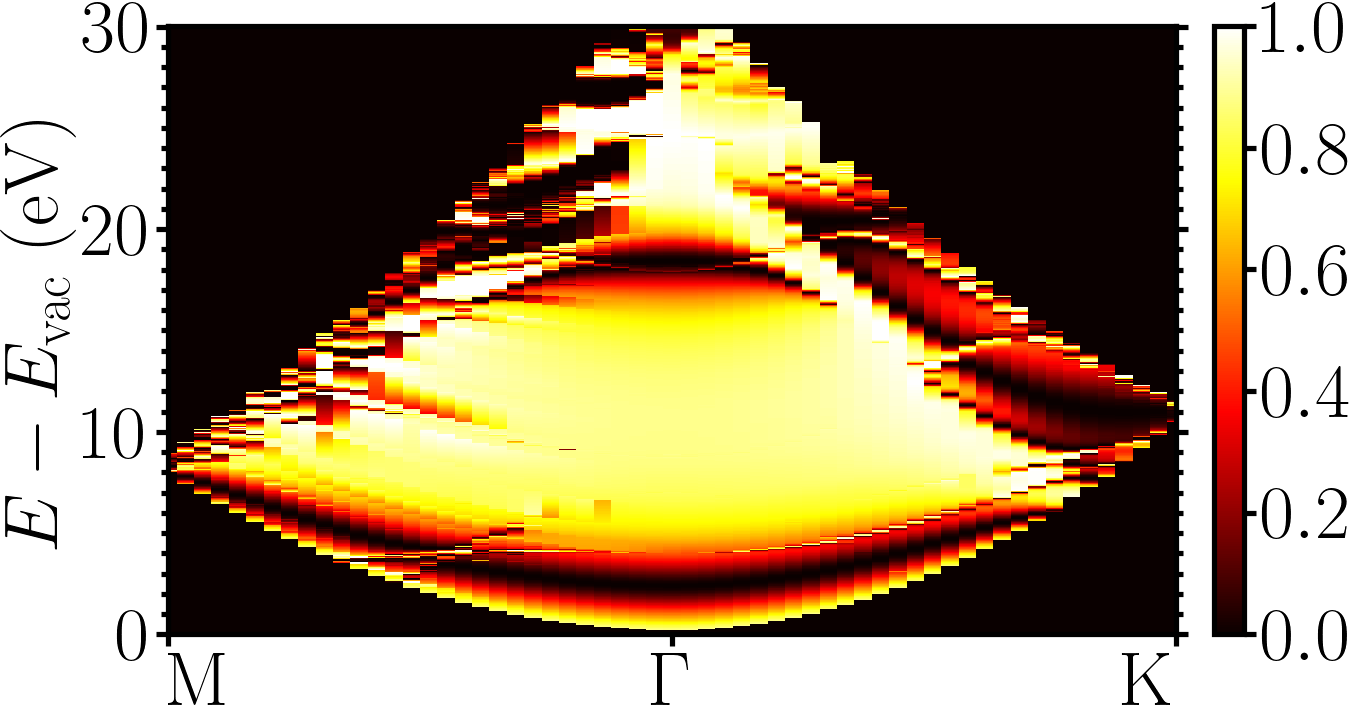}
\includegraphics[clip,trim={4.0cm} 0 {4.0cm} 0,height=0.154\textwidth]{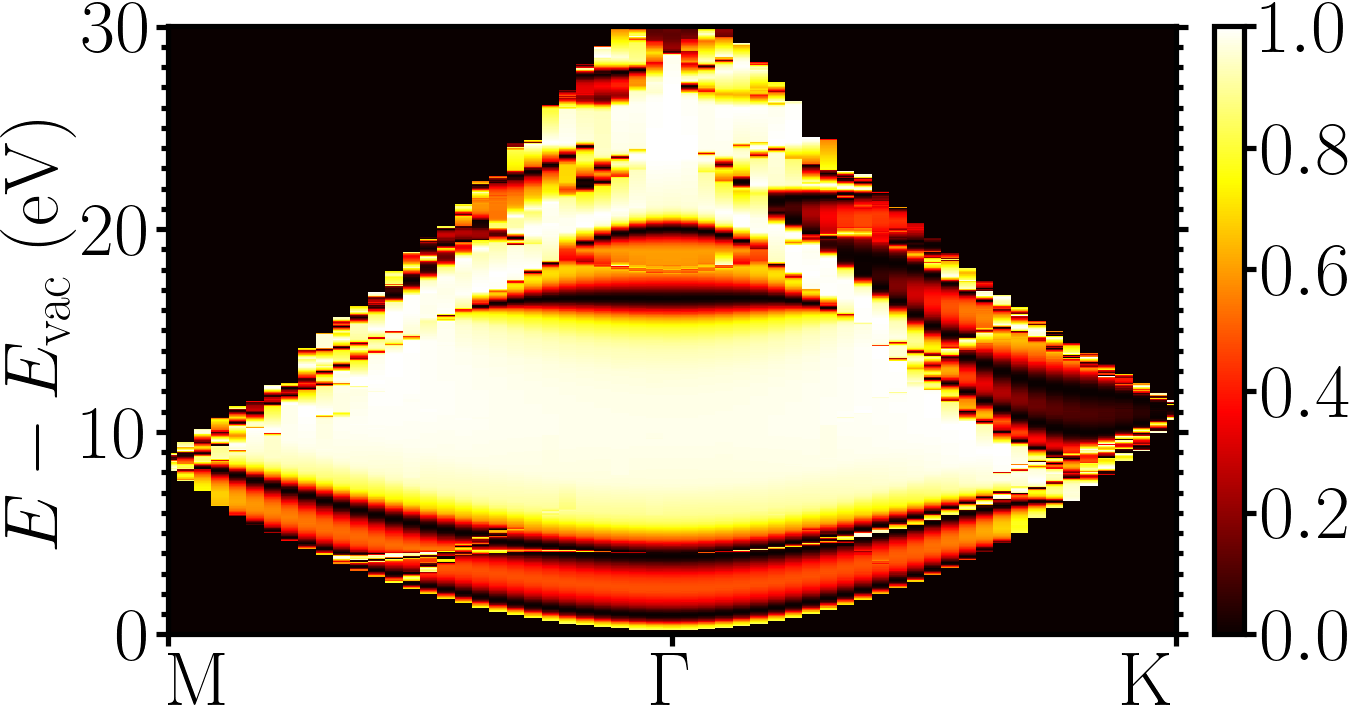}	
\includegraphics[clip,trim={4.0cm} 0 0 0,height=0.154\textwidth]{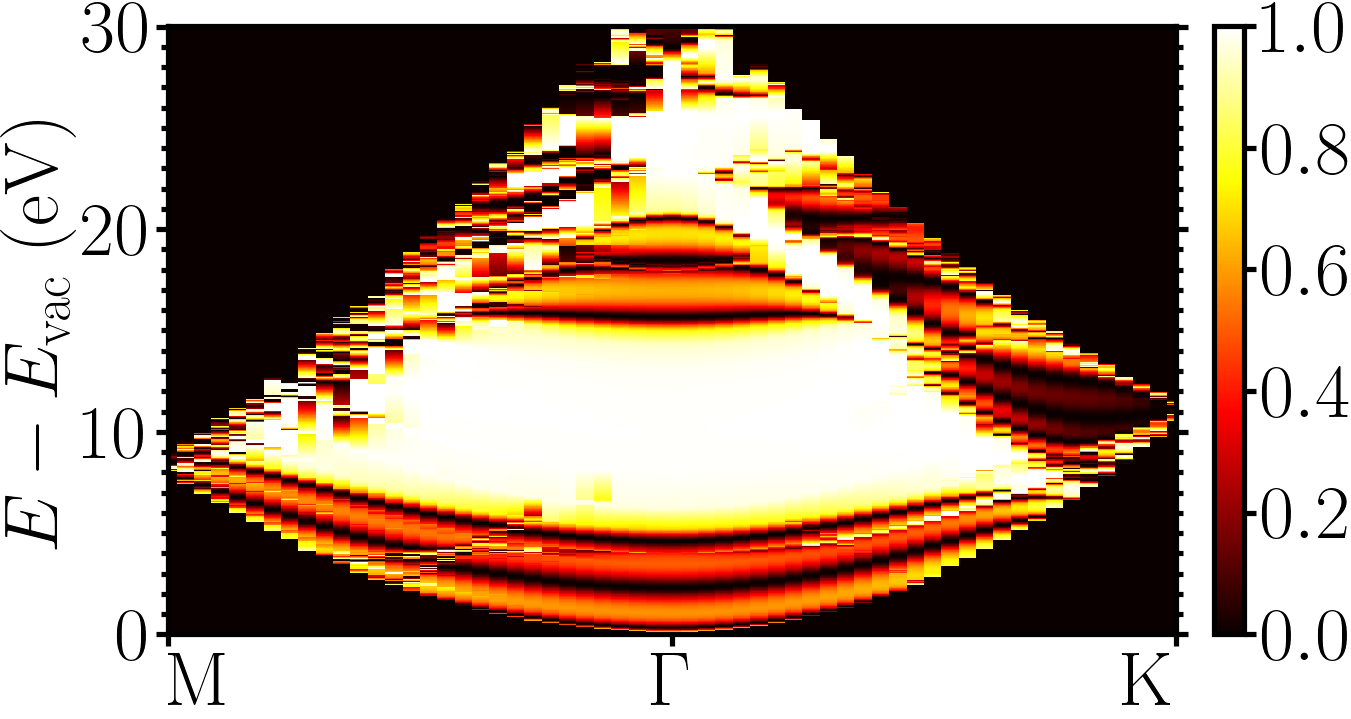}\\
\includegraphics[clip,trim=0 0 {4.0cm} 0,height=0.154\textwidth]{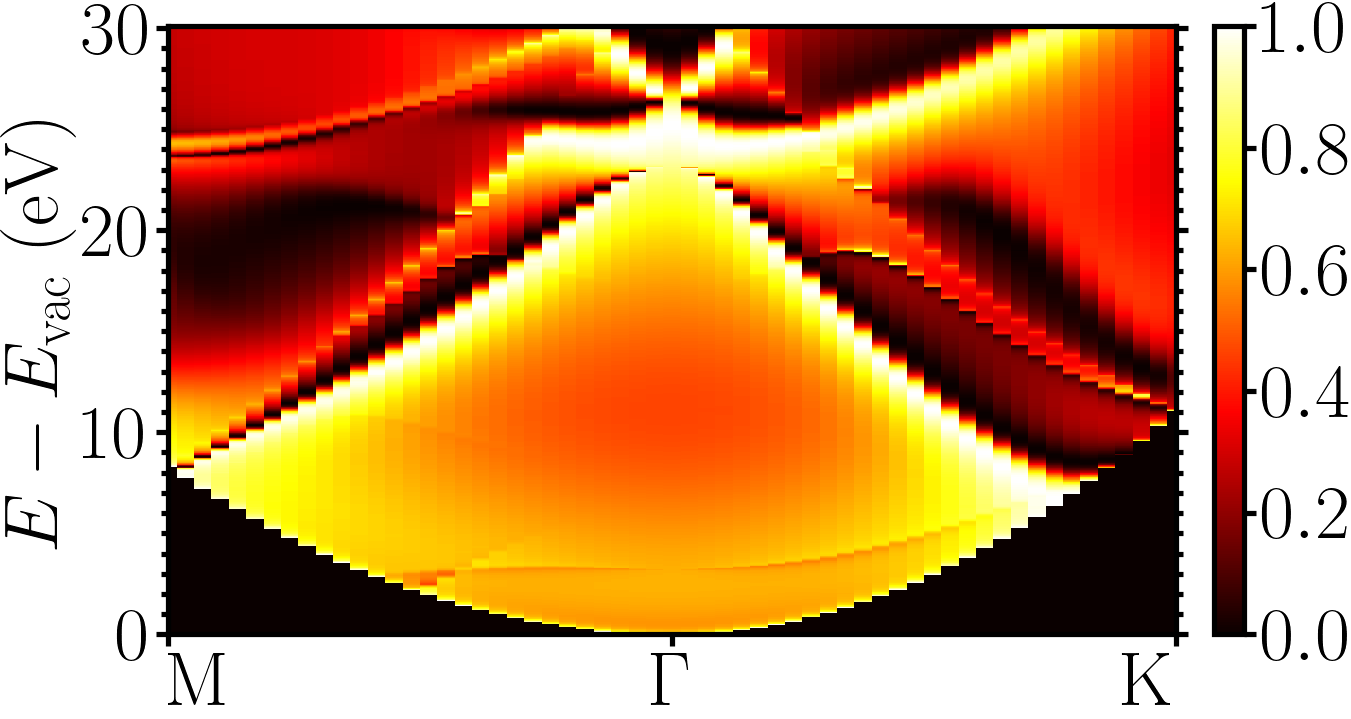}
\includegraphics[clip,trim={4.0cm} 0 {4.0cm} 0,height=0.154\textwidth]{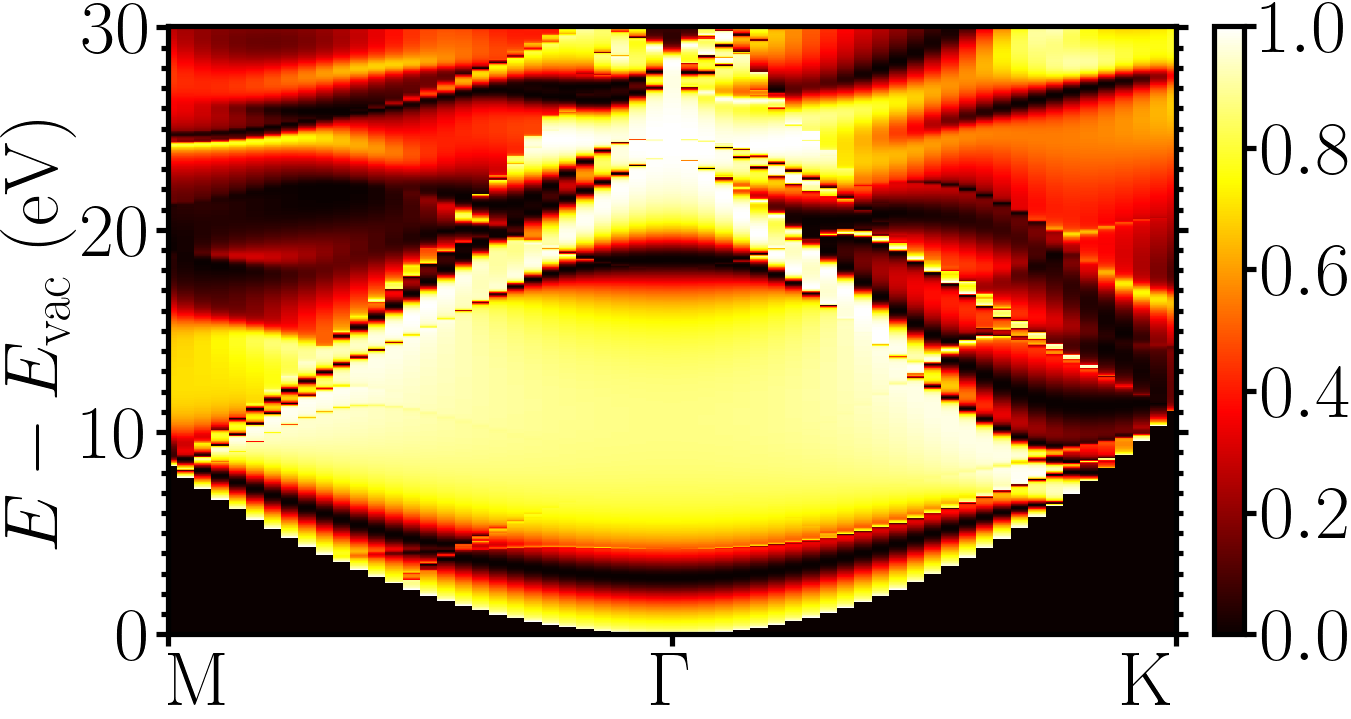}
\includegraphics[clip,trim={4.0cm} 0 {4.0cm} 0,height=0.154\textwidth]{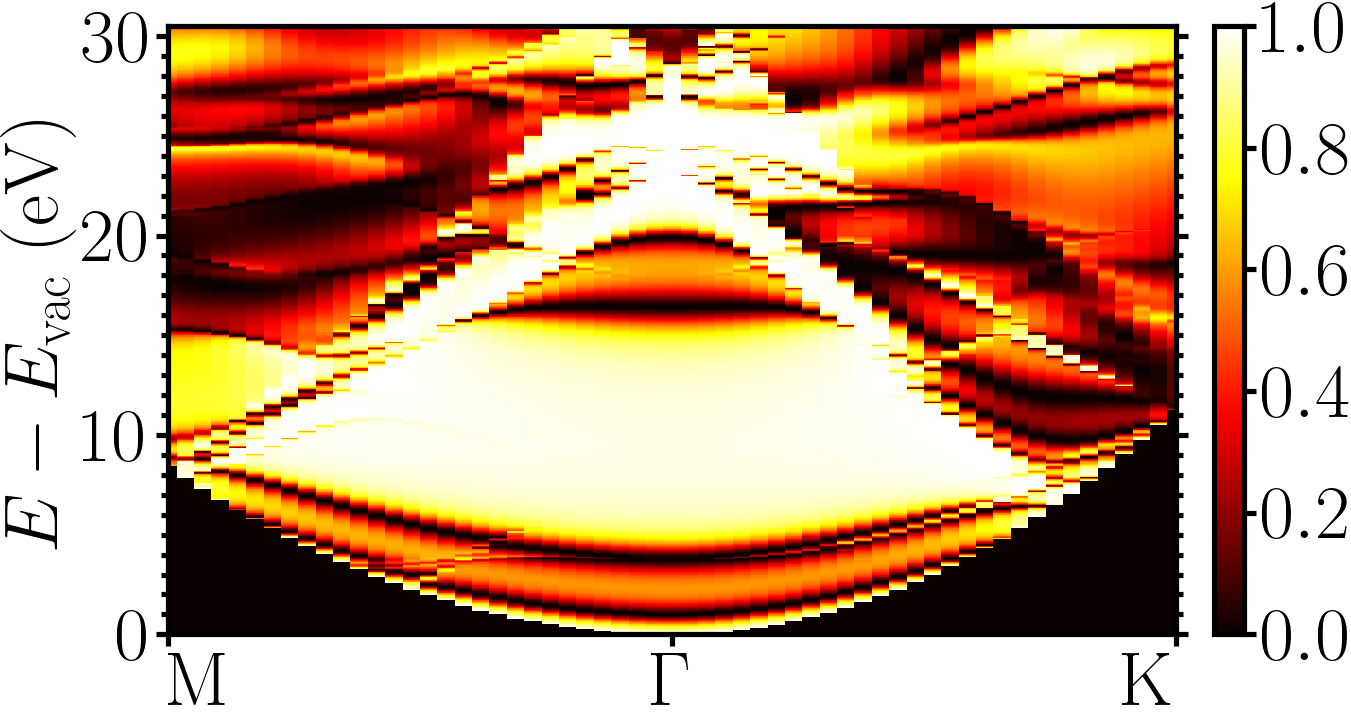}
\includegraphics[clip,trim={4.0cm} 0 0 0,height=0.154\textwidth]{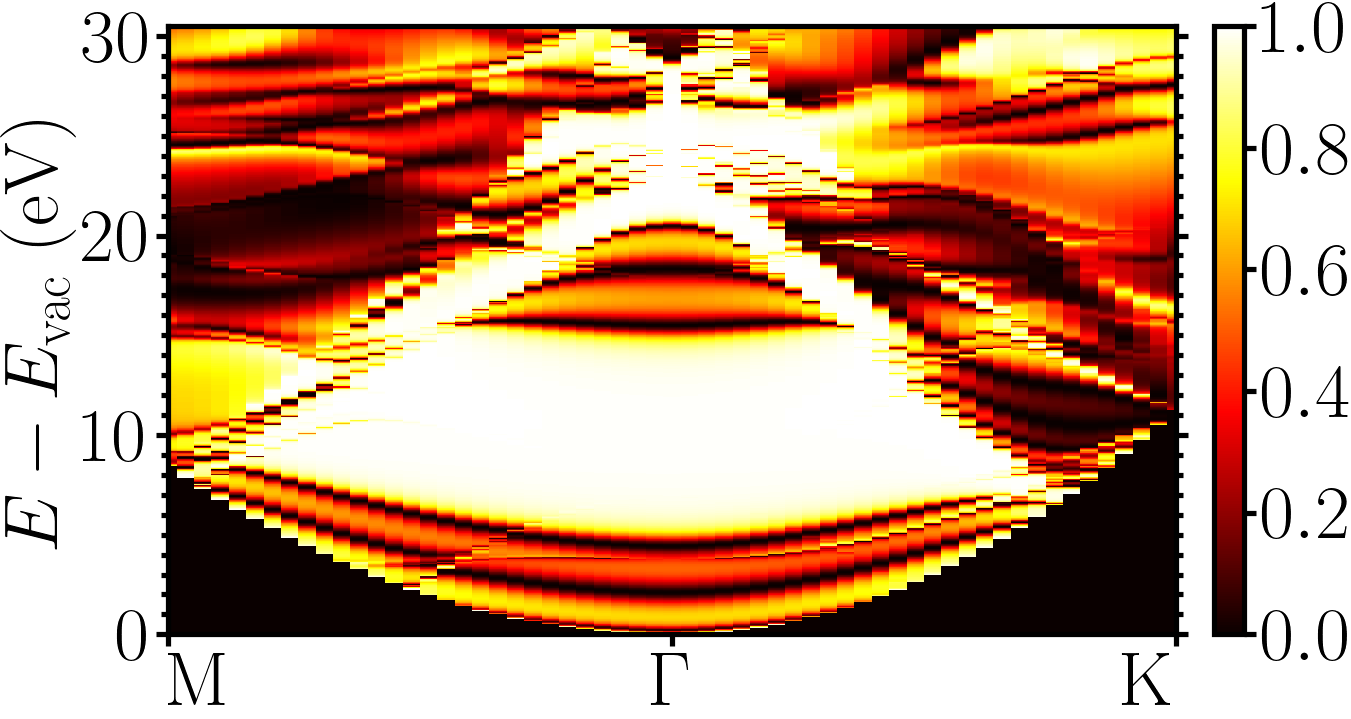}\\
\includegraphics[clip,trim=0 0 {4.0cm} 0,height=0.154\textwidth]{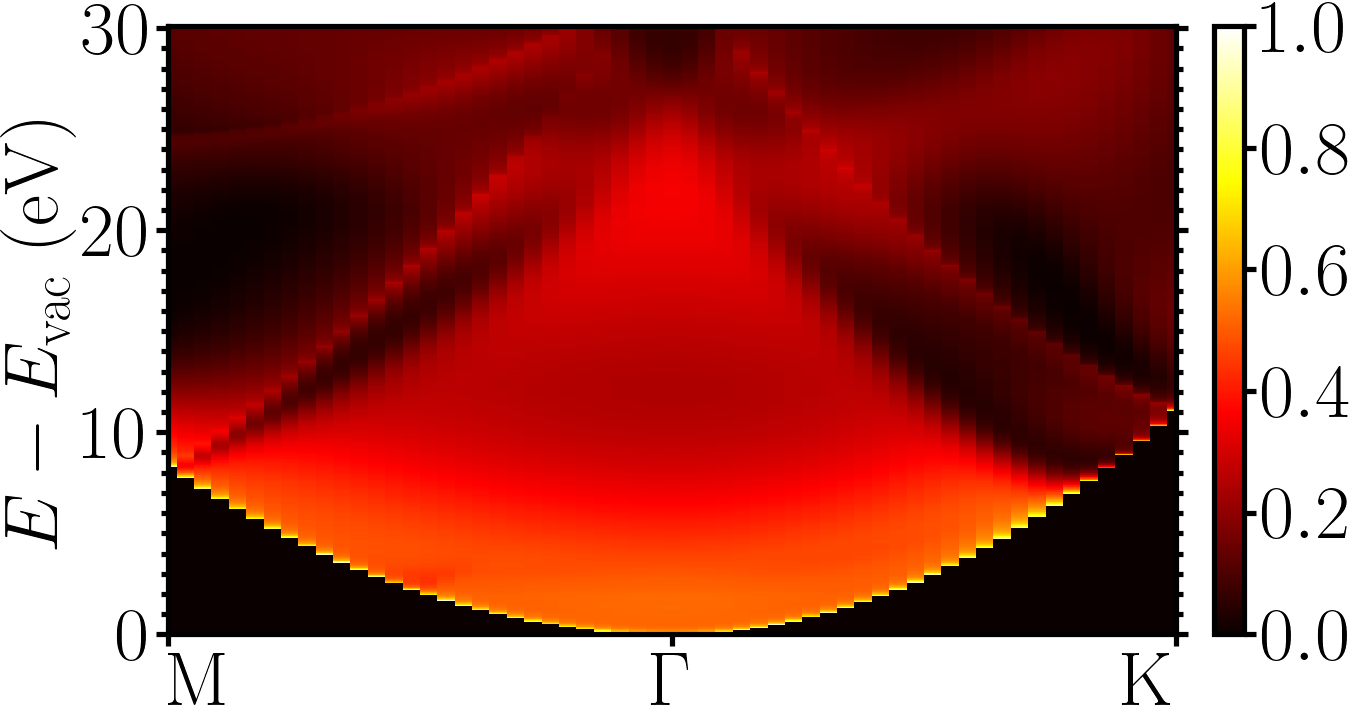}
\includegraphics[clip,trim={4.0cm} 0 {4.0cm} 0,height=0.154\textwidth]{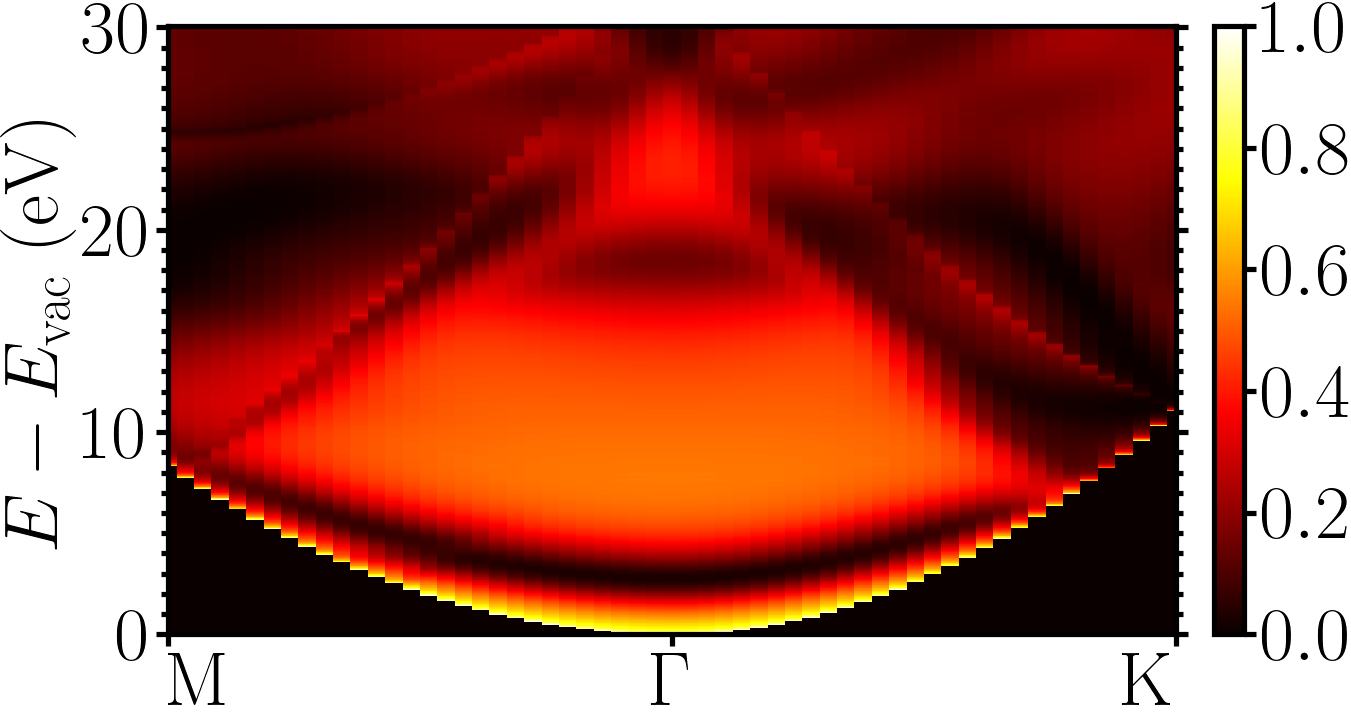}
\includegraphics[clip,trim={4.0cm} 0 {4.0cm} 0,height=0.154\textwidth]{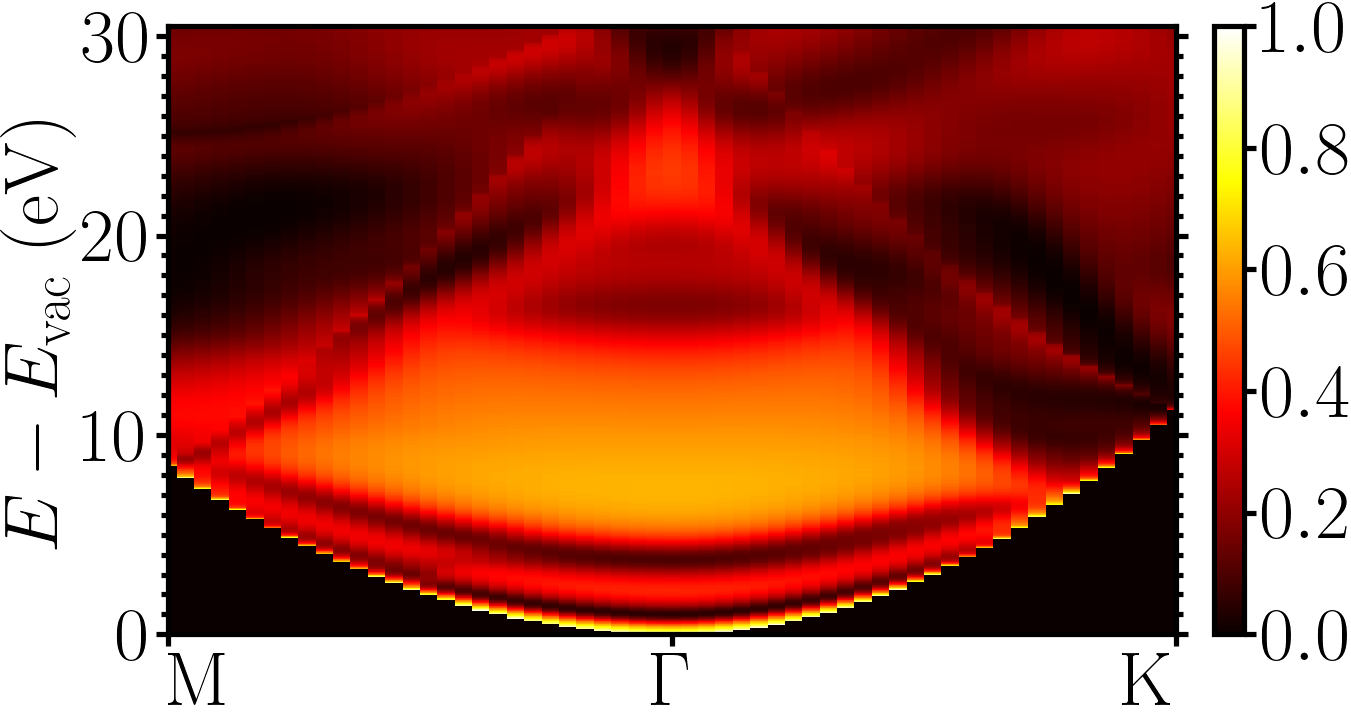}
\includegraphics[clip,trim={4.0cm} 0 0 0,height=0.154\textwidth]{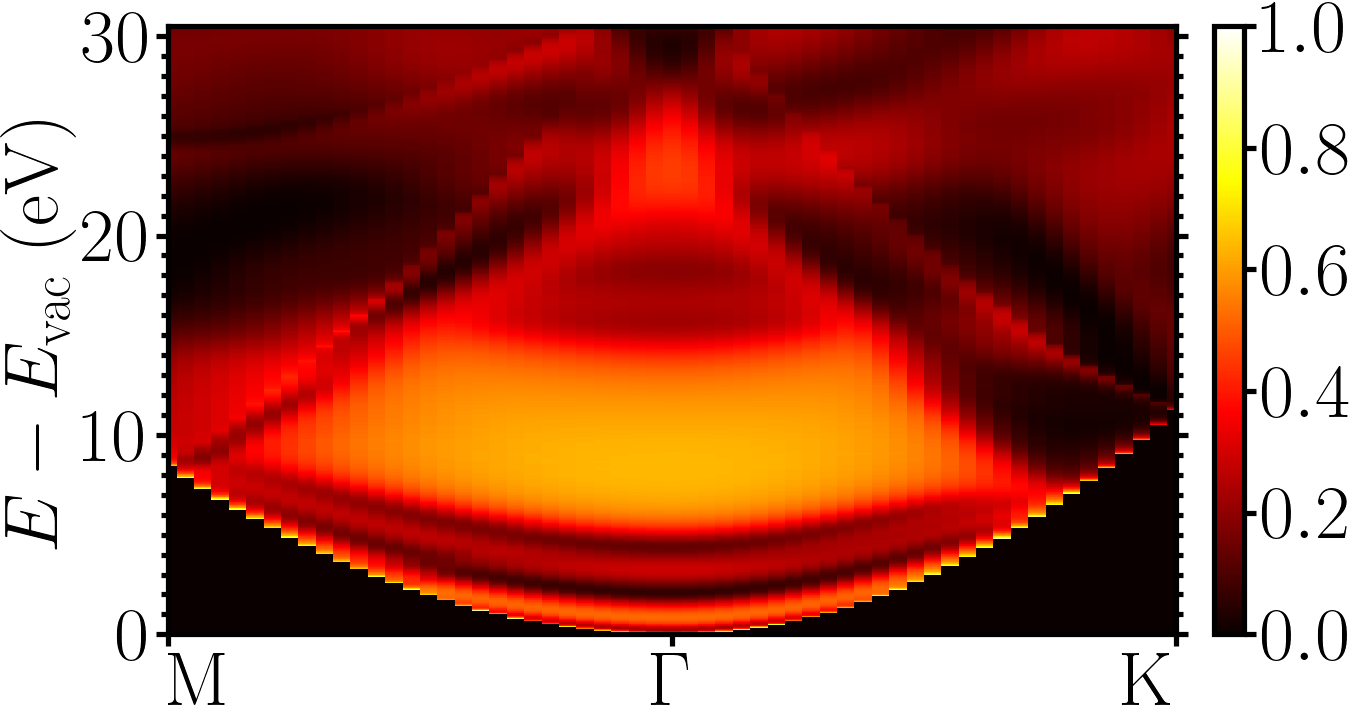}\\
\hspace{0.225\textwidth}
		\includegraphics[clip,trim=0 0 {4.0cm} 0,height=0.154\textwidth]{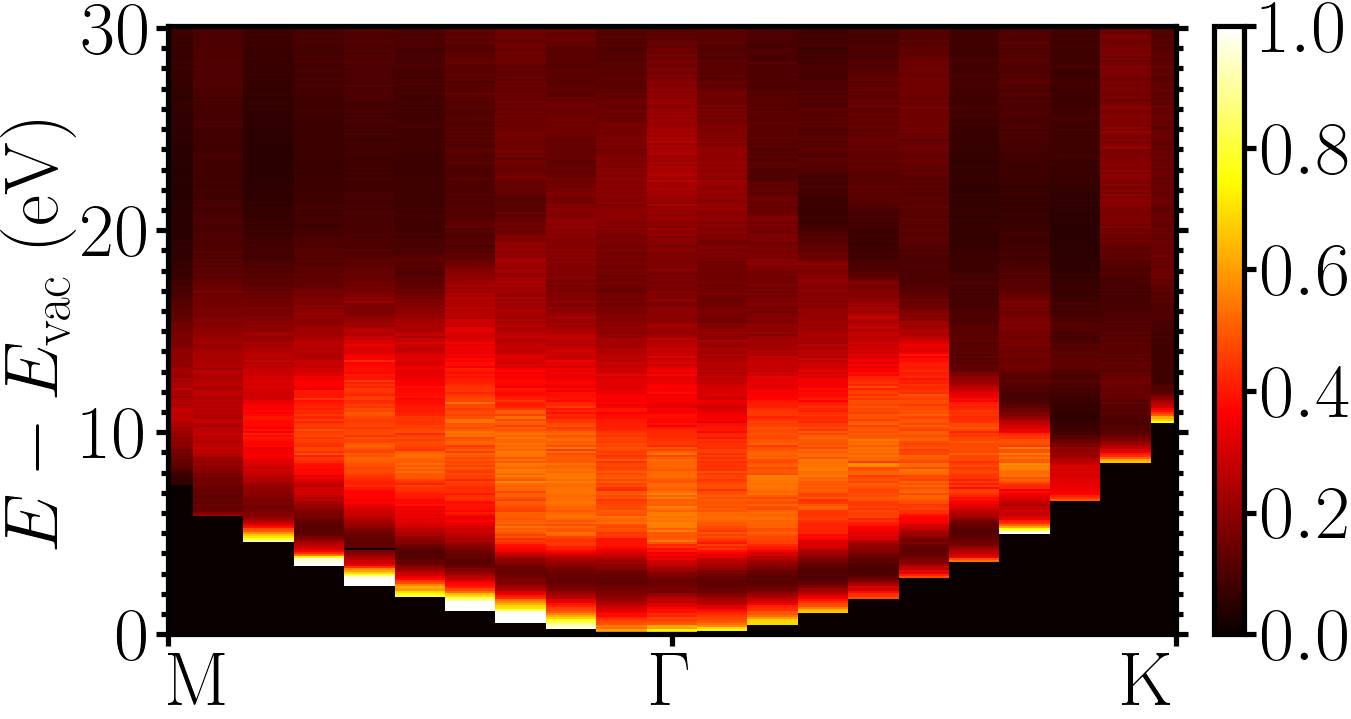}
		\includegraphics[clip,trim={4.0cm} 0 {4.0cm} 0,height=0.154\textwidth]{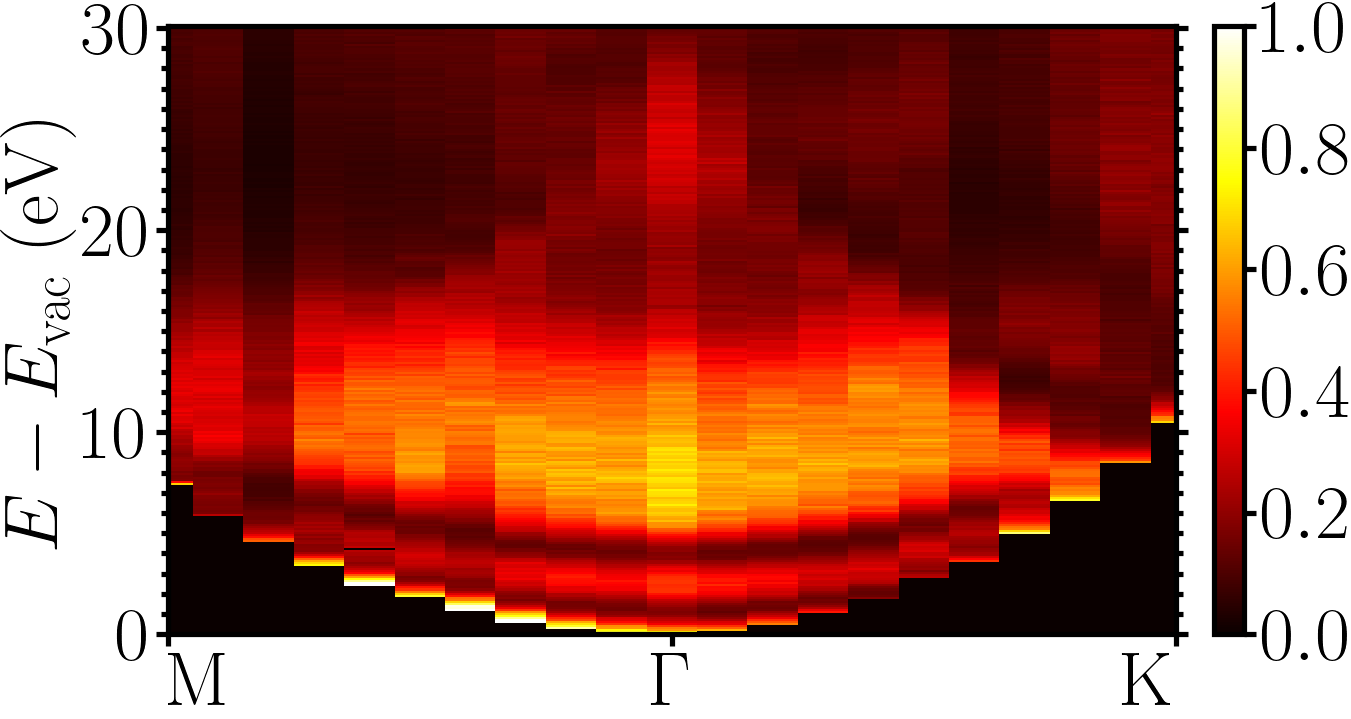}
		\includegraphics[clip,trim={4.0cm} 0 0 0,height=0.154\textwidth]{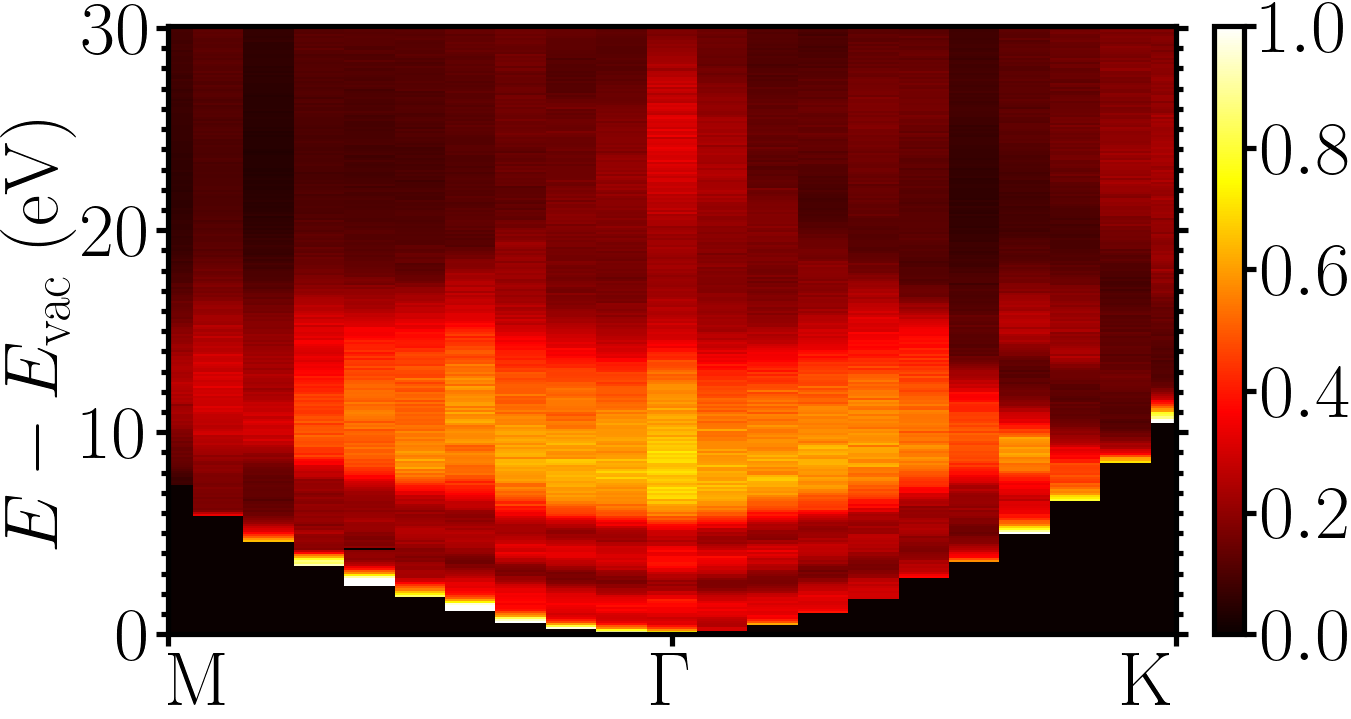}%
\caption{
ARRES of freestanding FLG for incident electron beams with in-plane wave vectors along $M\Gamma K$.
The number of layers increases from 1 to 4, from left to right.
Each row corresponds to different data: without inelastic effects via the Bloch wave matching method (first row), without inelastic effects using the APW method (second row),
with inelastic effects using the APW method (third row), and measurement of graphene on silicon carbide (SiC) from Ref.~\cite{jobst_nanoscale_2015} (fourth row) which is 
normalized as in Fig.~\ref{niflg} at the $\Gamma$ point and for different incidence angles to the simulated reflectivities with inelastic effects at maxima significantly above the mirror boundary. 
The landing energies on the vertical axes are considered with respect to the energy in vacuum $E_{\mathrm{vac}}$.
}
\label{angles}
\end{figure*}
The result of our simulations is presented in Fig.~\ref{angles}, complemented by experimental data of FLG on SiC~\cite{jobst_nanoscale_2015}.
All the columns corresponding to 2--4 layers in Fig.~\ref{angles} exhibit local minima (up to the wide maximum which starts to appear near $\SI{7}{\eV}$ at the $\Gamma$ point) that allow us to count layers of graphene, similarly to the normal incidence data displayed in Fig.~\ref{niflg}.
The minima are clearly visible not only at the $\Gamma$ point but also further away.
There is a less visible minimum very close to zero landing energy in the case of the four-layer graphene reflectivities, Fig.~\ref{angles} (fourth column). 
The inclusion of inelastic effects makes it less recognizable.
Nevertheless, it has been experimentally observed~\cite{jobst_nanoscale_2015}.
As foreseeable, one such resonance band is visible for bilayer graphene. Trilayer graphene has two bands that disperse upward, and four-layer graphene has three bands that eventually become less recognizable when they touch the mirror boundary, defining a landing energy threshold above which scattering can occur.
This threshold is given by the in-plane component of the wave vector, as noted in Appendix~\ref{sec:matching}. The upper boundary of the bottom black regions in Fig.~\ref{angles} is defined by this threshold.
Another notable feature that aligns well with the experimental results for 2--4 layers is that the region of high reflectivity spans approximately between \SI{7}{\eV} and \SI{15}{\eV} (shown in bright yellowish colors). This feature is well-known to correspond to the band gap in unoccupied states.
There are subtle features and pockets in reflectivity in the same regions for both freestanding FLG simulations and the experiment with a substrate.
This lends support to the fact stated in Ref.~\cite{jobst_nanoscale_2015} that the interaction of interlayer states with the SiC substrate is negligible in the studied region of landing energies and impact angles.

\section{Momentum-resolved EELS simulations}\label{sec:momentum.resolved.EELS}

Momentum transfer (MT) resolved EELS simulations are a prerequisite for the optical potential calculation in the dielectric formalism. 
The unsuitability of different approximations has already been briefly discussed in Ref.~\cite{ToF_2021}; e.g., independent-particle approximation is insufficient and the random phase approximation (RPA) is acceptable to describe collective phenomena. 
Many electron microscopes allow for a selection of the collected MT in 
EELS. 
In such cases, simulations can provide complementary data or finer MT resolution to gain more insight into the experimental results. 

\begin{figure*}
\centering
\includegraphics[height=0.22\textwidth]{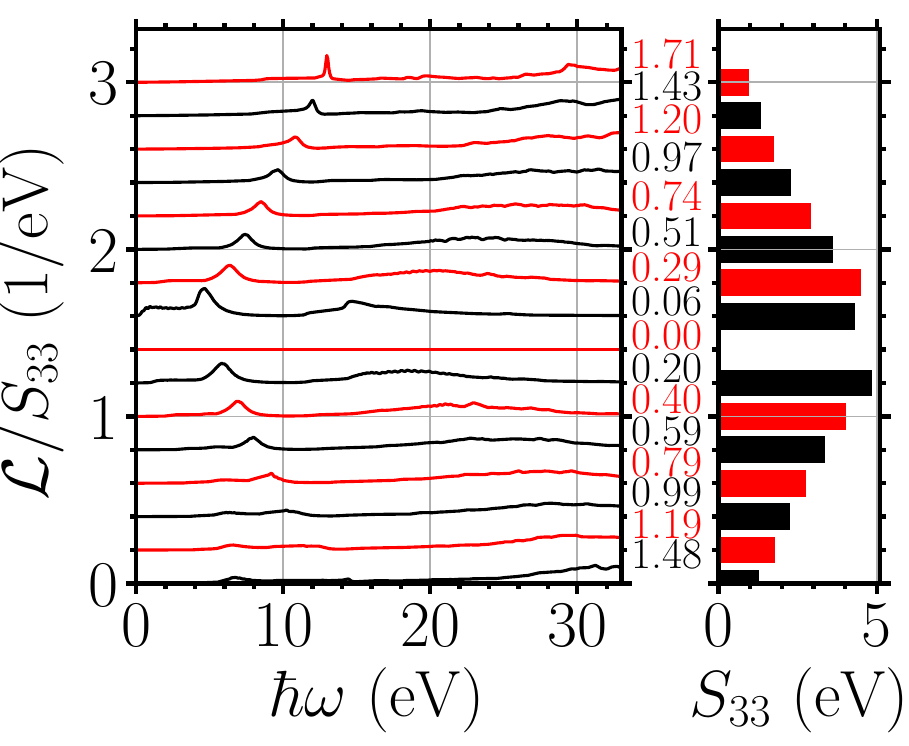}
\includegraphics[clip,trim={2.0cm} 0 0 0,height=0.22\textwidth]{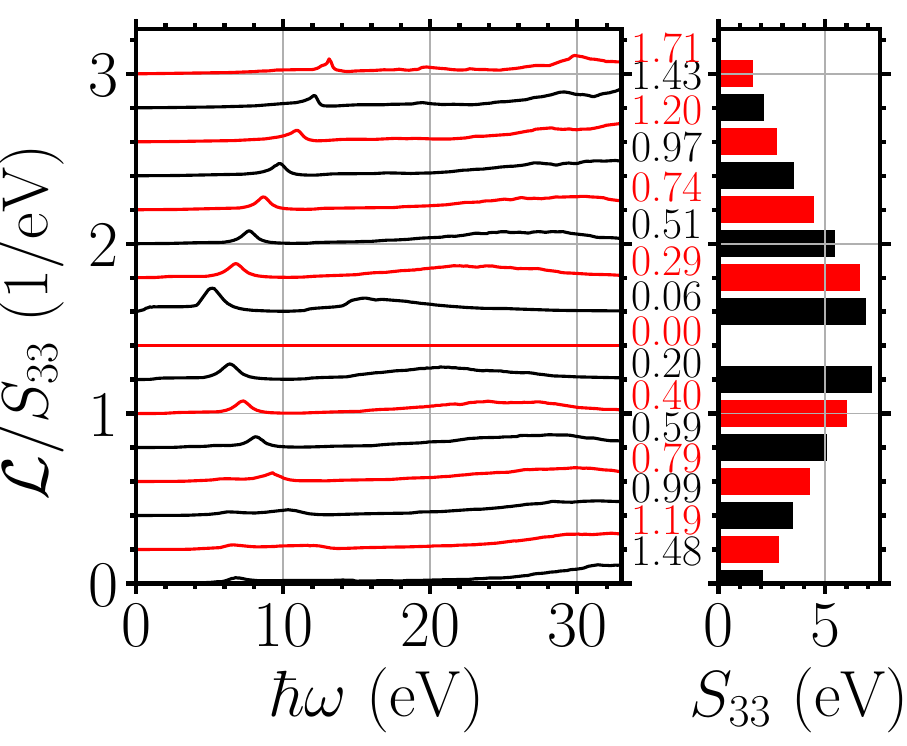}
\includegraphics[clip,trim={2.0cm} 0 0 0,height=0.22\textwidth]{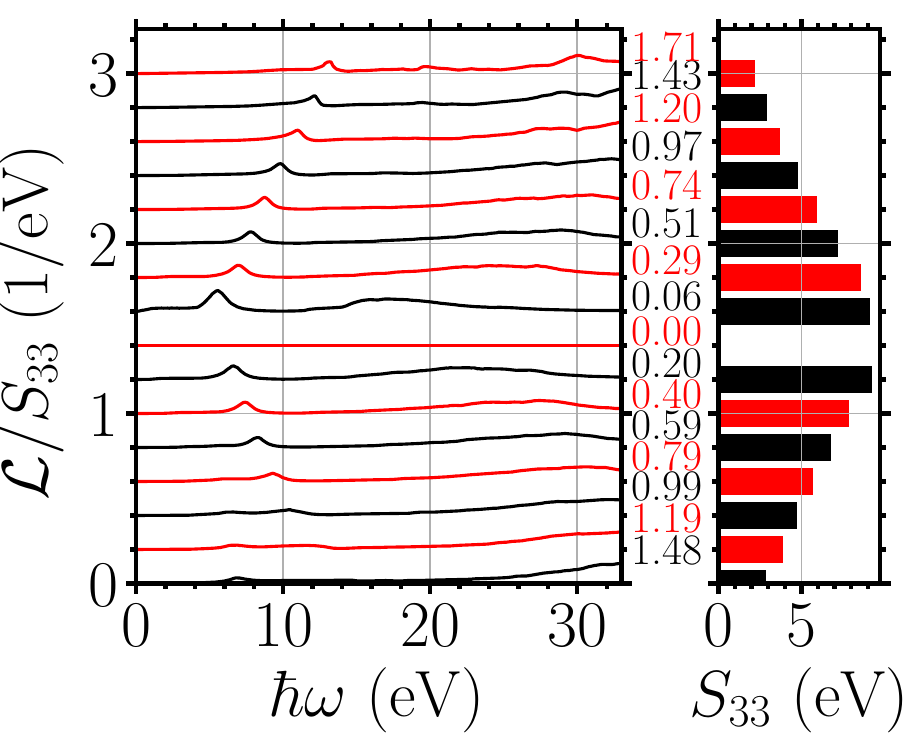}
\includegraphics[clip,trim={2.0cm} 0 0 0,height=0.22\textwidth]{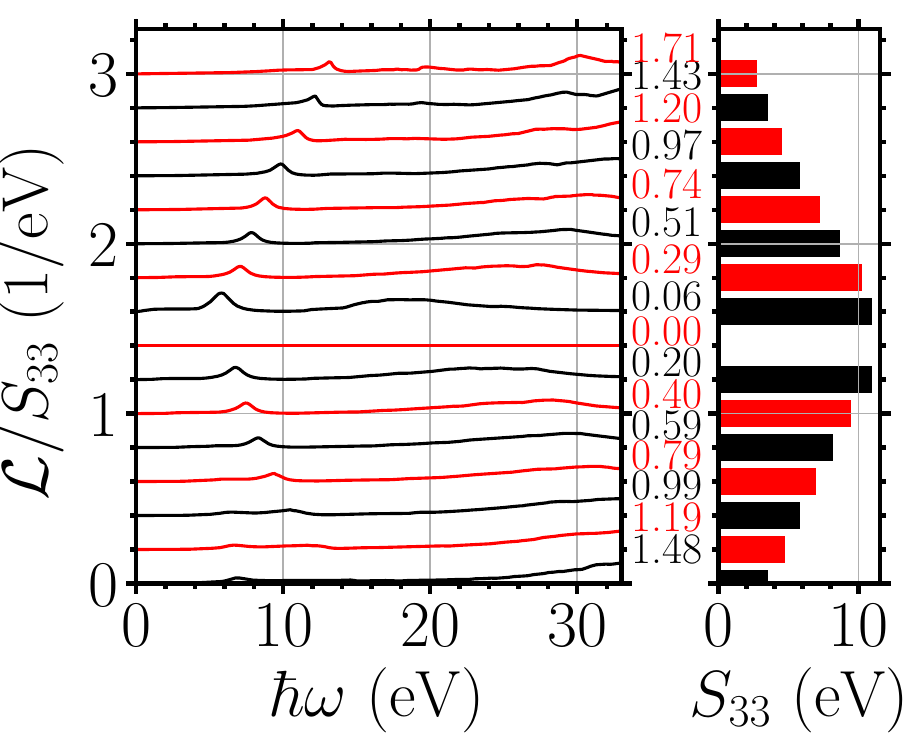}
\caption{Momentum-resolved EELS simulation for freestanding graphene for 1--4 layers, left to right, and for in-plane momenta along $M\Gamma K$ ($K$ point top, $M$ point bottom, the corresponding values of the momenta are in units of~\SI{}{\angstrom}$^{-1}$). The energy loss function $\mathcal{L} = \ii\left\{-1/\varepsilon\right\}$ is normalized by its integral $S_{33}$ for the losses $\hbar\omega$ in the range from 
0 to \SI{33}{\eV} 
for better visualization. The simulations were obtained using the Dyson equation, Eq.~\eqref{eqn:dyson-like}, approach (\texttt{Yambo}).
}
\label{fig:eels}
\end{figure*}

The theoretical momentum-resolved electron energy loss (EEL) spectra presented in Fig.~\ref{fig:eels} were obtained using MBPT on top of the DFT as implemented in the \texttt{Yambo} code 
and \texttt{yambopy}~\cite{marini_yambo:_2009,yambo_2019}.
The calculations utilized RPA with Hartree kernel
and with the inclusion of local-field effects~(LFEs).
Furthermore, because the system under study is a 2D material, both random integration method and cutoff Coulomb potentials were applied.
See Appendix~\ref{sec:compdeta_eels} for more details regarding convergence and values of important parameters.

The typically experimentally sampled segments of a path in reciprocal space are along $\Gamma M$ and $\Gamma K$, i.e., the same as in Sec.~\ref{sec:ares}.
Corresponding simulations are presented up to MT $\approx 1.7\, \si{\angstrom} ^{-1}$ for the $K$ point and \mbox{$\approx 1.5\, \si{\angstrom} ^{-1}$} for the $M$ point, see Fig.~\ref{fig:eels}.

The most intensive electronic excitations are the \mbox{$\pi$-plasmon}, with a peak occurring at approximately \SI{4}{\eV} and the $\pi+\sigma$-plasmon at about \SI{14}{\eV}; both energy-loss values
provided correspond to MT close to zero.
It is important to note that there has been a debate surrounding the interpretation of these excitations as plasmons; see Ref.~\cite{nazarov.NJP17.2015}.
There is a criterion to distinguish true collective phenomena, plasmons, from other features in the spectra.
The condition to be attained at the locations of plasmon peaks is $\Re\{\varepsilon\} = 0$~\cite{nazarov.NJP17.2015}.
Because $\Re\{1/\varepsilon\}$ converges slowly when compared to the imaginary part in the MBPT-RPA calculations, the zero value is not always attained but a pronounced local minimum is present instead.
This is essentially the same as the data in the top segment of Fig.~2 in Ref.~\cite{nazarov.NJP17.2015}.
Such data are not included here in order to present a reasonable amount of figures.

It can be seen that the position of local maxima in EELS of the plasmons shifts to higher values of energy loss with increasing MT, revealing so-called dispersion relations.
Furthermore, the monolayer graphene spectrum shows a split of $\pi$-plasmon into two branches along $\Gamma M$, already reported in simulations in Ref.~\cite{Despoja_PRB87_2013}.
Similar split is present also in the case of all remaining FLG EEL spectra.
Also, the position of each of the plasmons, as read off at the $\Gamma$ point, shifts to higher energy losses with increasing layer count.
The simulations presented in Fig.~\ref{fig:eels} demonstrate that the \mbox{$\pi+\sigma$}-plasmon peak is broader for three and four layers of graphene.
Furthermore, the authors of Ref.~\cite{Idrobo_UM180_2017} have noted the presence of a peak ``shoulder'' in the $\pi + \sigma$-plasmon fall-off region at larger losses at $\SI{25}{\eV}$ in the case of three and four layers of graphene, which bears resemblance to the EELS of graphite.
For a broader range of energy losses and of MT along $\Gamma M$ and $\Gamma A$
see Fig.~\ref{fig:elf_heatmaps}.
The aforementioned simulations were obtained through the use of the Liouville-Lanczos approach as implemented in the \texttt{turboEELS} code.

\begin{figure}[h]
\centering
\includegraphics[width=0.95\linewidth]{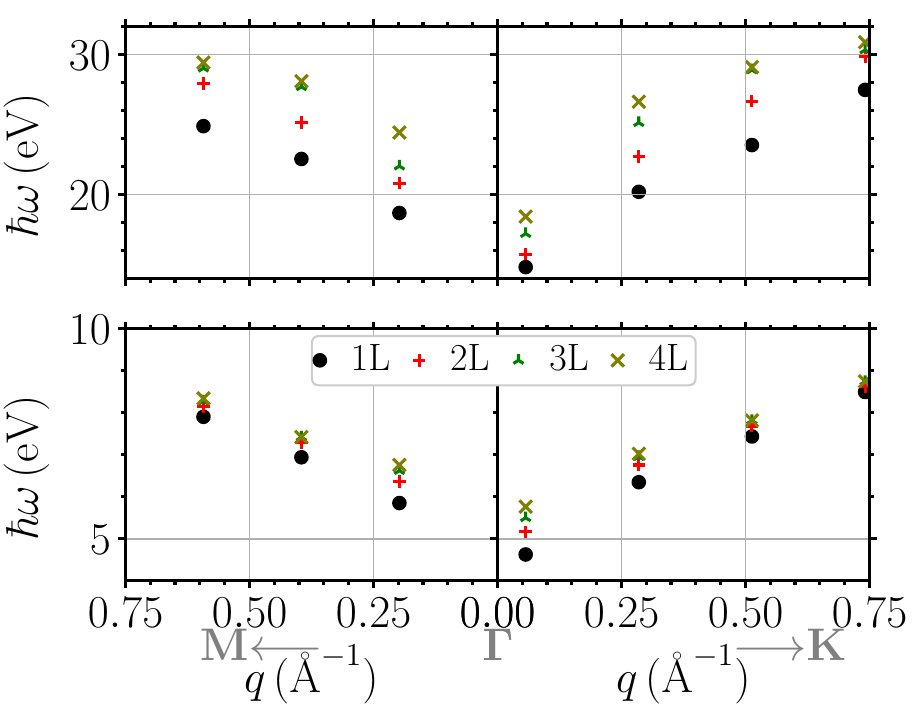}
\caption{
Dispersion relation of both plasmon peaks, wherein the energy loss values $\hbar\omega$ of the peak maxima are plotted as a function of the magnitude $q$ of MT along $M\Gamma K$ for \mbox{1--4} layers corresponding to the data in Fig.~\ref{fig:eels}, the $\pi$-plasmon dispersion (bottom) and the $\pi+\sigma$-plasmon dispersion (top).
}
\label{fig:eels.dispersion}
\end{figure}

Details of the plasmon dispersion relations in the case of both plasmon peaks along the two segments in reciprocal space are displayed in Fig.~\ref{fig:eels.dispersion}.
It shows that a maximal spread of peak positions due to a different number of layers appears up to MT values \SI{0.2}{\angstrom}$^{-1}$.
This is in agreement with the experimental and simulated data in Ref.~\cite{Wachsmuth_PRB90_2014} (their Figs.~1 and~2).

\section{EELS experiment and simulations at the $\Gamma$ point}\label{sec:EELS.at.Gamma}

A realistic EELS experiment uses a finite size of aperture which transmits the electrons with a certain range of axial angles.
Consequently, a typical measured EELS is contributed by processes with a certain range of MT, and it can be represented as a weighted sum over different values of MT.
Any comparison with simulations has to take the finite size of the apertures in the electron microscope into account even in the optical limit, i.e., at the $\Gamma$ point.
Thus we calculate momentum-resolved EELS around the $\Gamma$ point.
 The discussed dependence of plasmon peak positions on the number of layers in Sec.~\ref{sec:momentum.resolved.EELS} is verified and measurable even for very small MT given by the aperture in TEM. The details of the experimental and simulation procedures are provided below.

 FLG was fabricated in a chemical vapor deposition~(CVD) furnace with methane and hydrogen precursors according to the recipe in Ref.~\cite{Xuesong}.
 The CVD graphene was then transferred using wet PMMA transfer (PMMA dissolved in $\SI{4}{\percent}$ anisole and deposited by spin coating) onto 1500 mesh TEM grid.
 The PMMA was removed from graphene using $\SI{99}{\percent}$ acetic acid for 5 h.

TEM and EELS was performed with TEM FEI Titan equipped with a GIF Quantum spectrometer operated at \SI{60}{\kilo\eV}.
EELS was measured in the diffraction mode of the TEM, where the area on the sample was defined by the selective area aperture and the collection angle was defined by the camera length and the spectrometer entrance aperture reading \SI{0.2}{\milli\radian} and corresponding to $\SI{0.026}{\angstrom}^{-1}$.
The challenge presented by using the smallest pinhole aperture and the largest camera length is that the Bragg spots cannot be recorded simultaneously, which precludes their use for calibrating the magnitude of $\bq$.
Furthermore, the smallest machine step for fine-tuning the central position of the aperture was insufficiently precise.
Considering also the beam path stability as another source of combined uncertainty, we estimate the maximum MT to be $\SI{0.06}{\angstrom}^{-1}$.
The dispersion of the spectrometer was set to \SI{0.025}{\eV}/pixel.
The measurement was done at the \mbox{$\Gamma$ point}.
The raw spectra were normalized by dividing by the integral intensity of the zero-loss peak (the energy window for integration was set from \SI{-3}{\eV} to \SI{+3}{\eV}) and divided by the pixel size of \SI{0.025}{\eV} so the counts were transformed to a quantity proportional to the loss probability density in units~$\SI{}{\eV}^{-1}$ (referred to as the energy loss function $\mathcal{L}$). Finally, we have subtracted the EEL spectrum recorded on the vacuum to remove the background consisting of the elastically scattered electrons.
The uncertainties in Table \ref{tab:plasmons} correspond to the FWHM of the experimental zero-loss peak of the respective EEL spectra. 

\begin{figure}[h]
\centering
\includegraphics[width=0.95\linewidth]{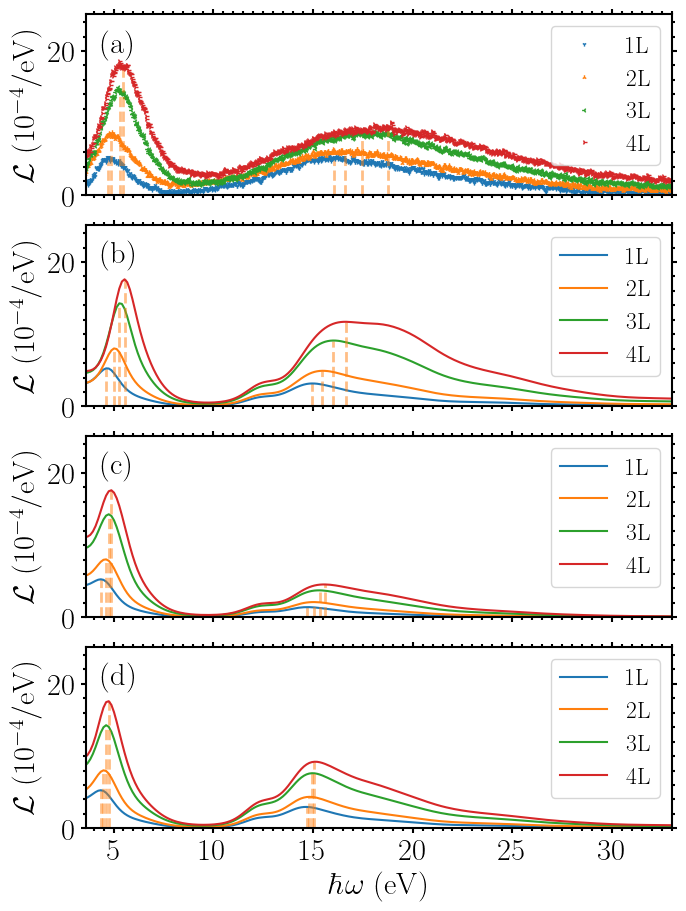}\\
\includegraphics[width=0.95\linewidth]{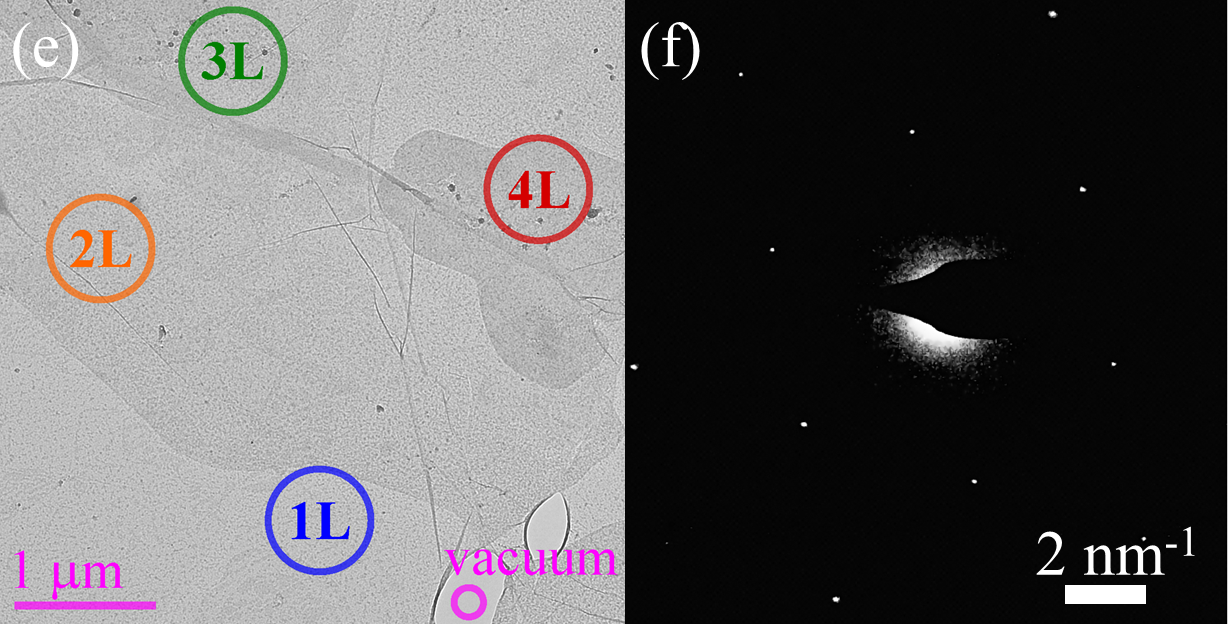}
\caption{The experimental EELS (a) compared with appropriately scaled and weighted simulated ELF~$\mathcal{L}$ of the FLG at the vicinity of the $\Gamma$ point as a function of losses $\hbar\omega$. The simulated ELF is weighted 
by areas of annuli around the given MTs (b), the kinematic factor from \mbox{Eq. (8)} from Ref.~\cite{Guandalini_CMMH_2024} (c), and 
the kinematic factor inspired by 
Eq.~(3.29) from Ref.~\cite{Mills} (d).
The simulations are normalized to the experimental EELS at maxima of the $\pi$-plasmon.
The vertical dashed (orange) lines indicate the positions of the peak maxima.
TEM micrograph of freestanding FLG sample with ROIs for 1--4 layers, 1L--4L respectively, and the hole for vacuum reference, where EELS were measured (e).
Diffractogram from a region with the same contrast as 1L (f). 
}
\label{fig:eels.measurement}
\end{figure}

 A comparison of RPA simulations of FLG loss spectra with our EELS experiment is presented in~Fig.~\ref{fig:eels.measurement}.
The measured loss spectra are presented in Fig.~\ref{fig:eels.measurement}(a). They exhibit two clear peaks attributed to the $\pi$-plasmon (lower energy) and $\pi+\sigma$-plasmon (higher energy). 
The vertical dashed (orange) lines indicate the positions of the peak maxima,
which were determined from denoised data (not presented). The blue shift of the peak positions with increasing layer count is clearly visible and the corresponding energy loss values for the $\pi$-plasmon and the $\pi+\sigma$-plasmon are presented in Table~\ref{tab:plasmons}. 

Figures~\ref{fig:eels.measurement}(b)--\ref{fig:eels.measurement}(d) illustrate the outcomes of RPA simulations conducted using the Liouville-Lanczos approach.
The figures demonstrate the impact of varying angular correction factors on \mbox{the energy loss function (ELF)}.
The peak positions are indicated by vertical dashed (orange) lines for the simulated spectra as well.
It should be noted that the positions of the simulated plasmon peaks were extracted after the as-simulated spectra were convolved with a variable Gaussian kernel in order to mimic the experimental smearing.
The kernel width linearly increases with energy loss, with the 0th-order term equal to~0 and the 1st-order term coefficient equal to 0.04 as implemented in \texttt{varconvolve}~\cite{varconvolve.2016}.

Figure~\ref{fig:eels.measurement}(b) presents the ELF calculated by weighting ELFs for nine different MTs with an increasing distance from the $\Gamma$ point (where the spectrum vanishes) along~$\Gamma M$ up to the maximum MT value of $\SI{0.06}{\angstrom}^{-1}$.
The increment was set to approximately $\SI{0.0063}{\angstrom}^{-1}$. 
The formula (A.12) from Ref.~\cite{nazarov.NJP17.2015} was applied prior to the weights, calculated simply as areas of annuli, with each annulus containing a single computed MT.
The blue shift of the $\pi$-plasmon and $\pi+\sigma$-plasmon with increasing layer count is clearly visible once more. 

Figure~\ref{fig:eels.measurement}(c) shows the combined effect (product) of the aforementioned weights and the following angular correction weighting factor.
The factor is defined by $q_{\paral}/(q_{\paral}^2+q_z^2)^2$, as detailed in 
\mbox{Eq. (8)} from Ref.~\cite{Guandalini_CMMH_2024}, \mbox{$q_z \approx (2m \omega + \hbar q_{\paral}^2)/2\hbar k_i$}, where $k_i$ denotes the magnitude of the wave vector of the incoming electron and $m$ stands for its rest mass.
The first term of $q_z$ is dominant and can be written as $\omega /v_0$ and $v_0\approx 0.45 c$, representing the velocity of the incident electrons ($c$ being the speed of light).
This factor  differs from the Lorentzian factor $1/(q_{\paral}^2+q_z^2)$ employed in the Supplemental Material of Ref.~\cite{wachsmuth_PRB88} due to the incorporation of the quasi-2D dielectric function~\cite{nazarov.NJP17.2015} (derived from the corresponding density-response function and 2D Coulomb potential).
The intensity of the $\pi + \sigma$-plasmon is overdamped for the kinematic factor leading to Fig.~\ref{fig:eels.measurement}(c).
The suppression of the $\pi + \sigma$-plasmon in our simulations is more pronounced than in Ref.~\cite{wachsmuth_PRB88}, in which the Lorentzian prefactor was used.
It is attributed to the presence of the angular correction factor, which contains $\omega$ in the fourth power in the denominator.
This leads to a noticeable decrease in the height of peaks situated at larger losses in comparison to peaks in regions with smaller losses. 
The incorporation of relativistic factors~\cite{Schattschneider_relativistic} could in principle enhance the intensity of $\pi + \sigma$-plasmon in comparison to the intensity of $\pi$-plasmon as desired.
The implementation of the relativistic factor, as described in \mbox{Eq. (4)} from Ref.~\cite{Segui}, 
did not result in a discernible change of the peak height for our velocity of the incident electrons $v_0=0.45c$.

Figure~\ref{fig:eels.measurement}(d) presents an attempt to correct the formerly mentioned overdamping utilizing a kinematic factor inspired by Eq.~(3.29) from Ref.~\cite{Mills} given by 
\mbox{$(c_1q_{\paral}^2+c_2q_z^2)/q_{\paral}(q_{\paral}^2+q_z^2)^2$}, where $c_1$ and $c_2$ are constants.
If we set $c_1=1$ and $c_2=0$, then the previous kinematic factor, used in Fig.~\ref{fig:eels.measurement}(c), is obtained.
It is one of the manifestations of a close analogy between the (transmission) EELS double differential cross section of a 2D material and the high-resolution EELS cross section.
A choice of $c_1=0.15$ and $c_2=0.85$, displayed in Fig.~\ref{fig:eels.measurement}(d), yields results that are comparable to those of the experiment in terms of the magnitudes of the peaks associated with the $\pi + \sigma$-plasmon.

The regions of interest (ROI) from which the spectra were obtained are displayed in Fig.~\ref{fig:eels.measurement}(e) and denoted by $n$L, where $n$ is the number of layers.
The micrograph in Fig.~\ref{fig:eels.measurement}(e) was obtained using TEM.
The ROI designated as ``vacuum'' conforms to a hole that is utilized for background subtraction.
Figure~\ref{fig:eels.measurement}(f) depicts a diffractogram from a point in a region with the same contrast as the 1L ROI in Fig.~\ref{fig:eels.measurement}(e). The diffractogram corroborates the expected six-fold symmetry of graphene. The denoising of the diffractogram was achieved through the utilization of the anisotropic diffusion plugin Ref.~\cite{aniso_diffusion} within the \texttt{Fiji} software~\cite{fiji}.

Figures~\ref{fig:eels.measurement}(a) and ~\ref{fig:eels.measurement}(d) illustrate a good qualitative and a reasonable quantitative agreement between the experimental and simulation results.
However, there are some discrepancies in the positions of the maxima, their relative heights and the shapes of the peaks.
The discrepancy between the experiment and simulation can be attributed to both experimental and theoretical factors.
On the experimental side, not all examined parts of the graphene exhibited the same degree of symmetry as the one shown in Fig.~\ref{fig:eels.measurement}(f).
This indicates that the CVD graphene contains defects and possible contamination.
It should be noted that scattering on defects such as grain boundaries is not included in the simulations.
On the theoretical side, more effects can contribute to the observed discrepancies.
To begin, Figs.~\ref{fig:eels.measurement}(b)--\ref{fig:eels.measurement}(d) demonstrate the considerable impact of the selected kinematic factor on the processed simulated spectra, particularly in the context of 2D materials.
For further insight, please refer to Ref.~\cite{Guandalini_CMMH_2024}.
Moreover, a recent publication~\cite{Guandalini_grBSE} demonstrated that the incorporation of electron-electron repulsion via GW corrections results in an almost rigid blue shift of the $\pi$-plasmon peak. Furthermore, the inclusion of excitonic effects via the \mbox{Bethe-Salpeter equation (BSE)} produces a momentum-dependent red shift that increases with MT before reaching a saturation point.
The combined effect of GW and BSE gives rise to an overall blue shift with respect to the RPA result of approximately \SI{0.25}{\eV} in the vicinity of the $\Gamma$ point along~$\Gamma M$.
Furthermore, the incorporation of the electron-hole interaction by BSE also results in a reshaping of the peak. 
While alternative \mbox{\textit{ab initio}} methods for simulating dielectric function, such as the \mbox{GW + BSE}, may be employed, this would entail a significant increase in computational effort. 
Furthermore, incorporating additional effects, such as multiple scattering or the tensorial character of the dielectric function (thereby accounting for the anisotropy of ELF), could potentially enhance the agreement with experimental data.
However, this is beyond the scope of the present paper.
Generally, RPA gives reasonable agreement with the experiment, as evidenced by Figs.~\ref{angles}, \ref{fig:eels.measurement}, or~\ref{imfp_compare}.

The blue shift of the plasmons with increasing layer count also agrees with the experimental and theoretical findings, see Table~\ref{tab:plasmons} for comparison with some previous publications.
It can be seen from Table~\ref{tab:plasmons} that the differences from our results are almost negligible when compared to the spread of values in the literature.
It can be concluded that RPA, which employs either the Dyson equation or the Liouville-Lanczos approach, is an efficient method for capturing the primary features of EELS observed in~FLG.
\begin{table}[h]
\begin{ruledtabular}
\begin{tabular}{clll}
Layer & Source  & $\hbar\omega_{\pi} (\si{\eV})$ & $\hbar\omega_{\pi+\sigma} (\si{\eV})$
\\
count &        &                       &
\\\hline
One & Simulation (ours)   & $4.6$ & $14.6$ 
\\
  & Experiment (ours)    & $4.7\pm 0.9$ & $16.1\pm 0.9$ 
\\
  & Experiment~\cite{Liou_PRB91_2015}  & $4.0$ & $13.5$
\\
  & Experiment~\cite{Idrobo_UM180_2017}  & $4.9$ & $15.4$
\\
  & Experiment~\cite{Lu_PRB80_2009} & $5.0$ & $14$
\\
  & Experiment~\cite{wachsmuth_PRB88} & $5.0$ & $15.3$
\\
  & Simulation~\cite{Djordjevic_UM184_2018} & $4.1$ & $14.0$
\\
  & Simulation~\cite{eberlein_plasmon_2008}  & $4.9$ & $15.1$
\\
  & Simulation~\cite{Despoja_PRB87_2013}  & $4.3$ & $14.0$
\\
  & Simulation~\cite{Mowbray_PSSB251_2014}  & $4.9$ & $14.7$
\\
  & Simulation~\cite{nazarov.NJP17.2015}  & $5.0$ & $15.5$
\\
  & Simulation~\cite{nazarov.NJP17.2015}  & $5.0$ & $17.2$
\\
  & Simulation~\cite{wachsmuth_PRB88}  & $4.0$ & $14.2$
\\\hline
Two & Simulation (ours)    & $4.9$ & $14.8$
\\
  & Experiment (ours)    & $4.9\pm 0.8$ & $16.6\pm 0.8$ 
\\
  & Experiment~\cite{Idrobo_UM180_2017}  & $5.1$ & $15.6$
\\
  & Experiment~\cite{Lu_PRB80_2009}  & $5.4$ & $18$
\\
  & Experiment~\cite{Wachsmuth_PRB90_2014}  & $6.3$ & $23$
\\
  & Simulation~\cite{eberlein_plasmon_2008}  & $5.4$ & $15.8$
\\
  & Simulation~\cite{nazarov.NJP17.2015}  & $5.6$ & $19.0$
\\
  & Simulation~\cite{nazarov.NJP17.2015}  & $5.6$ & $17.4$
\\
  & Simulation~\cite{Pisarra_PRB93_2016}  & $5.8$ & $17.4$
\\\hline
Three & Simulation (ours)   & $5.2$ & $15.1$
\\
  & Experiment (ours)    & $5.3\pm 0.8$ & $17.5\pm 0.8$
\\
  & Experiment~\cite{Idrobo_UM180_2017}  & $5.1$ & $16.5$ 
\\
  & Experiment~\cite{Lu_PRB80_2009}  & $6.2$ & $25$
\\
  & Experiment~\cite{Wachsmuth_PRB90_2014}  & $6.7$ & $24$
\\
  & Simulation~\cite{eberlein_plasmon_2008}  & $5.9$ & $18.6$
\\\hline
Four & Simulation (ours)    & $5.3$ & $16.0$
\\
  & Experiment (ours)    & $5.5\pm 0.8$ & $18.7\pm 0.8$ 
\\
  & Experiment~\cite{Idrobo_UM180_2017}  & $5.2$ & $18.1$ 
\\
  & Experiment~\cite{Lu_PRB80_2009}  & $6.2$ & $25$
\\
  & Experiment~\cite{Wachsmuth_PRB90_2014}  & $6.8$ & $25$
\end{tabular}
\end{ruledtabular}
\caption{Comparison of our results with literature.
Energy loss values $\hbar\omega\ (\si{\eV})$ corresponding to the $\pi$- and $\pi+\sigma$-plasmon.
The theoretical energy loss value is determined as close as possible to the $\Gamma$ point. However, the MT is $\approx \SI{0.3}{\angstrom}^{-1}$ in Ref.~\cite{Wachsmuth_PRB90_2014} and $\approx\SI{0.1}{\angstrom^{-1}}$ in Ref.~\cite{Lu_PRB80_2009} (angle of incidence $\SI{53}{\degree}$) and Ref.~\cite{Mowbray_PSSB251_2014}.
Our data are read along the direction $\Gamma M$ from 
ELFs obtained from RPA using the Liouville-Lanczos approach at MT~$\SI{0.03}{\angstrom}^{-1}$ and the formula (A.12) from Ref.~\cite{nazarov.NJP17.2015}.
FWHM of zero loss peaks are used as an estimate of our experimental errors.
}\label{tab:plasmons}
\end{table}

\section{Optical potential and IMFP}\label{sec:v.optical.and.IMFP}

The momentum-resolved EELS simulations can be used to estimate the optical potential. The optical potential allows us to include the inelastic effects in the simulations of the low-energy electron reflectivities.
This phenomenological approach was initialized by Slater in Ref.~\cite{Slater1937}.
The isotropic formula~\eqref{eqn:lifetime.elf.rpa.isotropic} is implemented for the optical potential $V_{\mathrm{opt}}$.
It follows from Penn~\cite{Penn1976} [his Eq.~(2)] and is rederived in Appendix~\ref{sec:optical}.
Related computational details of the optical potential are presented in Appendixes~\ref{sec:inelastic} and \ref{sec:compdeta_vopt}.

\begin{figure}[h]\begin{center}
\includegraphics[width=0.95\linewidth]{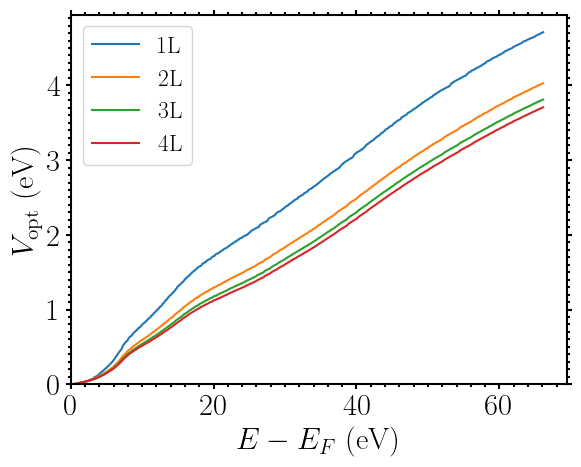}
\caption{
The energy dependence (with respect to the Fermi energy $E_F$) of the optical potential $V_{\mathrm{opt}}$ per layer for 1-4 layers of graphene.
The interpolation formula~\eqref{eqn:interpolation_formula} was applied.
}\label{fig:optical_potential_compare}\end{center}\end{figure}

The energy dependence of the optical potential $V_{\mathrm{opt}}$ is often approximated by a linear function, see Ref.~\cite{gao_inelastic_2015} and related references therein.
Here, we calculate $V_{\mathrm{opt}}(E)$ by integrating $\mathcal{L}$ over the momentum transfer and energy losses.
As a result, the dependence will be somewhat sensitive to EELS features such as plasmons.
The dependence of the optical potential $V_{\mathrm{opt}}$ on energy is presented in Fig.~\ref{fig:optical_potential_compare}.
A comparison of the dependence of $V_{\mathrm{opt}}$ per layer is shown for the different number of layers.
The dependence of $V_{\mathrm{opt}}$ for four-layer graphene was normalized in a final step (as described in Appendix~\ref{sec:compdeta_vopt}) using the dependence of $V_{\mathrm{opt}}$ for the graphite overlayer presented in Ref.~\cite{barrett_elastic_2005}.
The two regions of steeper slopes of $V_{\mathrm{opt}}(E)$ within the energy range from 0 to \SI{20}{\eV} 
reflect the positions of the plasmon peaks.
However, this increase is not very pronounced because the integral in Eq.~\eqref{eqn:lifetime.elf.rpa.isotropic} together with the dispersion of the plasmon peaks smooths it out. 
Therefore, the resulting energy dependence is fairly linear.
The comparison shown in Fig.~\ref{fig:optical_potential_compare} indicates that $V_{\mathrm{opt}}$ per layer decreases as the number of layers of graphene increases.
The higher value of $V_{\mathrm{opt}}$ per layer for monolayer graphene could be attributed to more significant surface effects than in FLG with more layers, which is expected to converge to the graphite $V_{\mathrm{opt}}$.

\begin{figure}[h]\begin{center}
\includegraphics[width=0.95\linewidth]{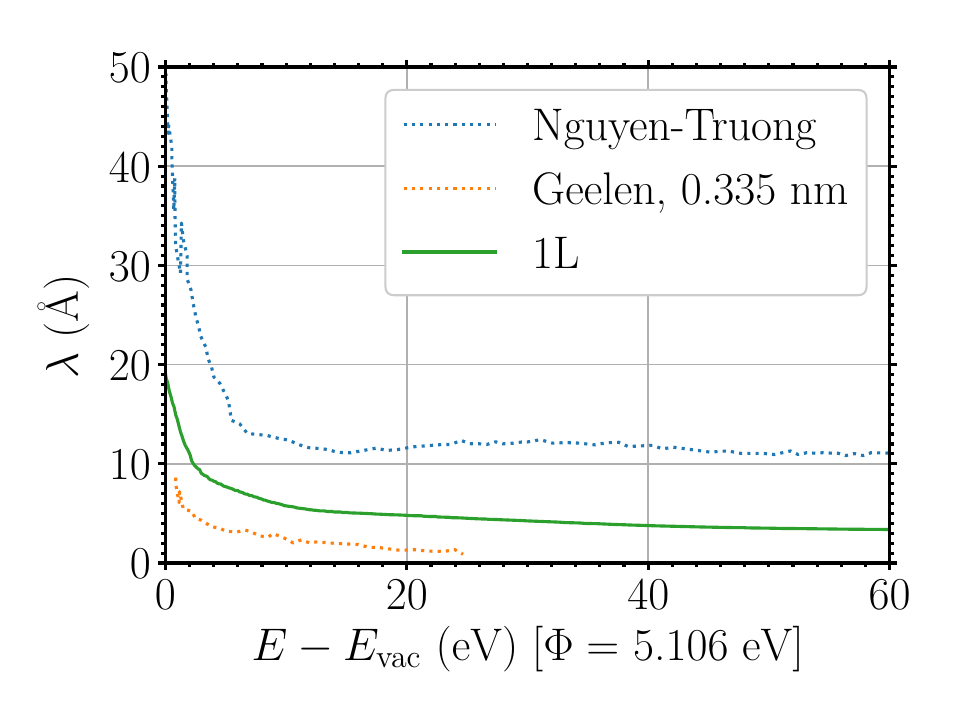}
\caption{
Comparison of the IMFP $\lambda$ as a function of energy (with respect to the energy in vacuum $E_{\mathrm{vac}}$, $\Phi$ denoting the work function) for the monolayer graphene obtained from literature~\cite{nguyen-truong_low-energy_2020} 
(experimental orange and simulated along $\Gamma M$ blue dotted lines labeled by Geelen and Nguyen-Truong, respectively) 
and our results (green solid line) calculated from $V_{\mathrm{opt}}$ displayed in Fig.~\ref{fig:optical_potential_compare}.
}\label{imfp_compare}\end{center}\end{figure}

IMFP is inversely proportional to $V_{\mathrm{opt}}$, see Eq.~\eqref{eqn:lifetime.elf.rpa.isotropic}, and since IMFP is of general interest, we display it for graphene in Fig.~\ref{imfp_compare}.
It can be seen that the IMFP 
based on the interpolation formula~\eqref{eqn:interpolation_formula}, which combines Eq.~\eqref{eqn:lifetime.elf.rpa.isotropic} along $\Gamma A$ and $\Gamma M$ with ELF obtained by the Liouville-Lanczos approach with LFE (see Appendix~\ref{sec:compdeta_eels}), 
lies between the simulated dependence of Nguyen-Truong and the experimental dependence of Geelen. The IMFP of Nguyen-Truong was calculated without LFE and 
with the upper limit on energy losses in Eq.~\eqref{eqn:lifetime.elf.rpa.isotropic} replaced by $E/2$ and based only on the $\Gamma M$ direction.
The alternative limit~$E/2$ is justified by the consideration of exchange effects in Ref.~\cite{nguyen-truong_low-energy_2020}. 
On the other hand, it is criticized in Ref.~\cite{bourke_electron_2012} and it is argued that it is not a valid limit especially for plasmon excitations, which are the dominant source of inelastic losses in the integration domain here.
Therefore, for consistency, the standard upper limit $E-E_F$ is used in our calculations.
We note that several experimental data sets are scattered around the Nguyen-Truong result, but some are well below the values of Ref.~\cite{nguyen-truong_low-energy_2020}, e.g., Refs.~\cite{barrett_elastic_2005,geelen_nonuniversal_2019}.
The reason for the almost order of magnitude lower IMFP in Ref.~\cite{geelen_nonuniversal_2019} can be explained by the fact that the experimental data are closer to a related physical quantity - the so-called attenuation length, which also includes some elastic processes, see, e.g., Ref.~\cite{Chen_1992} for a comparison of these physical quantities.

\section{Summary and Conclusion}\label{sec:conclusion} 
We have presented a study of electron reflection spectroscopy of few-layer graphene using a 
multistep framework.
Initially, the EELS data were simulated using DFT and MBPT. 
Subsequently, the resulting spectra were employed to construct the optical potential and IMFP. Finally, the optical potential was utilized to incorporate inelastic effects into simulations of low-energy electron reflectivity.

Two methods were applied for the purpose of conducting reflectivity simulations.
The APW method and the simple supercell Bloch wave approach were employed, both with and without the inclusion of inelastic effects.
Three different cases of reflectivity spectra were simulated and compared to experimental data: normal incidence on a varying number of layers of FLG, normal incidence on different stacking types of TBG and TDBG, and general incidence of in-plane vectors along the $M\Gamma K$ path on~FLG.
It has been demonstrated that the incorporation of inelastic effects within the supercell Bloch wave approach results in overdamping, whereas the APW method yields results that are in good agreement with experimental data. 
The results obtained without inelastic effects agree for both methods but 
they are not suited for comparison with the experiment. 
The simulated angle-resolved reflectivity spectra also reasonably agree with experimental results from literature.
When comparing reflectivity spectra for regions of TBG and TDBG with different stacking types, the simulations indicate high contrast for certain ranges of the landing energy. This may be used in experiments to distinguish areas with different stacking types in SEM and also for the landing energies below \SI{30}{\eV}.

The intermediate results, namely the simulated loss functions of EELS and optical potential, are of interest on their own. The simulated EELS data have been compared to our experimental results and have been found to exhibit a reasonable degree of agreement. 
Good agreement can be regarded as a certain justification of the obtained optical potential. 

It can be concluded that low-energy electron reflectivity techniques (both normal incidence reflectivity and ARRES), interpreted with insight from simulations, are invaluable for the characterization of 2D materials.
These methods can serve as a crucial quality control step during the production process of 2D materials, enabling the accurate determination of key parameters such as the number of layers and their precise ordering with good spatial resolution.

\begin{acknowledgments}
This paper was supported by the Czech Science Foundation (Grant
No. GA22–34286S) and the Spanish Ministry 
of Science and Innovation (Grants No. PID2022-139230NB-I00 and No. PID2022-138750NB-C22). 
We are grateful to S. J. van der Molen and P. Neu from Leiden University for providing the experimental ARRES data published in Ref.~\cite{jobst_nanoscale_2015}.
The authors thank the following individuals for their expertise and
assistance throughout the experimental part of the study carried out in ISI CAS, namely F. Mika, J. S\'{y}kora and F. Hrub\'{y}.
A.P. thanks to R. D\'{i}ez Mui\~{n}o for his invitation to DIPC
which enabled the development of this common scientific project with E.K.
A.P. is thankful to the organizers of the \texttt{Yambo} computational school at ICTP for fruitful discussions.
A.P. and M.Z. acknowledge kind support on the \texttt{Yambo} forum and helpful feedback from 
D. Varsano, C. Attaccalite and D. Sangalli.
A.P., M.Z. and E.K. express their gratitude to V. Nazarov for clarifying discussions about the quasi-2D dielectric function.
M.H. acknowledges the project CzechNanoLab (MEYS CR, Project No. LM2023051) supporting the CEITEC Nano laboratories, where the EELS measurement was performed. 
M.H. and V.K. acknowledge support from Brno University of Technology (Project No.~FSI-S-23-8336). ACC and MV were supported by Institutional Subsidy for Long-Term Conceptual Development of a Research Organization granted to the Czech Metrology Institute by the Ministry of Industry and Trade of the Czech Republic.

\end{acknowledgments}

\appendix
\section{Supercell Bloch wave matching}\label{sec:matching}

While the variational embedding method (Appendix~\ref{sec:compdeta_refl}) offers a
universal solution to the low energy electron diffraction (LEED) problem, it is instructive
to compare it with the supercell Bloch wave matching approach, which has the advantage that
it can use as an input the output of any standard plane-wave band structure code. Here, we
briefly describe the principle of this approach.

Consider a plane wave $\exp(i\bq\br)$ incident onto the slab located between the boundaries
$z=z_L$ and $z=z_R$; see Fig.~\ref{scheme}. To the left of $z_L$ and to the right of $z_R$,
the potential is constant, thus the scattering wave function $\psi(\br)$ in the two semi-infinite
half-spaces is a sum of propagating and evanescent plane waves, each of which satisfies the
Schr\"odinger equation for the energy $E=(\hbar q)^2/2m$. In the left half-space the propagating
waves are the incident and reflected waves
\begin{equation}
\psi_L(\br)=\ee^{i\bq\br}+\sum_{\bG}r_{\bG}\ee^{i(\bq^{\paral}+\bG)\br^{\paral}-iq_{\bG}^\perp z}, \label{psiL}
\end{equation}
and in the right half-space it is the transmitted waves  
\begin{equation}
\psi_R(\br)=\sum_{\bG}t_{\bG}\ee^{i(\bq^{\paral}+\bG)\br^{\paral}+iq_{\bG}^\perp z}. \label{psiR}
\end{equation}
Here $\bq=\bq^{\paral}+q_{\mathbf{0}}^\perp\hat{\mathbf z}$, and $\bG$ are the surface reciprocal lattice vectors. 
The normal projection of the wave vectors of the secondary beams are $q_{\bG}^\perp=\sqrt{q^{\perp 2}_{\mathbf{0}}-|\bG|^2-2\bq^{\paral}\bG}$,
as follows from the requirement that the energy of the outgoing beams be equal to $E$.
Between the boundaries the wave function is sought as a linear combination of
the Bloch eigenstates $\psi_{\bk_n}^\lambda$ of the supercell whose energies are equal to 
$E$:
\begin{equation}
\psi_S(\br)=\sum_{n=1}^{D}f_n\psi_{\bk_n}^\lambda(\br). \label{psiS}
\end{equation}
The supercell eigenfunctions are labeled by the 3D Bloch vector
$\bk_n=\bk^{\paral}_{n}+k_n^\perp\hat{\mathbf z}$ and the band index $\lambda$, but because the
degenerate states belong to the same band in the following we drop the band index. All the
waves involved have the same surface projection of the Bloch vector: $\bk^{\paral}_{n}=\bq^{\paral}$.
For the normal incidence, $\bq^{\paral}=\mathbf{0}$, it follows from the the time-reversal symmetry
that the normal components of the two Bloch vectors in Eq.~(\ref{psiS}) are related as
$k_1^\perp=-k_2^\perp$. For a mirror symmetric slab this relation holds for any angle of
incidence.

Obviously, this method is not applicable to the energies that fall into the forbidden gaps
of the supercell band structure, which is its important limitation. Nevertheless, for a
sufficiently thick supercell the set of accessible energies is sufficiently dense to yield
a reasonable picture of the reflectivity curve.

The representation (\ref{psiS}) must match the plane wave
expansions (\ref{psiL}) and (\ref{psiR}) both in value and in slope at the boundaries. Here we consider
the simplest case of sufficiently low energies, at which the secondary beams do not
arise (hence the dark regions in Fig.~\ref{angles}). In this case, all $q_{\bG}^\perp$
except $q_{\mathbf{0}}^\perp$ are purely imaginary, and for $z_L$ and $z_R$, sufficiently far
from the crystal boundaries, the evanescent waves can be neglected, and in Eqs.~(\ref{psiL})
and~(\ref{psiR}) there remain only one reflected and one transmitted wave, respectively.
Thus, the two vacuum half-spaces are described by the two coefficients, $r_{\mathbf{0}}$ and $t_{\mathbf{0}}$.

The scattering wave function must satisfy the four matching conditions (continuity of
function and derivative at both boundaries): \mbox{$\psi_B(\br^{\paral},z_B)=\psi_S(\br^{\paral},z_B)$}, 
\mbox{$\partial_z\psi_B(\br^{\paral},z_B)=\partial_z\psi_S(\br^{\paral},z_B)$}, where \mbox{$B=L,R$}. 
Imposing uniqueness of the solution implies that at most four independent coefficients enter the four matching conditions; these are $r_{\mathbf{0}}$, $t_{\mathbf{0}}$ and $f_1$ and $f_2$ of Eq.~(\ref{psiS}).
As a result, the degeneracy is fixed to $D=2$. 
Using the Laue representation of the supercell eigenstates
$\psi_{\bk}(\br)=\ee^{i\bk^{\paral}\br^{\paral}}\sum_{\bG}\phi_{\bk}(\bG,z)\ee^{i\bG\br^{\paral}}$, we reduce
the matching problem to the $4\times 4$ system of linear equations:
\begin{equation*}
\begin{pmatrix} 1\\[5pt] iq_{\mathbf{0}}^\perp\\[5pt] 0 \\[5pt] 0 \end{pmatrix} =
\begin{pmatrix} -1 & \f_{\bk_1}^L & \f_{\bk_2}^L & 0\\[3pt] 
iq_{\mathbf{0}}^\perp & \partial_z\f_{\bk_1}^L & \partial_z \f_{\bk_2}^L & 0\\[3pt]  
0 & \f_{\bk_1}^R & \f_{\bk_2}^R & -\ee^{i q_{\mathbf{0}}^\perp z_R}\\[3pt] 
0 & \partial_z\f_{\bk_1}^R & \partial_z \f_{\bk_2}^R & -i q_{\mathbf{0}}^\perp \ee^{iq_{\mathbf{0}}^\perp z_R}
\end{pmatrix}
\begin{pmatrix} r_{\mathbf{0}}\\[6pt] f_1\\[6pt] f_2\\[6pt] t_{\mathbf{0}} \end{pmatrix} ,
\label{emab}
\end{equation*}
where $\phi_{\bk_n}^B = \phi_{\bk_n}(\mathbf{0},z_B)$ and $z_L=0$. Further details can be found in Ref.~\cite{mcclain}.

\section{EELS and optical potential}\label{sec:optical}
The energy losses occurring during inelastic scattering lead to a decrease of the intensity, damping of electronic density at the original energy.
It can effectively be achieved by adding an imaginary component to the Hamiltonian $H$. 
This imaginary term can be considered at different levels of sophistication.
The imaginary part of the self-energy $\Sigma$, correcting energy eigenvalues of the ground state, is a natural candidate within the Green's function formalism.
It is possible to calculate \textit{ab initio} $\Sigma$ directly using MBPT.
However, a simpler \textit{ab initio} model is used and the imaginary component is expressed by means of the dielectric function $\varepsilon$ using RPA, see Appendixes ~\ref{sec:rpa} and~\ref{sec:compdeta_eels}. 
The inelastic scattering can be characterized by
the optical potential $V_{\mathrm{opt}}$, the lifetime $\tau$, or the inelastic mean free path $\lambda$. These quantities differ by various prefactors only, e.g. $V_{\mathrm{opt}} = \hbar/2\tau$.

Let us consider a probe electron with initial momentum~$\bk_i$ scattering with the electrons in solid with energy below the Fermi energy~$E_F$ and acquiring final momentum~$\bk_f$.
Its inverse lifetime $\tau^{-1}$ can be expressed in terms of the RPA dielectric function $\varepsilon$ as follows~\cite{Echenique_CP251_2000}
\begin{equation}
\frac{1}{\tau} = \frac{2 e}{\hbar}\int\limits'\frac{\d^3\bq}{(2\pi)^3}V_{\bq}\ii\left\{\frac{-1}{\varepsilon(\bq, \omega)}\right\}
,\label{eqn:lifetime.elf.rpa}\end{equation}
where $V_{\bq}$ stands for the Fourier transform of the interaction potential, $\bq = \bk_i - \bk_f$ being the momentum transfer (in this appendix, the restriction of $\bq$ to the first Brillouin zone is not imposed), $\hbar\omega = E_{\bk_i} - E_{\bk_f}$ denotes the energy loss with $E_\bk = (\hbar k)^2/(2m)$ for the energy of the initial and final state and $e$ is the elementary charge.
In addition, the prime in the integration indicates the following restrictions:
$0 < \hbar\omega < E_{\bk_i} - E_F$, where the second inequality is a consequence of the fact that the probe electron cannot be scattered to the occupied state in the Fermi sea because of the Pauli exclusion principle.
Furthermore, the prime also accounts for the conservation of energy, which is often emphasized using the delta function in the integrand and the energy conservation law,
$$0 = C_E = \hbar\omega - [E_{\bk_i} - E_{\bk_i - \bq}] = \hbar\omega - \frac{\hbar^2}{2m}[2k_iqx - q^2],$$
where $x = \cos\theta$ with $\theta$ denoting the angle between $\bk_i$ and~$\bq$.
In the special case of the Lindhard dielectric function the region of momentum space that contributes to the integral is presented in Ref.~\cite{quinn_range_1962} with the boundaries given by the \textit{Pascal lima\c{c}on}. For the general case, the domain of integration is sketched in Ref.~\cite{gersten_theory_1970} using appropriately adapted coordinates.

The above formula~\eqref{eqn:lifetime.elf.rpa} uses the so-called bulk term, which identifies ELF $\mathcal{L}$ with $\ii\left\{-1/\varepsilon\right\}$.
Since the DFT calculations effectively are bulk calculations (even in the case of a slab in the computational cell), the bulk term is used and we merely note more elaborate extensions to semi-infinite dielectric with a surface do exist~\cite{Gersten_PR188_1969}.

An alternative formula for $\tau^{-1}$ can be derived in case of 
an isotropic dielectric.
First, let us introduce a dummy integral over $\omega$, the value of which is already fixed by the energy conservation law.
The above restrictions on the final energy $E_{\bk_f}~\geq~E_F$ define the limits of the integration over $\omega$: 
$$\frac{1}{\tau} =
\frac{2e}{\hbar}\int\limits^{E_{\bk_i} - E_F}_0\d\left(\hbar\omega\right)\int\limits\frac{\d^3\bq}{(2\pi)^3}
V_{\bq}\mathcal{L}(\bq, \omega)\,\delta\left(C_E\right).$$

If the material is isotropic the integral in $\bq$ space is best performed in spherical coordinates.
Because the cosine function (of real variable) satisfies $|x| = |\cos\theta| \leq 1$, the following inequality may be derived
\begin{equation*}
0 = C_E\Rightarrow \frac{2m}{\hbar}\omega = 2k_iqx - q^2 \leq 2k_iq - q^2
.\end{equation*}
This implies that the allowed $q$ must lie within the interval $[q_-, q_+]$, where $\hbar q_{\pm} = \sqrt{2mE_{\bk_i}}\pm\sqrt{2m(E_{\bk_i} - \hbar\omega)}$ are the roots of the above inequality.
Such a restriction ensures $x\leq 1$ for a solution $x(q)$ of $C_E=0$.

Finally, the assumption of an 
isotropic medium implies that the integrand depends only on $q$, and the angular integration can be easily performed.
The integration over the azimuthal angle is trivial and yields a multiplicative factor $2\pi$.
The integration over polar angle $\theta$ requires one more step--switching to integration over $x\hbar^2k_iq/m$.
The corresponding substitution means that the $\delta(C_E)$ is integrated to unity and an extra multiplicative factor $m/(\hbar^2k_iq)$ appears due to the substitution.
Hence the final result is 
\begin{equation}
\frac{1}{\tau} = \frac{2V_{\mathrm{opt}}}{\hbar} = \frac{v}{\lambda} = 
\frac{b}{k_i}
\int\limits^{E_{\bk_i} - E_F}_0\d\left(\hbar\omega\right)\int\limits^{q_+}_{q_-}\frac{\d q}{q}\mathcal{L}(q, \omega)
,\label{eqn:lifetime.elf.rpa.isotropic}\end{equation}
where the symbols have the following meaning:
$v$~stands for velocity $\hbar k_i/m$, $b$~is a constant equal to 
$me^2/(2\pi^2\hbar^3\varepsilon_0)$
and $\varepsilon_0$~is the vacuum permittivity.
The energy-loss function $\mathcal{L}$ is given by the bulk formula.

Such a simplified formula in Eq.~\eqref{eqn:lifetime.elf.rpa.isotropic} is often used to acquire the energy profile of IMFP 
even for crystalline materials with MT along a single direction.

\section{Inclusion of inelastic effects}\label{sec:inelastic}

The supercell Bloch wave approach outlined in Appendix~\ref{sec:matching}
can be extended to include inelastic effects as described in Ref.~\cite{gao_inelastic_2015}.
In their study, the authors employed a linear energy dependence of the optical potential. 
This approach can be improved by using the optical potential derived from the ELF simulation, rather than assuming its linear energy dependence.
\begin{figure}[h]
\includegraphics[width=0.95\linewidth]{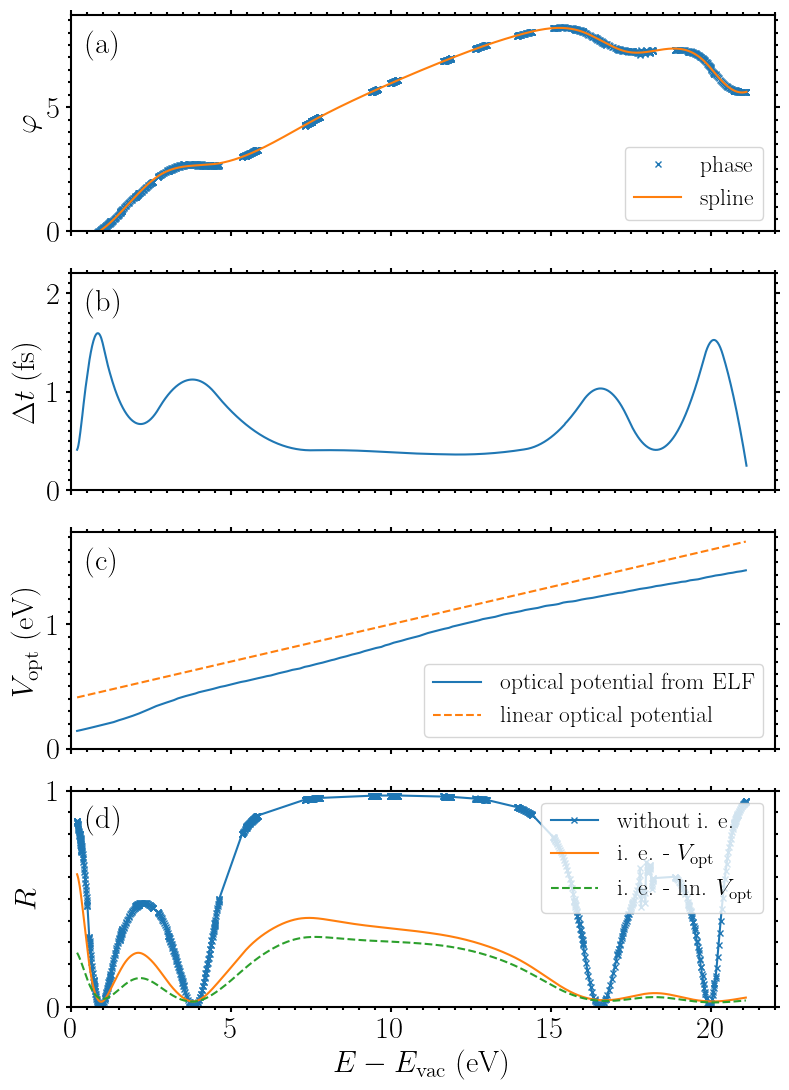}
\centering
\caption{Main steps in the inclusion of inelastic effects for the freestanding slab of a three-layer graphene: the phase $\varphi$ (blue symbols) together with the corresponding spline (orange solid line) are displayed in (a), the dwell time $\Delta t$ is shown in~(b), the optical potentials $V_{\mathrm{opt}}$ obtained from ELF (blue solid line) and the linear one (orange dashed line) are presented in~(c), the reflectivities $R$ without inelastic effects as obtained from the simple supercell Bloch wave approach (blue solid line with symbols) are shown in~(d) alongside the reflectivities with inelastic effects utilizing $V_{\mathrm{opt}}$ obtained from ELF (orange solid line) and the reflectivities with inelastic effects utilizing the linear $V_{\mathrm{opt}}$ (green dashed line).
$E_{\mathrm{vac}}$ denotes energy in vacuum.
}
\label{comparison}
\end{figure}

It is instructive to compare the straightforward approach, in which the optical
 potential is included into the Hamiltonian~\cite{krasovskii} with the physically
 intuitive procedure of Ref.~\cite{gao_inelastic_2015} based on the concept of absorbing
 potential. Its main steps are illustrated in Fig.~\ref{comparison}: First, the phase shift
 $\varphi$ is computed between the incident and reflected wave. Its energy derivative yields
 the dwell time $\Delta t_{\alpha}$ for a state $\alpha$ with energy
 $E_{\alpha}$, see Ref.~\cite{gao_inelastic_2015, merzbacher}:
$$
\Delta t_{\alpha} = \hbar \left( \frac{d\varphi}{dE}\right)_{E_{\alpha}}-2\int_0^{z_E}\frac{\d z}{\sqrt{2[E_{\alpha}-V(z)]/m}},
$$
where $V(z)$ is the potential averaged over the surface plane, $m$ is the electron rest
mass, and $z_E$ is the nominal boundary of the scatterer, which is here placed one a.u.
to the left of the leftmost atomic layer.
The current absorbed in the scatterer is proportional to the time spent inside the scatterer.
The imaginary potential $-iV_{\mathrm{opt}}$ implies a characteristic decay time $\tau =\hbar/2V_{\mathrm{opt}}$, hence the reflected wave is attenuated by the factor $\exp (-\Delta t_{\alpha}/\tau)$.
It should be noted that for the method to be consistent 
the same procedure should be applied to the transmitted wave, and the current conservation should be ensured:
The sum of the reflected, transmitted, and total absorbed current should equal the incident current.
Apparently, arbitrarily chosen boundaries may lead to a violation of the current conservation.

\begin{figure}[h]
\includegraphics[width=0.92\linewidth]{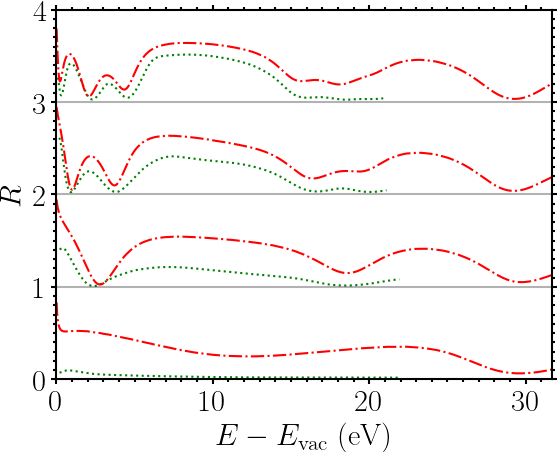}
\centering
\caption{Comparison of the inclusion of inelastic effects into normal incidence FLG reflectivity $R$ for the freestanding slab of $n = $ 1 to 4 layers, from bottom to top.
This is achieved using the approach described in Appendix~\ref{sec:inelastic} (green dotted lines) and with the implementation in the APW method (red dotted-dashed lines).
The calculated reflectivities as functions of landing energy ($E_{\mathrm{vac}}$ denotes energy in vacuum) are shifted vertically by $n - 1$ for the sake of clarity and they are separated by horizontal gray solid lines.
}
\label{comparison_with_ISOLDA}
\end{figure}

The inclusion of inelastic effects as described in this appendix is compared with the results obtained using the APW method in Fig.~\ref{comparison_with_ISOLDA}.
The edge of the slab $z_E$ was redefined as one half of the interlayer distance farther to the left of the left-most atomic layer.
This modification ensures the absorbing slab regions become consistent with those in the APW method model.

The optical potentials shown in Fig.~\ref{fig:optical_potential_compare} are used in both methods. 
However, the approach of Ref.~\cite{gao_inelastic_2015} seems to overdamp the reflectivities, especially in the case of monolayer graphene.
The dependencies of the reflectivities obtained using the approach of this appendix are also limited to the energy range from 0 to \SI{22}{\eV}, since an unphysical peak appears in the reflectivities for higher landing energies.
This spurious peak is caused by the fact that the dwell time acquires negative values, and the advice of Ref.~\cite{gao_inelastic_2015} to truncate all dwell times to zero didn't solve the problem. 
Therefore, reflectivities with inelastic effects included in Figs.~\ref{niflg} and~\ref{angles} are comprised by the APW method only. 
The parameters of the linear dependence of $V_{\mathrm{opt}}$ on energy can be regarded as fitting parameters and correspondingly adjusted to improve agreement with the experimental data.
Then the approach of this appendix describes well the inclusion of inelastic effects in comparison with experiment; see Ref.~\cite{gao_inelastic_2015}
for the case of graphene on copper.

\section{Ground state calculations}\label{sec:struc_opt}

Several norm-conserving pseudopotentials were tested.
The DFT calculation of ground state structures, including relaxation, were accomplished with the \textsc{Quantum ESPRESSO} software package~\cite{QE-2009,QE-2017}.
Convergence tests were performed with respect to the kinetic energy cutoff for wavefunctions (\texttt{ecutwfc}) in the range from \SIrange{40}{120}{Ry} with a step of \SI{10}{Ry} and with respect to the k-point grid starting with $12\times 12 \times 1$ Monkhorst-Pack (MP) grid and ending with $22\times 22 \times 1$ MP grid with a step of 2 in the first two grid parameters.
Several quantities were compared and their differences from the highest tested values--the differences in the total energy in self-consistent DFT calculations were considered converged if their absolute values were smaller than \SI{5}{\meV}.

After that, a relaxation of the considered structures was carried out. It was accomplished in two runs. First, a relaxation of the bulk structure (\texttt{vc-relax}) which produces an in-plane lattice constant.
Second, a relaxation of the vertical coordinates (\texttt{relax}) which gives relaxed interlayer spacings.
The bulk relaxation led to comparable values of the lattice constant for all pseudopotentials. However, 
the relaxed interlayer spacings were obviously overestimated for two tested Perdew-Burke-Ernzerhof pseudopotentials,
\SI{3.84}{\angstrom} and \SI{4.24}{\angstrom}.

The selected optimized norm-conserving Vanderbilt pseudopotential in the scalar-relativistic LDA version~\cite{Hamann} gives both structural parameters in very good agreement with experiment.
A~relaxation of the bulk structure gives an in-plane lattice constant \SI{2.4486}{\angstrom} which agrees well with experiments \SI[separate-uncertainty = true]{2.45(4)}{\angstrom} 
or \SI[separate-uncertainty = true]{2.4589(5)}{\angstrom}, reported in Refs.~\cite{ndiaye_structure_2008,baskin}, respectively.
Relaxation along the $z$ axis (see Appendix~\ref{sec:matching}) produces an interlayer distance for AB-stacking \SI{3.3129}{\angstrom} being in good agreement with an experiment~\cite{baskin}, \SI{3.3378}{\angstrom} at \SI{78}{\kelvin}, too.
Other relaxed parameters, e.g. for non-AB-stacking, are in good agreement with the literature, too.

Relaxation of AA-stacked bilayer graphene gives the interlayer distance \SI{3.6073}{\angstrom}.
This is in good agreement with the experiment in Ref.~\cite{lee_growth_2008}, where the value for the interlayer distance was $\approx$\SI{3.55}{\angstrom}.
It also compares well with other DFT simulations, see Ref.~\cite{campanera_density_2007}, where the interlayer separation of AA-stacked graphite is calculated to be \SI{3.591}{\angstrom} (for BA-stacked graphite it is \SI{3.321}{\angstrom}).
The validity of LDA calculations was examined in Ref.~\cite{uchida_atomic_2014} by employing the van der Waals functional with essentially identical results for the AA interlayer distance \SI{3.56}{\angstrom} and AB interlayer distance \SI{3.31}{\angstrom}.
In Ref.~\cite{uchida_atomic_2014}, the atomic and electronic structure of TBG was clarified by means of large-scale DFT simulations. This is a formidable task due to the large supercell, which consists of more than 10\,000 atoms. Therefore, it is reasonable to study TBGs by means of an analytical approach~\cite{neek_membrane_2014} or tight-binding calculations~\cite{brihuega_unraveling_2012}. 
The distance between two consecutive graphite planes using vdW-DF2 is $\SI{3.31}{\angstrom}$ (BA stacking) and $\SI{3.496}{\angstrom}$ (AA stacking) and a TBG's average distance $\SI{3.408}{\angstrom}$ at the magic angle $\Theta = \SI{1.08}{\degree}$ is reported in Ref.~\cite{Lucignano_PRB99_2019}.

Regarding TDBG, the relaxed interlayer distance between the surface and inner layers is \SI{3.3110}{\angstrom} and between two inner layers is \SI{3.6018}{\angstrom}, which compares well with the values from Ref.~\cite{haddadi_moire_2020}: \SI{3.364}{\angstrom} and \SI{3.647}{\angstrom}, respectively.

Hence, this pseudopotential is the best choice for our calculations both at the DFT level and at the MBPT level with \texttt{Yambo}, even if we take into consideration the well known fact that LDA tends to overbind in general.

Gaussian smearing of \SI{0.001}{\rydberg} is utilized. The plane
wave cutoff energy (\texttt{ecutwfc}) is set to \SI{90}{\rydberg}; the kinetic energy cutoff for charge density and potential (\texttt{ecutrho}) equals to \SI{360}{\rydberg}. The MP grid for the self-consistent ground state calculation is $14\times 14 \times 1$.
Number of bands is equal to 180.
The simulation size in the $z$ direction is minimized 
in such a way that the minimum distance from the surface layers to the 
supercell boundary is at least \SI{15}{\angstrom}.

\section{Computational details - Reflectivities}\label{sec:compdeta_refl}

The same relaxed values for the lattice constant and the interlayer distance were used in the APW method.
Values of the work function~$\Phi$ extracted from \textsc{Quantum ESPRESSO} simulations were also utilized as input parameters in the APW method, see Table~\ref{tab:work}.
\begin{table}[h]
\begin{ruledtabular}
\begin{tabular}{ccccc}
Number of layers & One & Two & Three & Four\\
\\\hline
$\Phi$ & \SI{5.11}{\eV} & \SI{4.08}{\eV} & \SI{4.61}{\eV} & \SI{4.61}{\eV}
\end{tabular}
\end{ruledtabular}
\caption{Work function $\Phi$ as obtained using \textsc{Quantum ESPRESSO} for FLG and employed in the APW method .
}\label{tab:work}
\end{table} 
The absorbing slab was enlarged by $n$-times the interlayer distance for the $n$-layer graphene.

A more accurate and consistent approach to provide the reflectivity data is offered by the band-structure formalism based on APWs.
The LEED problem for the freestanding films is solved with the variational embedding method of Ref.~\cite{Eugeneregularization}.
The space is divided into two semi-infinite vacuum halfspaces separated by the scattering region containing the graphene multilayers.
The LEED wave function $\Phi_{\rm LEED}$ is defined by the surface-parallel Bloch vector $\bkp$.
In the scattering region, $\Phi_{\rm LEED}$ is sought as a superposition of the Bloch eigenfunctions $\psi^{\lambda}_{\bkp}$ of an auxiliary three-dimensional $z$-periodic
crystal, which contains the scattering region as a part of the unit cell.
The supercell lattice constant perpendicular to the graphene plane was $c=\SI{15}{\angstrom}$ for the monolayer and \SI{20.7}{\angstrom} for the thicker films.
The wave functions $\psi^{\lambda}_{\bkp}$ are obtained with the full-potential APW method of Ref.~\cite{KSS1999}.
The linear combination of $\psi^{\lambda}_{\bkp}$ is constructed so as to best satisfy the Schr\"odinger equation in the scattering region and to match at its boundaries the function and derivative of the plane-wave expansions in the vacuum half-spaces. 
The basis set comprised around 200 $\psi$ functions with energies up to at least \SI{80}{\eV} above the Fermi energy.
Then the all-electron wave function is expanded in terms of plane waves with the wave vectors within the sphere of $G=\SI{11}{\au}^{-1}$, which corresponds to 11985 plane waves for the monolayer and 16969 for the larger supercell. For the matching of $\Phi_{\rm LEED}$ at the boundaries of the scattering region 19 surface reciprocal vectors were taken into account.

\section{RPA method}\label{sec:rpa}
Since the RPA approach is employed to simulate the dielectric function, a rather brief description of RPA is included in this Appendix.
This is to make the present paper more readable and self-contained.

The relation between the (retarded) inverse dielectric function $\varepsilon$ and the density response function (polarizibility) $\chi$, is best written in the reciprocal space, i.e., as a relation between Fourier transforms of these two quantities.
Let us index the Fourier coefficients by the reciprocal lattice vectors~$\mathbf{G}, \mathbf{G}'$ and denote the Coulomb kernel by $V$. Then the following formula can be written
\begin{equation}
\varepsilon^{-1}_{\mathbf{G}\mathbf{G}'}(\mathbf{q},\omega) = \delta_{\mathbf{G}\mathbf{G}'} + V_\mathbf{G}(\mathbf{q})\chi_{\mathbf{G}\mathbf{G}'}(\mathbf{q},\omega);
\label{eqn:epschi}
\end{equation}
see Appendix~L of Ref.~\cite{stefanucci_nonequilibrium_2013}.
Of course, both quantities are considered as functions of MT~$\mathbf{q}$ and energy loss\footnote{In this Appendix, the reduced Planck constant $\hbar$ is set to unity in order to simplify expressions and keep the notation in close alignment with the references~\cite{stefanucci_nonequilibrium_2013, cudazzo_local-field_2013, marini_optical_nodate,lin_mathematical_2019,Hybertsen,altland_simonds_book} cited in this section.}~$\omega$.

The macroscopic dielectric function is reciprocal of the $\mathbf{G}=\mathbf{G}'=\mathbf{0}$ component on the left-hand side~of~Eq.~\eqref{eqn:epschi}.
Generally, for inhomogeneous systems, the off-diagonal elements can have significant contribution to $\varepsilon^{-1}_{\mathbf{0}\mathbf{0}}$; the associated phenomena are called crystal local-field effects--more information can be found in Ref.~\cite{cudazzo_local-field_2013}.
The polarizibility $\chi$ in Eq.~\eqref{eqn:epschi} is given by a Dyson equation
\begin{equation}
\chi \approx \chi_0 + \chi_0 V \chi,
\label{eqn:dyson-like}\end{equation}
where the symbol $\approx$ reflects our neglecting the exchange-correlation kernel in RPA.
$\chi_0$ is the so-called irreducible polarizibility or the non-interacting density response function given by the effective one-body non-interacting (Kohn-Sham) theory. 

There are several ways to obtain the irreducible polarizibility.
One approach is a convolution of Green's function with itself, see the Ph.D. thesis of the originator of the \texttt{Yambo} project~\cite{marini_optical_nodate}.
However, the ideas from the introductory textbook on DFT and the linear response theory~\cite{lin_mathematical_2019} are followed, where $\chi_0$ is introduced as a derivative of the electron density $\rho$ with respect to a local potential instead of the aforementioned convolution.
In other words, we intend to express differences of perturbed and unperturbed densities 
 within linear response theory, i.e., up to the first-order perturbation.

Let us consider $\rho(\br)$ as the diagonal of the density matrix $P(\br,\br)$. 
Let us write the density matrix using the contour integral
\begin{equation}
P = \frac{1}{2\pi i}\oint_{\mathscr{C}}f_{\beta}(\lambda-\mu)(\lambda-H)^{-1}\d \lambda,
\label{eqn:denmat}
\end{equation}
where $f_{\beta}(E)=(1+\exp(\beta E))^{-1}$ is the Fermi-Dirac distribution for inverse temperature~$\beta$, $\mu$ being the chemical potential, $H$ is the unperturbed Hamiltonian and $\mathscr{C}$ is a contour close to the real axis enclosing the entire spectrum of~$H$ and avoiding poles of the Fermi-Dirac distribution extended to the complex plane--the so-called Matsubara frequencies on the imaginary axis.

Now, let us consider a perturbed Hamiltonian $H_{\epsilon}=H+\epsilon W$ with the perturbation $W$ given by a Hermitian operator satisfying mild conditions such as $H$ boundedness, see Ref.~\cite{lin_mathematical_2019}.
Hence, the perturbed density matrix~$P_{\epsilon}$ can be expressed in a direct analog of Eq.~\eqref{eqn:denmat} simply by using the Green's function (resolvent)~$(\lambda-H_{\epsilon})^{-1}$ instead.
The conditions on $W$ ensure that the contour integral is well defined for sufficiently small $\epsilon$.

Let us compute $\mathcal{X}_0$, the derivative of the perturbed density matrix 
at $\epsilon=0$ in the direction~$W$.
Introducing a shorthand notation $f_p\equiv f(E_p)=f_{\beta}(E_p-\mu)$, we can write
\begin{eqnarray*}
P_{\epsilon}-P &=& \frac{1}{2\pi i}\oint_{\mathscr{C}}f(\lambda)\left[(\lambda-H_{\epsilon})^{-1}-(\lambda-H)^{-1}\right]\d \lambda\\
&=& \frac{\epsilon}{2\pi i}\oint_{\mathscr{C}}f(\lambda)\frac{1}{\lambda-H} W\frac{1}{\lambda-H}\d \lambda + \mathcal{O}(\epsilon^2)\\
&=& \epsilon\,\mathcal{X}_0 W + \mathcal{O}(\epsilon^2),
\end{eqnarray*}
where a standard resolvent identity and the Taylor serieslike expansion of the perturbed resolvent were used.
The spectral decomposition of the resolvents gives us
\begin{eqnarray*}
\mathcal{X}_0 W &=& \frac{1}{2\pi i}\oint_{\mathscr{C}}\sum_{p,q}f(\lambda)\frac{\ket{\psi_p}\mel{\psi_p}{W}{\psi_q}\bra{\psi_q}}{(\lambda-E_p)(\lambda-E_q)}\d \lambda \\
&=&\sum_{p,q} \frac{f_p-f_q}{E_p-E_q}\ket{\psi_p}\mel{\psi_p}{W}{\psi_q}\bra{\psi_q},
\end{eqnarray*}
where the Cauchy integral formula was used and the summation is over all orbital indices.
Let us consider the density response function $\chi_0$ as a diagonal of $\mathcal{X}_0$ and treat it as an integral kernel acting on a local potential $W$.
Writing $\chi_0$ in coordinate representation results in
\begin{equation*}
\chi_0(\br,\br')=\sum_{p,q} \frac{f_p(1-f_q)}{E_p-E_q}[\psi_p^{\ast}(\br)\psi_q(\br)\psi_q^{\ast}(\br')\psi_p(\br') + \text{c.c.}],
\end{equation*}
where the 
identity \mbox{$f_p-f_q=f_p(1-f_q)-f_q(1-f_p)$} was used and the summation indices were interchanged to get the second complex conjugate term; see, e.g., Ref.~\cite{Hybertsen}.

If the Hamiltonian depends explicitly on time, then one can do the same considerations as in Ref.~\cite{lin_mathematical_2019} (from time-dependent $P$ to time-dependent $\mathcal{X}_0W$ and perform a Fourier transform from the time domain to the frequency $\omega$ domain).
The above formula for $\chi_0$ transforms as follows:
$1/(E_p - E_q)$ is replaced by $1/(\omega + E_p - E_q +i\eta)$ in the first term and by $-1/(\omega - E_p + E_q +i\eta)$ in the c.c. term.
One can see alternative expressions, where the signs at $i\eta$ terms can be different, changed in the c.c. term only or in both terms.
This is due to competing definitions of response functions--retarded, advanced or time-ordered--leading to different signs of $i\eta$, see e.g. \cite{altland_simonds_book} for comparison.

Finally, it is necessary to perform the Fourier transform from the real space to the reciprocal space.
To arrive at an expression which can be implemented numerically for solids, Bloch waves 
are used with band index~$n$ and wave vector~$\bk$ as the one-particle states.
Then, several tricks can be performed, such as replacing integrals over the whole space by an integral over a unit cell followed by a sum over lattice vectors, assuming a time-inversion symmetry 
and constructing a basis 
using Kramer's theorem.
Finally, the following formula is obtained (see \texttt{Yambo} documentation or Ref.~\cite{marini_optical_nodate});
\begin{eqnarray*}
\chi_{{\textstyle\mathstrut}0_{\bg\bg'}}(\bq,\omega) = 2\lim_{\eta\rightarrow 0+}\sum_{nn'}\int_{\Omega^{\ast}}\frac{\d^3\bk}{(2\pi)^3}\tilde{\rho}_{nn'}^{\ast}(\bk,\bq,\bg)\\
\times\tilde{\rho}_{nn'}(\bk,\bq,\bg')f_{n'(\bk-\bq)}(1-f_{n\bk})\\
\times\biggl[\frac{1}{\omega-E_{n\bk}+E_{n'(\bk-\bq)}+i\eta}\quad\ \\
-\frac{1}{\omega-E_{n'(\bk-\bq)}+E_{n\bk}-i\eta}\biggr],
\end{eqnarray*}
where $\tilde{\rho}_{nn'}^{\ast}(\bk,\bq,\bg)=\int\limits \d^3\br u_{n\bk}^{\ast}(\br)u_{n'\overline{(\bk-\bq)}}(\br)\ee^{i(\bg+\bg_{kq})\br}$ stands for the screening matrix elements, with the over-line denoting the corresponding vector translated to the first Brillouin zone~$\Omega^{\ast}$, $\bg_{kq}$ their difference 
and $u$ the periodic part of Bloch wave.
Several comments are in order.
First, the above density response function is the time-ordered one (the sum of the greater and lesser components; see Ref.~\cite{stefanucci_nonequilibrium_2013}).
Second, one should be aware of the different response functions, however the information stored in them is essentially equivalent~\cite{altland_simonds_book}.

\section{Computational details - EELS}\label{sec:compdeta_eels}

The MBPT calculations utilize RPA using the Hartree kernel and including LFEs as already mentioned in Sec.~\ref{sec:momentum.resolved.EELS}.
Convergence tests lead to the following values of the relevant parameters of the calculations:
The k-point grid is $90\times 90 \times 1$ for computations along the $\Gamma M$ and $\Gamma K$ path, the supercell size in the direction perpendicular to graphene is 55~a.u., and the 
energy cutoff for expanding the wave functions (\texttt{FFTGvecs}) is not reduced when compared to \texttt{ecutwfc} and the value 90 Ry is used.
The maximal number of bands entering in the sum over states in the RPA response function (\texttt{BndsRnXd}) converge at~110.
The energy cutoff in the screening (\texttt{NGSBlkXd})
is converged 
to the final value of $12.728$ Ry.
The number of random q points in the Brillouin zone to perform Monte Carlo integration (\texttt{RandQpts}) is set to $3\times 10^6$ and number of G vectors of the integration method (\texttt{RandGvec}) is 100.
Electronic temperature (\texttt{ElecTemp}) 0.001 eV is applied. 
The number of energy steps (\texttt{ETStpsXd}) used is either 250 or 500.

The MP grid $30\times 30 \times 12$ is utilized to simulate dielectric function along the $q_z$ direction.
We note that it would be desirable to use denser MP grids to achieve stricter convergence criteria.
It is possible to shift the grid and use the double-grid method in \texttt{Yambo}.
However, this approach proved to be too demanding on our resources.
The more feasible MP grids were accepted since the corresponding results already compare well with literature, see, e.g., Table~\ref{tab:plasmons}, and one of our main goals--acquiring the optical potential from the dielectric function--is fairly insensitive to small discrepancies, caused by a little bit sparser grid, because of the integral in~Eq.~\eqref{eqn:lifetime.elf.rpa.isotropic}.

The formula for IMFP, Eq.~\eqref{eqn:lifetime.elf.rpa.isotropic}, requires ELF to be simulated for momentum transfers well outside the first Brillouin zone. 
The sampled momentum transfer $\bq$'s contained rather large gaps along the examined directions.
A possible way to overcome this problem without modifying MP grids is to fit the simulated data with an appropriate model such as the extended two-fluid hydrodynamical model~\cite{Fetter.AP88.1974,Jovanovic.PRB84.2011, Djordjevic_UM184_2018}. 
Then this model should interpolate over the gaps in $\bq$'s or to extrapolate to $\bq$ points inaccessible by simulation.
However, this approach proved to be very tricky, especially for larger magnitudes of $\bq$.
It turns out that the $q$ dependence has to be extended to some other parameters of the model to fit the \textit{ab initio} data well.
The second problem encountered was the following.
Given that the supercell approach is employed, it is necessary to address the spurious interaction of artificial copies inherent in periodic boundary conditions (i.e., the supercell is periodically repeated along the $z$ direction perpendicular to the atomically thin crystal).
To minimize such spurious interactions, one increases the height of the computational cell until the spectra converge. 
However, increasing the interlayer distance of the copies doesn't prevent the unwanted interaction at the optical limit for a sufficiently small MT~$\mathbf{q}_{\paral}$, see, e.g., Ref.~\cite{nazarov.NJP17.2015}.
\texttt{Yambo} tackles the spurious interaction at the computational level by truncating the electrostatic potential~\cite{Ismail.PRB73.2006,Rozzi.PRB73.2006,guandalini_efficient_2023} at an appropriate distance from a material with sheetlike geometry.
Such a truncation also speeds up the calculations by modifying the bare Coulombic potential. 
This is equivalent to setting the \texttt{Yambo} variable \texttt{CUTGeo} to the value \texttt{slab~z}.
Despite this cut providing the expected spectra near the $\Gamma$ point, see Fig.~\ref{fig:eels}, it turns out it introduces periodic oscillations of the spectra amplitude along the $\Gamma A$ direction.
We consider them an artifact and their removal proved problematic.

These large gaps in $\bq$'s and these oscillations were the reasons to use an alternative code: \texttt{turboEELS}.
It is an open source software component of \textsc{Quantum ESPRESSO}, which implements the Liouville-Lanczos approach for time-dependent density functional perturbation theory, see, e.g., Refs.~\cite{timrov_turboeelscode_2015,motornyi_electron_2020,timrov_electron_2013}.

No empty states are needed in this approach, RPA with LFE is available, and the algorithms work for uniformly spaced $\bq$ points even outside the first Brillouin zone.
However, the spurious interaction of the artificial copies is not addressed in \texttt{turboEELS} and will be discussed later.
The same MP grids as in our \texttt{Yambo} simulations were used.
The approximation \texttt{RPA\_with\_CLFE} was activated together with the \texttt{lanczos} calculator. 
A~convergence test was made for the number of Lanczos iterations--was set to 1750 for the in-plane momentum transfers and 3000 for the out-of-plane momentum transfers.
The extrapolation \texttt{osc} of the Lanczos coefficients was used--was set to 20\,000 in both cases.
The Lorentzian broadening parameter \texttt{epsil} is equal to \SI{0.013}{\rydberg}.

Two challenges of using \texttt{turboEELS} were found.
First, the calculations cannot be performed at the $\Gamma$ point.
Second, the spectra at the $K$ point displayed an incorrect position of the $\pi$-plasmon in the case of the $90 \times 90 \times 1$ grid.
This issue persisted even when we increased \texttt{itermax0} to~$7000$. 
The boundary of the first Brillouin zone and integer multiples of its distance from the $\Gamma$ point are also problematic in other directions.
A slight shift, lowering the magnitude of the critical \mbox{$\bq$ points} by $1~\%$ of the distance from the $\Gamma$ point to the boundary of the first Brillouin zone, led to correct results. 

\begin{figure}[h]\begin{center}
\includegraphics[width=0.95\linewidth]{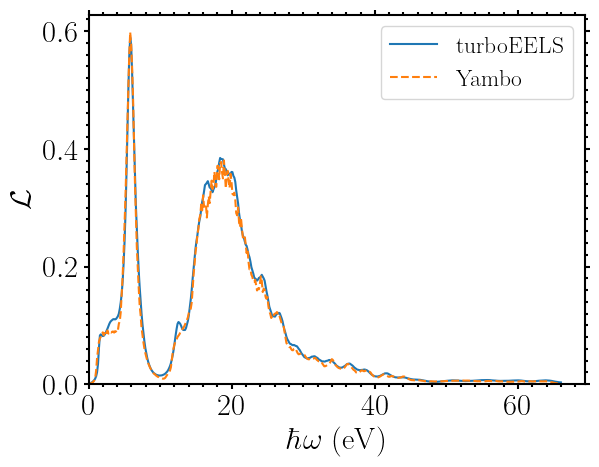}
\caption{
Comparison of the loss function~$\mathcal{L}$ dependence of graphene on the energy loss~$\hbar\omega$ for 
$q \doteq \SI{0.2}{\angstrom}^{-1}$
along $\Gamma M$ calculated using the Liouville-Lanczos approach (blue solid line), with the calculation 
obtained using the Dyson equation (orange dashed line).
}\label{eels_compare}\end{center}\end{figure}

A comparison of the energy loss function of graphene is presented for 
$q \doteq \SI{0.2}{\angstrom}^{-1}$
along $\Gamma M$ calculated using the Liouville-Lanczos approach as implemented in \texttt{turboEELS}, with previous calculations 
based on the solution of the Dyson equation as implemented in \texttt{Yambo} in Fig.~\ref{eels_compare}. The agreement between these approaches is excellent.

Furthermore, the validity of the \mbox{$f$-sum} rule was tested 
\begin{equation}
\frac{1}{C_f} = \frac{2}{\pi}\frac{1}{\omega^2_P}\int_0^{\infty} \omega \ii \left\{\frac{-1}{\varepsilon(\omega)}\right\}\d\omega \approx 1,
\label{eqn:f-sumrule}
\end{equation}
where $\omega_P=(4\pi n e^2/m)^{1/2}$, $n$ is the local (effective) 
number density of electrons, $m$ is the (effective) mass of the electron, $e$ is its charge, and $C_f$ just denotes the normalization multiplication factor for ELF.
This rule is applicable to all values of momentum transfer in general, see \cite{chantler_low-energy_2019}.
This sum rule is usually used to check the validity of an approximation in simulated dielectric data or the consistency of the measured data.
The validity of our simulated data was verified by implementing this integral and the mismatch is directly accessible also in the \texttt{turboEELS} logs. 
Our \texttt{turboEELS} results violate the \mbox{$f$-sum} rule by $\approx 10\, \%$ only in the neighbourhood of the $\Gamma$ point.
The \mbox{$f$-sum} rule is hugely violated for high momentum transfers.
In Ref.~\cite{sabio_f-sum_2008}, the \mbox{$f$-sum} rule is analyzed for two-dimensional systems.
Their finding is that the right-hand side of Eq.~\eqref{eqn:f-sumrule} has to be generalized to depend on $q$ even in the low-energy regime of undoped graphene in RPA.
We note that the aforementioned violation may be partly due to the fact that our calculated energy losses $\hbar\omega$ are limited to $66$~eV.

\begin{figure*}
\centering
		\includegraphics[clip,trim=0 {2.2cm} {5.0cm} 0,height=0.26\textwidth]{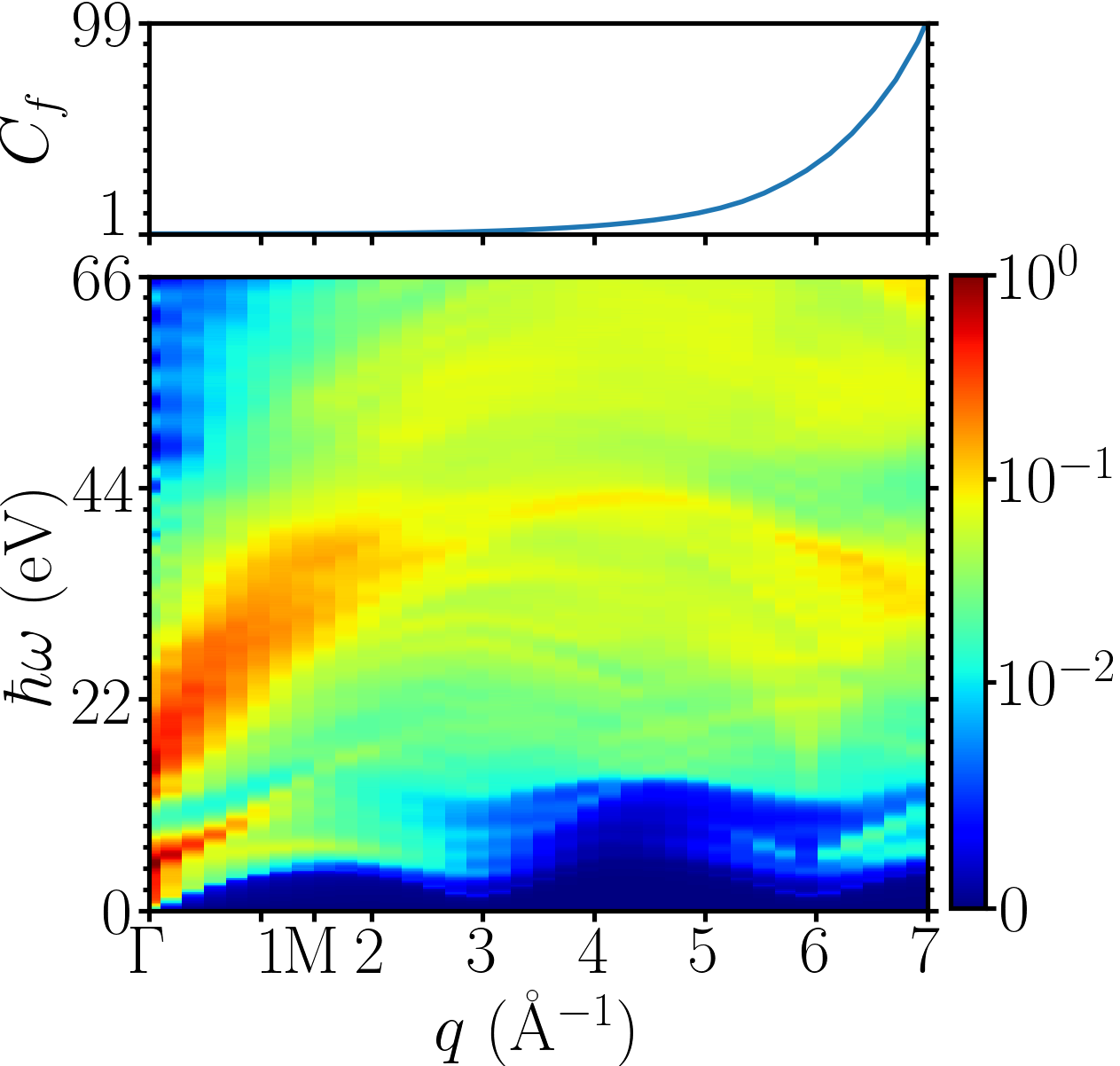}\hspace{0.1cm}
       \includegraphics[clip,trim={3.76cm} {2.2cm} {5.0cm} 0,height=0.26\textwidth]{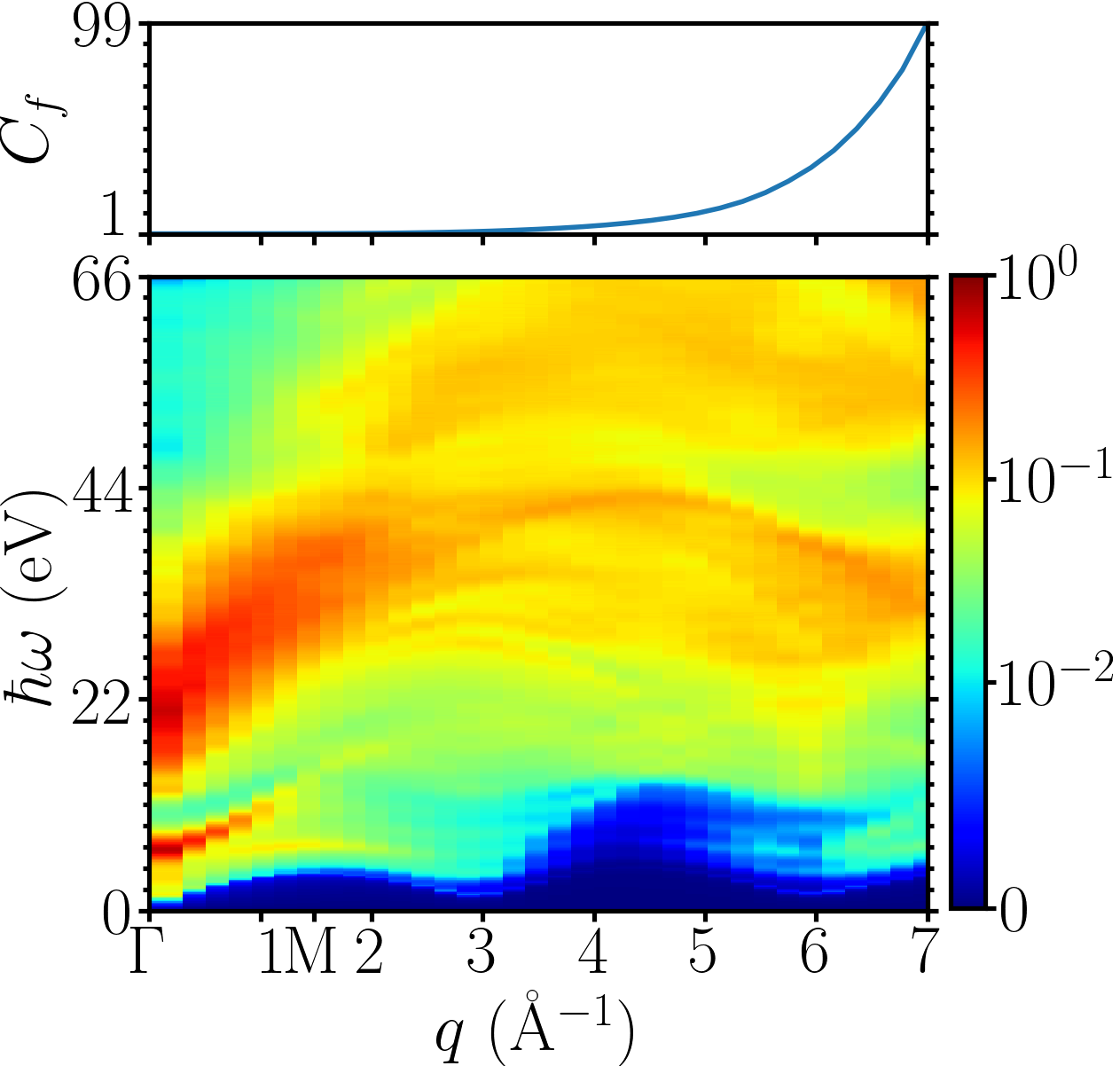}\hspace{0.1cm}		
		\includegraphics[clip,trim={3.76cm} {2.2cm} {5.0cm} 0,height=0.26\textwidth]{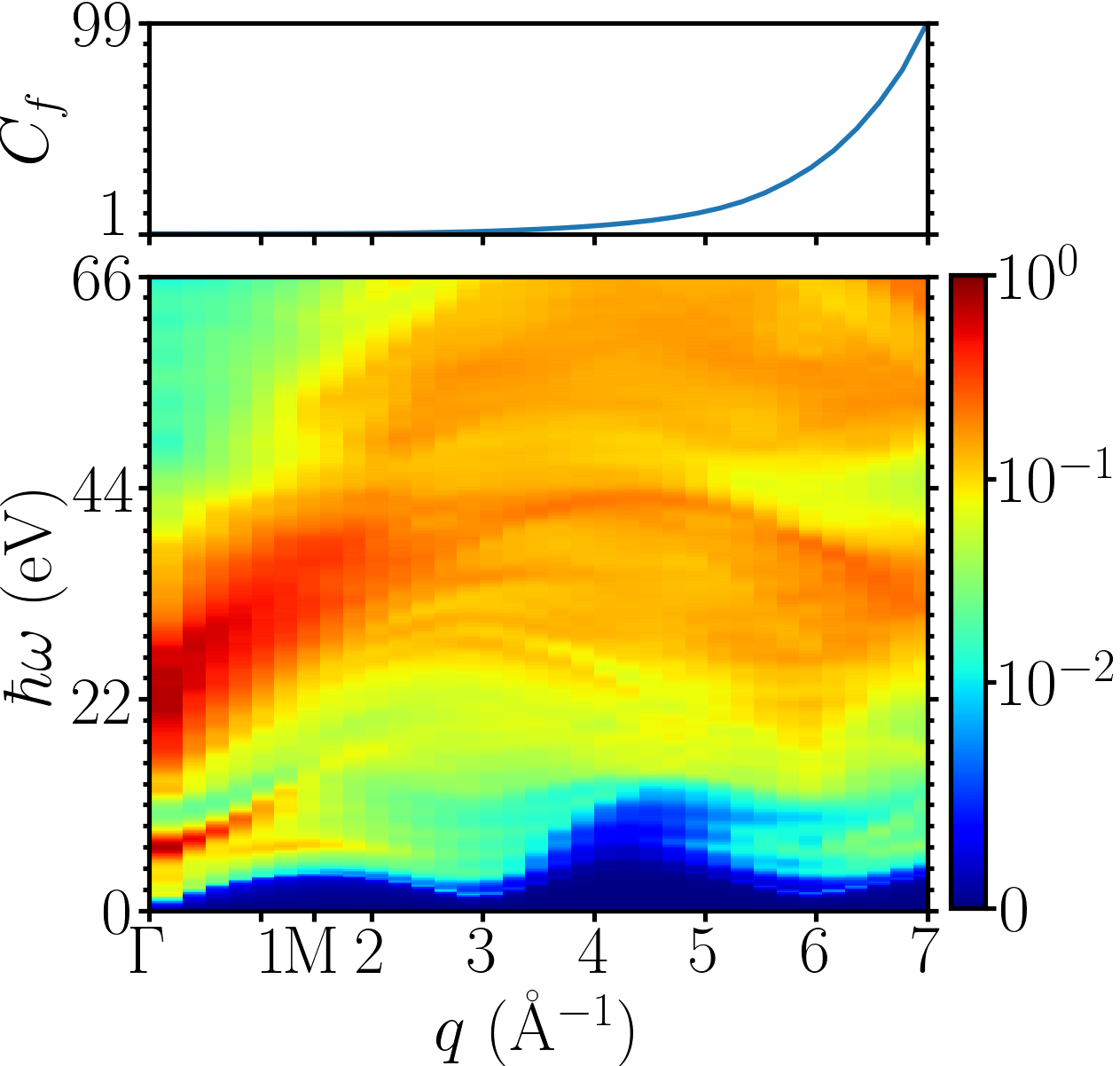}\hspace{0.1cm}
       \includegraphics[clip,trim={3.76cm} {2.2cm} 0 0,height=0.26\textwidth]{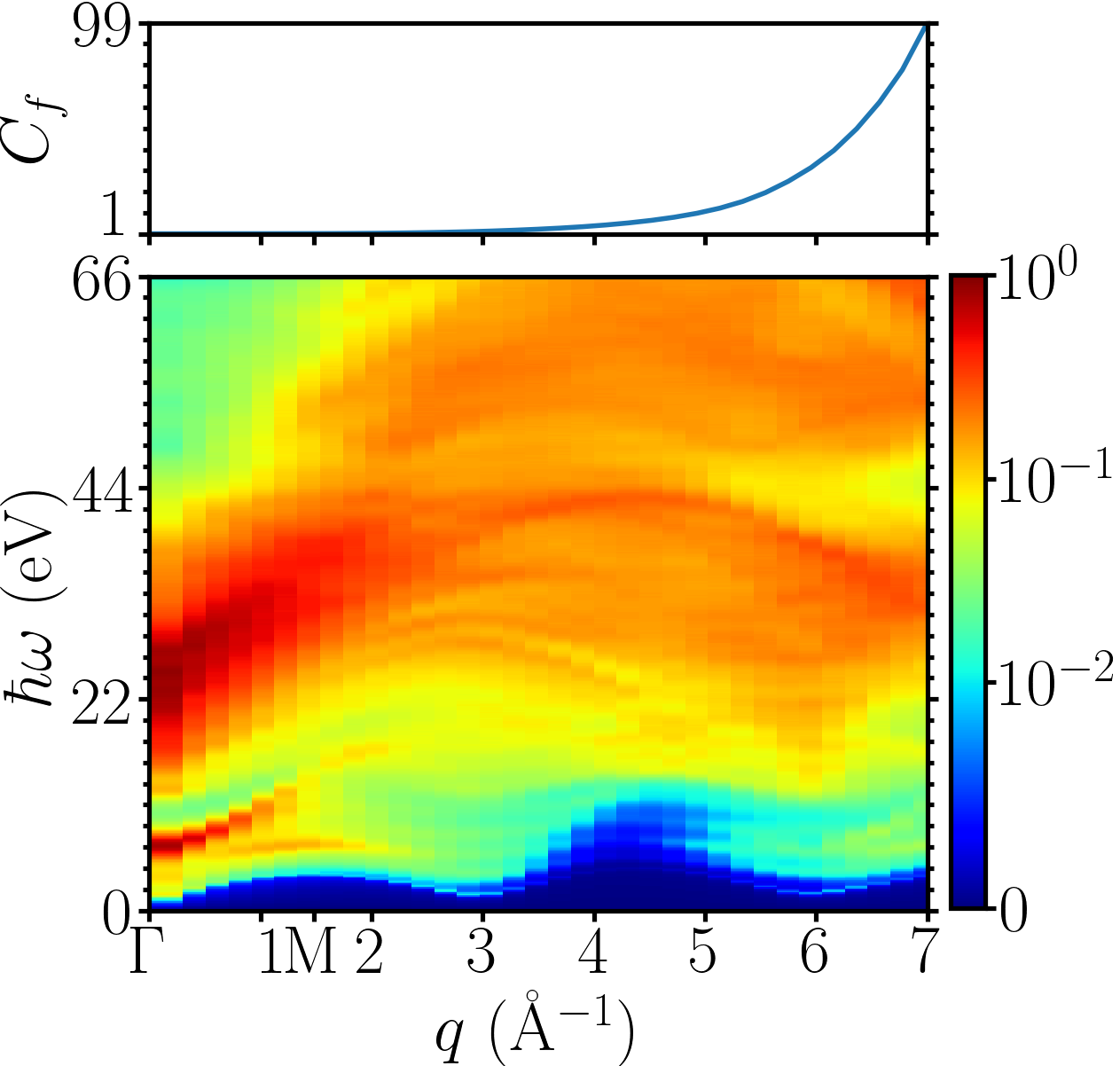}
\\
		\includegraphics[clip,trim=0 0 {5.0cm} 0,height=0.28\textwidth]{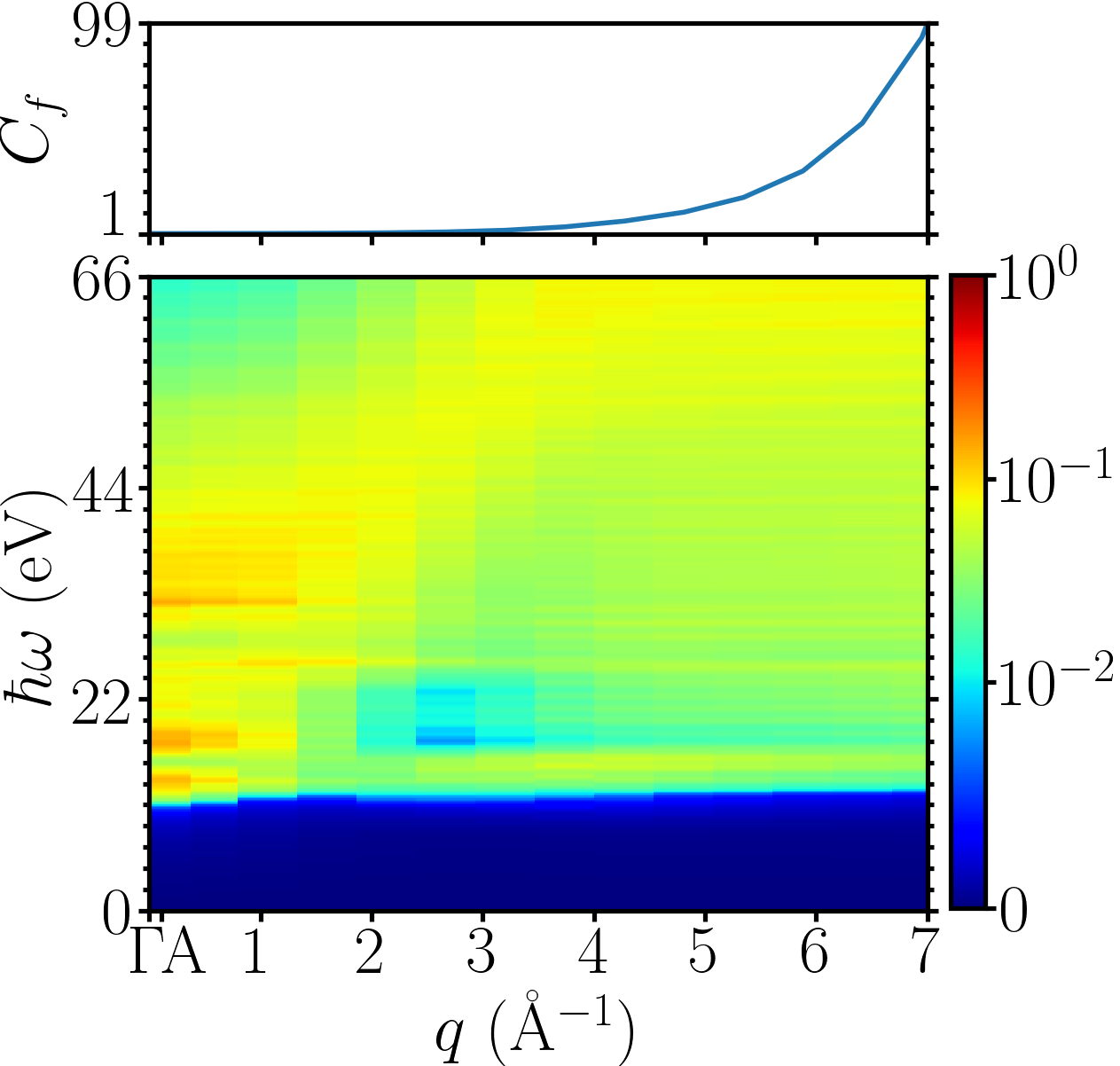}\hspace{0.1cm}
		\includegraphics[clip,trim={3.76cm} 0 {5.0cm} 0,height=0.28\textwidth]{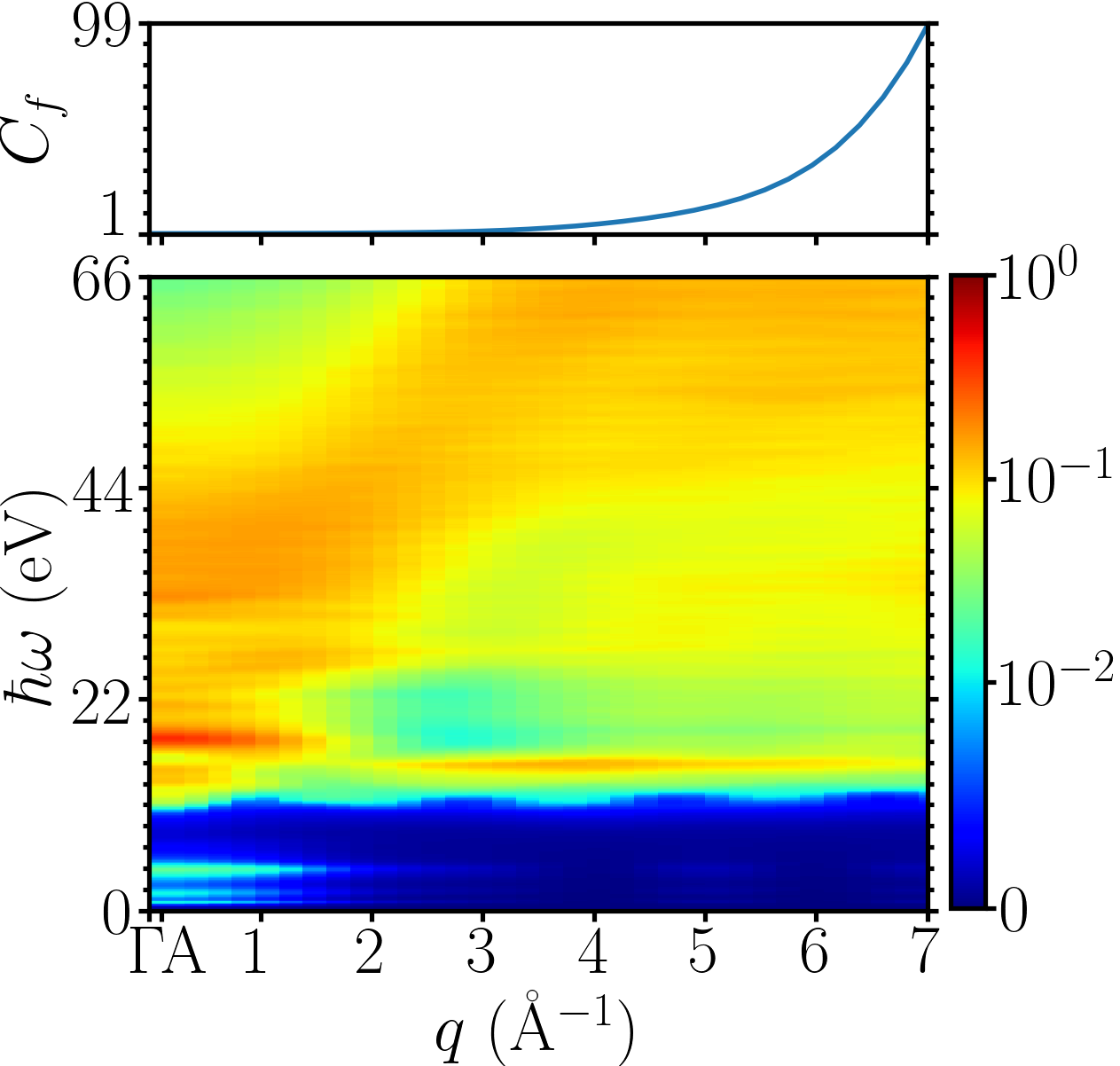}\hspace{0.1cm}
		\includegraphics[clip,trim={3.76cm} 0 {5.0cm} 0,height=0.28\textwidth]{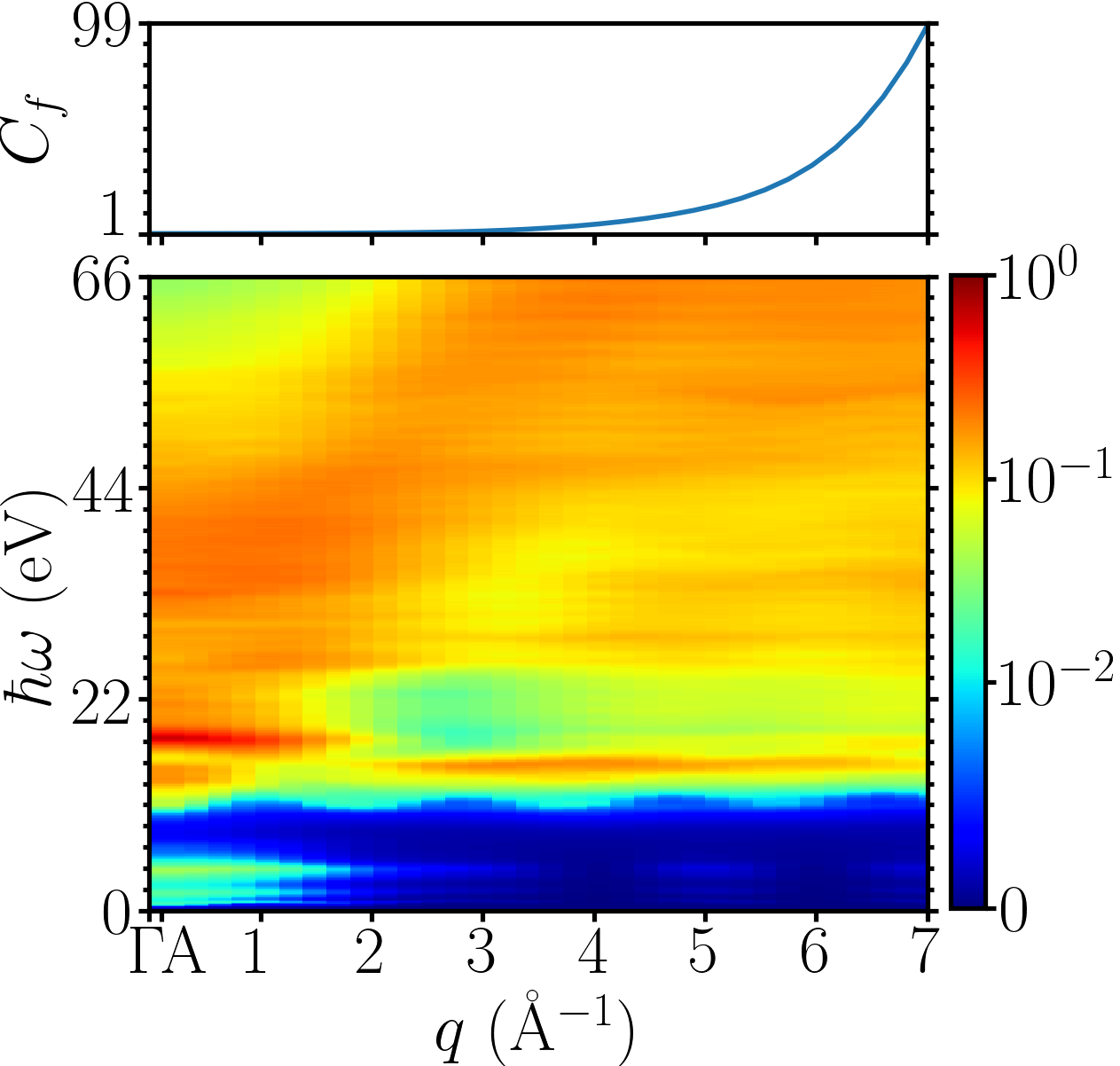}\hspace{0.1cm}
		\includegraphics[clip,trim={3.76cm} 0 0 0,height=0.28\textwidth]{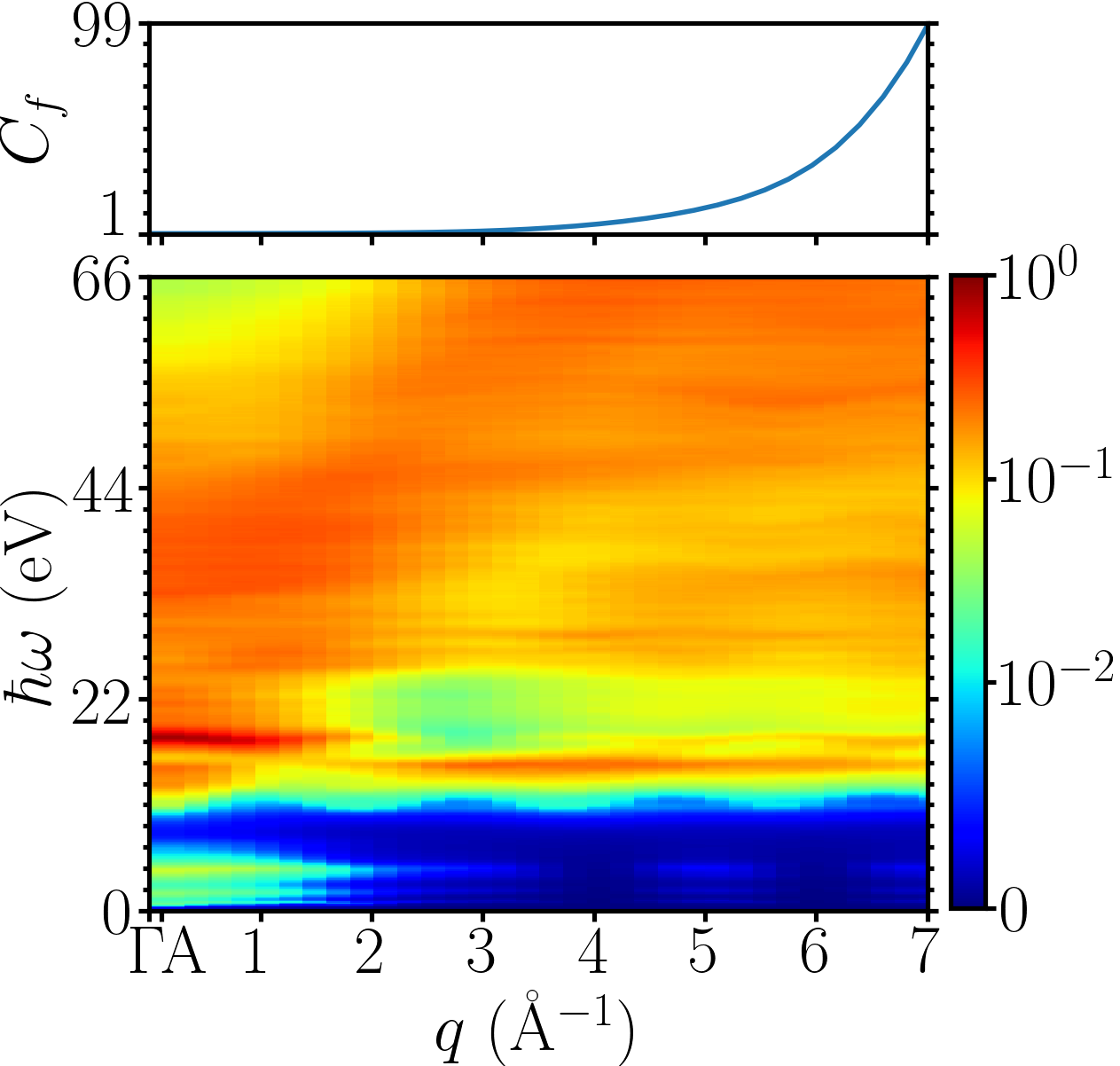}
\caption{Momentum-resolved EELS simulations of the freestanding graphene for one to four layers (left to right) along the \mbox{$\Gamma M$ (top)} and $\Gamma A$ (bottom) directions. The ELF is normalized by $C_f$ from Eq.~\eqref{eqn:f-sumrule}, displayed in the top parts, for the losses $\hbar\omega$ in the range from 
0 to \SI{66}{\eV}  
and the magnitude of MT $q$ in the range from 0 to \SI{7}{\angstrom^{-1}} 
for enhanced visual clarity.
The simulations were obtained using the Liouville-Lanczos approach (\texttt{turboEELS}).}
\label{fig:elf_heatmaps}
\end{figure*}

For completeness, a momentum-resolved EELS simulation of the freestanding graphene is also shown in Fig.~\ref{fig:elf_heatmaps}.
The extended range of energy losses and momentum transfers is obtained using \texttt{turboEELS}. The ELF is normalized using the \mbox{$f$-sum} rule to better visualize the dispersion for large MT, and the normalization multiplicative factor~$C_f$ is shown in the upper parts of the plots in Fig.~\ref{fig:elf_heatmaps}.
The upper part indicates the previously mentioned \mbox{$f$-sum} rule violation. Note that the width of the supercell in the $z$ direction increases with the number of layers to keep the minimum distance from the surface layers to the supercell boundary above the aforementioned threshold necessary for the calculation of reflectivities. 
The density-response function is known to be inversely proportional to the width of the supercell, see, e.g., Ref.~\cite{hambach_dis}. 
However, this effect is outweighed by the increasing number of scatterers with increasing number of layers.
Thus, the losses are higher for a larger number of layers, i.e., as we move from left to right in Fig.~\ref{fig:elf_heatmaps}.
Another fact is apparent from Fig.~\ref{fig:elf_heatmaps} immediately--the optical limit at the $\Gamma$ point would depend on the path chosen to approach the $\Gamma$ point, i.e., the limit is obviously different for the $\Gamma A$ and $\Gamma M$ directions.
This problem is related to the prior mentioned spurious interaction of artificial copies, which will be addressed in Appendix~\ref{sec:compdeta_vopt}. 

\section{Computational details - Optical potential}\label{sec:compdeta_vopt}

There are several ways to deal with this spurious interaction.
One is the selected-$G$ approach; see Ref.~\cite{giorgetti_electron_2020}. 
Another point is raised there, namely the question of the absolute amplitude of the spectrum is still open.
This is crucial for us to get a correctly scaled optical potential.
Another exact approach is presented in Ref.~\cite{nazarov.NJP17.2015}, even with a formula that gives a proper relation between the (quasi) 2D dielectric function and the dielectric function of the fictitious supercell 3D system. However, this formula is only applicable to MT satisfying $qa \ll 1$, where $a$ denotes the width of graphene.
For the large MT of our interest, one is referred to the full version of the theory, Eqs. (37)-(40) in Ref.~\cite{nazarov.NJP17.2015}.
Therefore, a Dyson solver that gives full microscopic information about $\chi_{\mathbf{G}\mathbf{G}'}$ would have to be used.
It is clear from Ref.~\cite{nazarov.NJP17.2015} that ELF of graphene approaches zero in the optical limit, which was also obtained in Refs.~\cite{Despoja_PRB87_2013,Mowbray_PSSB251_2014} using a different approach. 
This assertion has also been corroborated by recent experimental evidence~\cite{Guandalini_CMMH_2024}.
The following approach is adopted to mimic this 2D behavior of graphene ELF and to align the ELF obtained by \texttt{Yambo} with zero ELF at the $\Gamma$ point with the ELF obtained by \texttt{turboEELS}.
Since the ELF will be substituted in Eq.~\eqref{eqn:lifetime.elf.rpa.isotropic} to estimate the optical potential, ELF is set to zero at the $\Gamma$ point and linearly interpolated to the appropriately chosen closest $\bq$ point with nonzero ELF as obtained from \texttt{turboEELS}.
A distance of the nearest nonzero $\bq$-point from the $\Gamma$ point is of the order of one-tenth of $\si{\angstrom}^{-1}$. Reducing this distance by an order of magnitude results in approximately $\SI{10}{\percent}$ higher optical potential.

The trapezoidal integration rule was implemented to obtain the optical potential.
The simulation for the monolayer graphene shows that the dependence of the optical potential in the plane perpendicular to the $\Gamma A$ direction is nearly isotropic. 
Hence, the optical potential for 2--4 layers is simulated only along $\Gamma M$ and $\Gamma A$.
Based on this (an)isotropy, we propose the following interpolation formula, which combines the optical potentials obtained from Eq.~\eqref{eqn:lifetime.elf.rpa.isotropic} with the $\Gamma A$ loss function, $V_{\mathrm{opt}}^{\Gamma\mathrm{A}}$,
and 
with the $\Gamma M$ loss function, $V_{\mathrm{opt}}^{\Gamma\mathrm{M}}$, into one formula
\begin{equation}
V_{\mathrm{opt}} = \frac{1}{3}V_{\mathrm{opt}}^{\Gamma\mathrm{A}}+\frac{2}{3}V_{\mathrm{opt}}^{\Gamma\mathrm{M}}.
\label{eqn:interpolation_formula}
\end{equation}
The reasoning behind this interpolation formula is based on the division of the reciprocal space into two regions using a double cone. 
The axis of the cone is formed by the $\Gamma A$ direction, the 
$\bq$ points closer to the cone axis than to the $M\Gamma K$ plane
form the first region (interior of the cone) and the second region is its complement. 
The volumes of these two regions have the ratio $\frac{1}{3} : \frac{2}{3}$, which leads to the coefficients in Eq.~\eqref{eqn:interpolation_formula}.

Spectra were computed in two different supercells of different widths~$L$ for several $\bq$ points.
Applying scaling to ELF by the multiplicative factor $L/(n a)$ for $n$-layer graphene, spectra were in good agreement with each other.
The ``thickness'' of one graphene layer $a$ is an ambiguous parameter, which was fixed 
by normalizing the $V_{\mathrm{opt}}$ per layer for four-layer graphene to the optical potential of the graphite overlayer (at the loss \SI{23}{\eV} $V_{\mathrm{opt}} = \SI{1.25}{\eV}$) as presented in Ref.~\cite{barrett_elastic_2005}.

\section{Computational details - Resources}\label{sec:compdeta_resources}

Computations using \textsc{Quantum ESPRESSO}, \texttt{Yambo}, and \texttt{turboEELS} packages were performed on two systems installed at Czech Metrology Institute.
The older system consisted of 36 nodes each having two 8-core Intel Xeon E5-2650 processors running at 2.6 GHz.
Each node was equipped with a FDR Infiniband interconnect running at 54 GBit/s.
The total amount of RAM of the system was 4~TB. The operating system was the SUSE Linux Enterprise Server 12 SP1 running in a ScaleMP vSMP infrastructure.
A subset of calculations was performed on a system with a cluster consisting of 20 compute nodes, a big memory node, a GPU node, and a login node.
Each compute node used for calculation was equipped with two 32-core AMD EPYC 7543 processors running at 2.8 GHz, 512 GB RAM and a HDR Infiniband interconnect running at 200 GBit/s.
The operating system was Rocky Linux 9.2, with Warewulf and Slurm used for cluster and workload management.
A typical parallel calculation used 64 cores. 
The following python libraries were used for data processing: NumPy~\cite{harris2020array}, SciPy~\cite{2020SciPy-NMeth}, Matplotlib~\cite{Hunter:2007}.

\bibliography{abc.bib}

\end{document}